\documentclass[twoside,fleqn]{ActaStyle}
\usepackage[colorlinks=true, pdfstartview=FitV, linkcolor=black,
            citecolor=black, urlcolor=black]{hyperref}
\usepackage[dvips]{graphicx}
\usepackage{amsthm}
\usepackage{amsmath}
\usepackage{amssymb}
\usepackage{color}
\usepackage{framed}
\usepackage{hyperref}

\usepackage{fancyhdr}
\renewcommand{\subsectionmark}[1]{}
\newtheorem{definition}{Definition}
\newtheorem{theorem}{Theorem}

\renewcommand{\thesection}{\arabic{section}}
\renewcommand{\thesubsection}{\thesection.\arabic{subsection}}

\renewcommand{\thefigure}{\thesection.\arabic{figure}}
\renewcommand{\thetable}{\thesection.\arabic{table}}
\newcommand{\ket}[1]{\vert{#1}\rangle}
\newcommand{\bra}[1]{\langle{#1}\vert}

\def\authorheading{Matej Pivoluska and Martin Plesch}
\def\shorttitle{Device independent random number generation}

\begin{document}

\pagerange{600}{663}
\title{Device Independent Random Number Generation}

\author{Matej Pivoluska$^{a,}$\email{pivoluska@fi.muni.cz} and Martin Plesch$^{a,b,}$\email{martin.plesch@savba.sk}}
{$^{a}$ Faculty of Informatics, Masaryk University, Brno, Czech Republic\\
 $^{b}$ Institute of Physics, Slovak Academy of Sciences, Bratislava, Slovakia}

\abstract{Randomness is an invaluable resource in today's life with a broad use reaching from numerical simulations through randomized algorithms to cryptography. However, on the classical level no true randomness is available and even the use of simple quantum devices in a prepare-measure setting suffers from lack of stability and controllability. This gave rise to a group of quantum protocols that provide randomness certified by classical statistical tests -- Device Independent Quantum Random Number Generators. In this paper we review the most relevant results in this field, which allow the production of almost perfect randomness with help of quantum devices, supplemented with an arbitrary weak source of additional randomness. This is in fact the best one could hope for to achieve, as with no starting randomness (corresponding to no free will in a different concept) even a quantum world would have a fully deterministic description. }

\vspace{0.3cm}

\begin{minipage}{2.5cm}
\quad{\small {\sf KEYWORDS:}}
\end{minipage}
\begin{minipage}{9.5cm}
Quantum Information, Bell Inequalities, Device Independent Certification,
Non-locality, Violation of Local Causality, Randomness
\end{minipage}

\newpage
\tableofcontents

\setcounter{equation}{0} \setcounter{figure}{0} \setcounter{table}{0}\newpage
%%%%%%%%%%%%%%%%%%%%%
\section{Introduction}
%%%%%%%%%%%%%%%%%%%%
%
Randomness is one of the key concepts of modern science, finding many applications in both hard and soft sciences.
At the same time, it is a very controversial topic.
This fruitful controversy comes from the fact that the notion of ``randomness'' is not universally and uniquely defined
and in different fields of science can mean different things. In fact, it is even not clear whether
randomness shall be considered as an objective fact. In such a case there shall exist fundamentally unpredictable
processes in nature, which, if used correctly, could serve as perfect randomness generators.

The other approach says that randomness as such is only a subjective concept representing
incomplete knowledge about a process or system. In this approach no perfect randomness and randomness generators exist.
However it still makes sense to speak about randomness perceived by a given observer -- 
even a perfectly deterministic process can
be seen as perfectly random by an observer not having access to underlying information about the process.

In spite of this ambiguity, randomness has been shown to be an important resource with a variety
 of applications such as statistical sampling, numerical simulations,
algorithm design \cite{MotwaniRaghavan-Randomizedalgorithms-1995} and cryptography
\cite{MenezesHandbook} to name a few.

Random statistical sampling is a common way to avoid bias in deductions in cases such as marketing polls
or clinical trials of drugs. Sample space is here usually taken as small as possible, as it is directly connected with 
costs of the survey or the time needed for clinical test. It is thus crucial to cover homogenously different parameters 
of the underlying set with as few instances as possible. Even more, many parameters are not easily accessible and 
thus directly disallow proper selecting of the best sample space. Perfect randomness, if used correctly, was proven 
to overcome this complications in most of the common cases.

Numerical simulations of real world phenomena use randomness to
predict processes that are too complicated to be fully simulated. Common example are the weather forecasts -- as a chaotic phenomenon the reliability of their results exponentially depends on the preciseness of the starting point and the simulation itself. Perfect randomness helps to choose sample space for the simulation that brings the best results.

A slightly different example for utilization of randomness are \emph{randomized algorithms}. These involve a 
randomized component in their design and often have better performance and are easier to develop and analyze
in comparison to their deterministic counterparts.

However, the usefulness of randomness is perhaps most evident in the case of cryptography.
In order to break the symmetry between legitimate users of cryptographic protocols
and potential adversaries, the legitimate users have to be given some advantage.
The advantage is typically modeled as the knowledge of some randomly generated secret.
This is the reason why many cryptographic tasks, such as encryption (both the private and public key),
secret sharing or bit commitment require randomness for each use.

Typically, in most of the applications an access to a \emph{perfect random source} --
uniformly distributed bits independent of any other existing data -- is assumed. This assumption
is silently hidden in the analysis of the performance of algorithms and protocols -- random
number generators are assumed to produce uniform randomness.

Unfortunately, it turns out to be very difficult to show that given source of randomness is
sufficiently unpredictable and adding a requirement of
uniform distribution of its outputs given any other existing data seems downright
impossible. Consequently, this fact raises an important question of
whether or not \emph{weak random sources} --  non-uniform random processes
only partially independent of other relevant data -- can be effectively used
in different applications.

The problems that arise when one is forced to use a weak source of randomness
are well identified and have been extensively studied in classical information
processing. It has been shown that some information processing
tasks can be realized reasonably well with bounded weak sources of randomness
\cite{VaziraniVazirani-RandomPolynomialTime-1985, RennerWolf-Unconditionalauthenticityand-2003}.
However, many other tasks are infeasible without an access to an almost perfect random source
\cite{McInnes,DodisSpencer-(non)universalityofone-time-2002}.
Yet again, cryptography is the best example of how considering the use of weak sources
instead of nearly perfect sources of randomness can change
the picture. Already the classical result of Shannon \cite{Shannon49} shows that to obtain \emph{perfect secrecy}
in private key cryptography, communicating parties have to use perfectly random key.
Later it was analyzed whether at least some amount of secrecy can be salvaged in private key
encryption scenario with weakly random keys. The outcome of this effort was
that to guarantee secrecy, one has to use secret keys that are almost perfectly random
\cite{McInnes,BosleyDodis-DoesPrivacyRequire-2006}.
Even in practice, many security holes in the existing implementations of cryptographic protocols
can be traced back to imperfect random number generators. For example,
it has been estimated that around two out of every thousand RSA moduli used on the Internet
are insecure, as they share a factor with another RSA key \cite{LenstraHughesAugierEtAl-Ronwaswrong-2012}.
This points to an imperfect random number generator
used for the generation of large prime numbers used as the private key in RSA encryption.

However, up until recently (see \textit{e.g.} \cite{BoudaPivoluskaPleschEtAl-Weakrandomnessseriously-2012,
HuberPawlowski-Weakrandomnessin-2013}) 
there has been little analysis of the impact of weak randomness in
quantum information processing (QIP).
This is a surprising fact,  since randomness plays a vital role in a variety of quantum protocols.
A possible reason for this lack of research seems to be the
Copenhagen interpretation of Quantum theory, by which randomness is an objective
property of each quantum system.
The simplest manifestation of this randomness is a projective measurement of a qubit being
in a perfectly balanced superposition of two canonical basis states.
Outcome of such a measurement
is considered as fundamentally undetermined -- the outcomes are chosen at random during the measurement.
Therefore perfect randomness is essentially seen as an ever present and free resource in QIP.
In fact this line of thoughts culminated in commercially available quantum random number generators (QRNG)
\cite{-IDQuantique:-}.

In practice, however, perfect randomness cannot be expected even from measurement-based quantum
random number generators. What one can reasonably guarantee is only a relatively high entropy
of the outcomes of QRNG, which then requires post-processing \cite{Solcgravea-TestingofQuantum-2010,
FrauchigerRennerTroyer-Truerandomnessfrom-2013}.
Moreover, it has recently been shown that even the state of the art QRNGs
do not pass certain standard statistical tests for randomness \cite{Jakobsson-TheoryMethodsand-2014}.

Worse still, the relatively limited weakness of random bits produced by some implementations of QRNGs
can become much more severe if the QRNG is deliberately attacked by an adversary.
Such attacks range from changes of the device temperature, which affects the laser wavelength, 
leading to biased beam splitters, right through voltage changes in the electricity input.

For this reason we have to weaken the assumptions and ask,
if the randomness production is possible without knowing the precise specification
of the quantum devices.
Quantum protocols with unspecified black-box devices are called \emph{device independent protocols}.

The problem of device independent randomness production was first studied
in a setting, where the user of the protocol
starts with a uniformly distributed string and uses black-box quantum devices to produce
random string as large as possible -- hence the name -- \emph{device independent randomness
expansion}\cite{ColbeckKent-Privaterandomnessexpansion-2011,PironioAc'inMassarEtAl-Randomnumberscertified-2010,
vazirani2012certifiable,CoudronYuen-InfiniteRandomnessExpansion-2014,MillerShi-RobustProtocolsSecurely-2014}. 
The main idea behind using black-box devices for randomness expansion
is to use random bits as inputs and collect the outputs of the devices.
The fact that input-output statistics violate certain kind of
inequalities -- called Bell-type inequalities -- can certify that inner workings
of the black-boxes are genuinely quantum and therefore the outcomes are
fundamentally random.
We describe here the most relevant results in this area, allowing in the most elaborated setting an unbounded expansion. We also discuss a slightly stronger limitations for the adversary that allow easier and more efficient production of randomness.

Other group of protocols is devoted to a different version of device independent randomness expansion, often called also \emph{randomness amplification}\cite{ColbeckRenner-Freerandomnesscan-2012,GallegoMasanesEtAl-Fullrandomnessfrom-2013,BrandaoRamanathanGrudkaEtAl-RobustDevice-IndependentRandomness-2013,
BoudaPawlowskiPivoluskaEtAl-Device-independentrandomnessextraction-2014,
ChungShiWu-PhysicalRandomnessExtractors-2014}. 
Here the starting randomness is not provided as a short perfectly random string, but rather as a source of partially random bits. This source does not provide a uniform distribution, but their outputs fulfill some criteria given by conditional probability of individual output bits or entropy of the outcome as a whole.
Here we also review relevant work in the area, concluding with a protocol that can amplify any source of randomness
(\textit{i.e.} source that doesn't have deterministic outcome), at the cost of unbounded number of devices used.

The paper is organized as follows. In the second Section we define the notation and measures of weak random sources and present the results from classical theory. In the Section \ref{sec:Bell} we relate Bell inequalities and randomness production.
Sections \ref{sec:RandExp} and \ref{sec:RandAmplif} include the results for randomness expansion
and amplification using quantum devices, concluded with remarks in Section \ref{sec:Conclusion}.

\setcounter{equation}{0} \setcounter{figure}{0} \setcounter{table}{0}\newpage
%%%%%%%%%%%%%%%%%%%%%%%%%
\section{Preliminaries}
%%%%%%%%%%%%%%%%%%%%%%

This section serves as a technical introduction for the two main sections of this paper. We start by
formally defining the notion of weak random sources together with a short history of their most
studied types. In Subsection \ref{sec:Extractors}, randomness extractors -- algorithms for post-processing weak
random sources into almost perfect ones are introduced.
The last Subsection \ref{sec:QWS} discusses weak random sources
in the presence of quantum side information. Quantum proof extractors are shortly discussed as well.

%%%%%%%%%%%%%%%%%%%%%
\subsection{Notation}
%%%%%%%%%%%%%%%%%%%%%%%%%%%%%%%%%
In the rest of this paper
random variables are denoted by capital letters ($X,Y,Z,\dots$) and their domain by corresponding
calligraphic letters ($\mathcal{X},\mathcal{Y},\mathcal{Z},\dots$). Outcomes of these random variables
are then denoted by lower case letters $(x,y,z,\dots)$.
Uniform distribution over $s$-bit strings is denoted $U_s$.
The probability of the random variable $X$ having value $x$ is denoted
$\Pr[X=x]$ especially in the introductory part of the paper, later to shorten the notation
we use $p_X(x)$ and sometimes, when the random variable in question is clear from the context
we even drop the subscript and use simply $p(x)$. This notation extends to conditional 
probabilities and $\Pr[AB=ab\vert XY=xy]$ is often denoted $p_{AV|XY}(ab\vert xy)$ and sometimes even
$p(ab\vert xy)$,
mainly in the context of randomness production protocols, which is in accordance with the convention
used in the most of the relevant literature on the topic.

The reader is expected to be familiar with basics of quantum mechanics.
If this is not the case, introductory chapters of any of the well-known textbooks
on quantum information  \cite{NielsenChuang-QuantumComputationand-2004,Gruska-QuantumComputing-1999}
should be consulted.

%%%%%%%%%%%%%%%%%%%%%%%%%
\subsection{Introduction to weak randomness}
%%%%%%%%%%%%%%%%%%%%%%%%%

In this subsection we will first introduce the notion of weak randomness generally and then proceed
to a short review of  the most studied types of weak randomness sources.

We will take an operational approach and formally define randomness by random variables.
Because the nature of the randomness bias of the source is typically unknown,
it is insufficient to define a weak source by a random variable $X$ with a given probability distribution $p_X$.
Instead, we model weak randomness by a random variable with unknown probability distribution.
To guarantee at least some randomness we suppose that the probability distribution $p_X$
of the variable $X$ comes from a set ${\mathcal S}$; the level of randomness is then given by the properties of the set, or more specifically, by the property of the least random probability distribution(s) in the set.

If we say that a protocol or an algorithm uses randomness from a weak source $X$ of a given type, we mean that the
randomness is distributed according to an arbitrary (unknown) distribution ${p_X\in\mathcal S}$.
Analysis of a protocol or an algorithm with weak randomness then boils down to proving some desired
property in the worst case scenario. That is finding a set of distributions $\mathcal{W}\subseteq\mathcal{S}$,
for which the protocol or algorithm manifest the worst case performance,
followed by the proof of the desired property with the assumption that the randomness is distributed
according to a probability distribution $p_X\in\mathcal{W}$.
%This is especially well motivated in cryptography, where
%the security of the protocols has to be guaranteed in every possible scenario, as opposed to
%computational tasks, where often average complexity is sufficient.

Different types of weak randomness differ in the definition of the set $\mathcal{S}$.
The set $\mathcal{S}$ is usually given by a specific property
of allowed distributions, often motivated by the properties of the physical source, but in principle
any set of probability distributions can be seen as a weak randomness source.

Alternative way to arrive at the definition of a weak source $X$ by a set of possible
probability distributions $\mathcal{S}$ is to consider another randomness variable $E$ interpreted
as the information the relevant entity has about $X$.
%More precisely, because the classical physics is deterministic, randomness represents only our incomplete
%knowledge about the process in question.
 As an example consider a user
running statistical tests to determine the quality of his random number generator. If the generator passes
all the tests, then, according to the users knowledge, the random variable $X$ describing the outcomes of the
generator is very close to uniformly distributed.
However, in most applications the user's knowledge about the random number generator
is irrelevant. The random number generator output is typically required to be random against specific
entity -- examples being the adversary in cryptographic scenarios, or input data in randomized algorithms.
Statistical tests therefore correctly asses the usefulness of randomness only with an assumption
(which is very reasonable most of the time) that these
relevant entities do not have more precise information about the random number generator than the user.

The notion of weak randomness questions this assumption. In fact, the entities for whom the random data
used in applications needs to be unpredictable are assumed to have more information about it than the user.
This information can be characterized by a random variable $E$ and
the set $\mathcal{S}$ that contains distributions $p_{X\vert E=e}$ for each $e$. As the user doesn't know
the concrete piece of information $e$ and therefore the concrete distribution $p_{X\vert E=e}$ the source $X$ has
from the adversarial point of view, the algorithms or protocols have to work correctly with all of them.
Note that a string can be random for some entities and completely known for another entities.
Therefore, in this view randomness is not a property of a string, rather is it a subjective property of
a process creating the string. This is in a sharp contrast with algorithmic view of randomness
\cite{LiVitnyi-IntroductiontoKolmogorov-2008}, which
can be seen as measure of string's compressibility.

The view of weak randomness sources as randomness sources with side information available
to the adversary, although nicely illustrating the motivation behind the notion of weak randomness,
is rarely used in the classical (as opposed to quantum) literature about weak sources.
In fact, usually a definition presented before -- the adversary knows the distribution
$p_X\in\mathcal{S}$ of $X$  and the user doesn't -- is used in vast majority of the classical literature.
In fact however, these two notions are completely equivalent in classical world and if one would go through
a trouble of rephrasing all the classical results into the formalism where conditioned probability
distributions are used instead of the unconditioned ones, one would obtain identical results.

The situation gets much more complicated, when one acknowledges the existence of quantum mechanics.
According to the Copenhagen interpretation of quantum mechanics, randomness
is an objective property and there exist genuinely unpredictable events. Therefore, at least in theory, a device
producing objectively random outcomes -- \textit{i.e.} there exists no information in the
Universe that would help us predict them  -- can be constructed.
However to construct such device we would need perfect control over the quantum
devices -- a feat that is presently
not possible. Therefore, in practice our view of randomness produced by quantum devices is a mixture of two
qualitatively different types of randomness -- objective randomness coming from a genuinely unpredictable quantum
process and classical randomness coming from our imprecise implementation of these quantum processes.
Alternatively, quantum random number generators might leak the information about their outputs after
they have been produced, \textit{i.e.} during the post processing phase. Another possibility is that
the adversary might be a part of the process
producing the outcomes, as in the case of some applications like quantum key distribution
\cite{bb84,Renner2005a}.
In the light of this discussion, the view that the adversary might have more information about the random
process than the user is still valid even in the setting with quantum mechanics.

Another complication that comes with quantum mechanics in the picture is that the adversary is allowed
to hold quantum information (\textit{i.e.} quantum state $\rho_{E}$ instead of classical random variable $E$)
about $X$. It has been shown that such adversary is in some cases stronger than the adversary holding
classical information only. This is the reason why the side information understanding of the weak sources
is prevalent in quantum literature. We will define and discuss weak sources with quantum side information
in Subsection (\ref{sec:QWS}).

Another topic discussed in this section is the most common approach to tackle with the problem of weak randomness.
Because perfect randomness is expected by most of the applications, it is natural to attempt to
post--process  weak randomness into nearly perfect randomness, which can be subsequently
used in algorithms and protocols.
This process is called \emph{randomness extraction} and it is widely studied for all types of weak sources.

With respect to this task, we can divide weak sources of randomness into two classes -- extractable sources
and non-extractable sources. From extractable sources one can obtain by a \emph{deterministic} procedure
nearly perfect randomness. Although various examples of non-trivial extractable sources do exist
(see Subsection \ref{sec:VNS} or \cite{Trevisan,Kamp2} and references therein), most natural sources, for example defined
by the entropy of allowed distributions (see Subsections \ref{sec:SVS} and \ref{sec:MESources}), are non-extractable.
In such cases \emph{non-deterministic} randomness extractors (see Subsection \ref{sec:Extractors} ) have to be used.

In what follows we will introduce some of the most studied types of weak randomness sources.

%%%%%%%%%%%%%%%%%
\subsubsection{Von Neumann sources}\label{sec:VNS}
%%%%%%%%%%%%%%%%%

Historically, the first consideration of weak random source is due to von Neumann~\cite{vonNeumann}.
The so called von Neumann source produces a string of equally biased independent coin flips.

\begin{definition}[Von Neumann source]
The von Neumann source is defined as a sequence $X_1,X_2,\dots$ of binary independent
random variables with fixed but unknown bias. That is,
$\forall i\in\mathbb{N},\Pr[X_i=0] = \varepsilon$ and $\Pr[X_i=1] = 1-\varepsilon$,
for some (unknown) $0\leq\varepsilon\leq 1$.
\end{definition}

\begin{figure}[tb]
\centering
\includegraphics[width=11cm,clip]{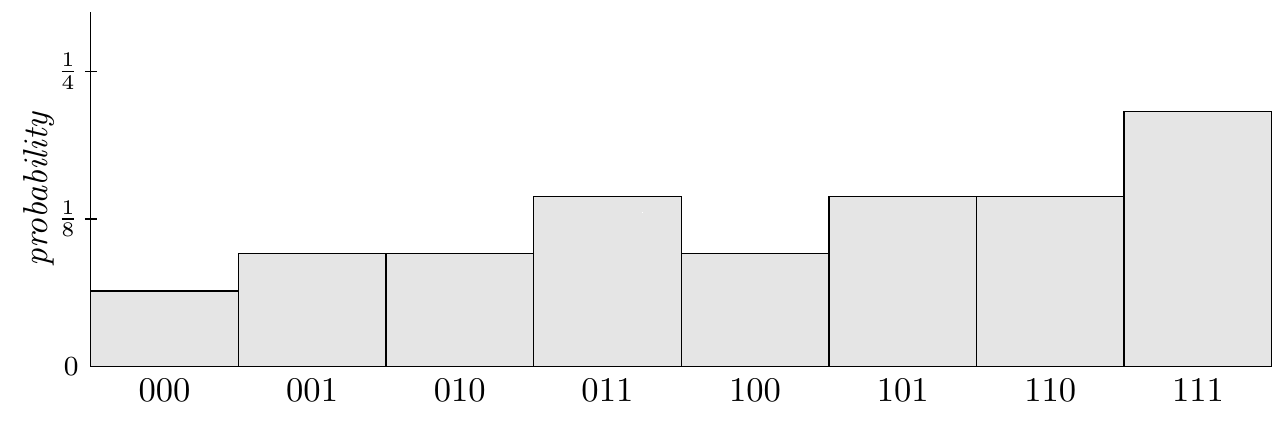}
	\caption{\it\small An example of distribution of von Neumann source.
This concrete distribution has $\varepsilon = \frac{2}{5}$. The other distributions from the source have different values of $\varepsilon$.
However, for each distribution from a von Neumann source it holds that strings with the equal number
of 0's appear with the same probability.}
	\label{fig:VNS}
\end{figure}

The parallel with our general definition of the weak sources as sets of probability distributions over the same domain
is clear, when we consider $n$-bit string produced by a von Neumann source.
Such string can be described by a set of joint random variables
$Y = X_1\dots X_n$, characterized by a parameter $\varepsilon$ (see Fig.~\ref{fig:VNS}).

\begin{figure}[!bp]
\centering
\includegraphics[scale = 0.8]{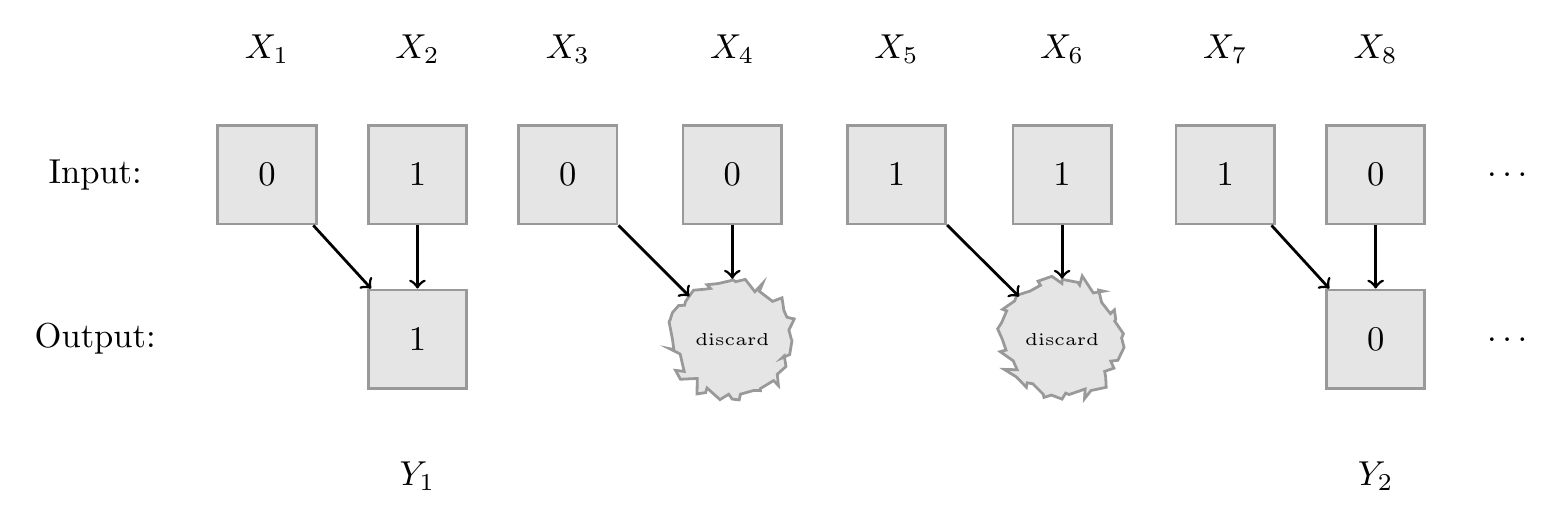}
\caption{\it\small A schematic drawing of von Neumann extractor. Random output bits are generated from pairs of input bits. Pairs with equal values are discarded, while the second bit of a non-equal pair is used as the output.}
\label{fig:vonNeumann}
\end{figure}

In his paper, von Neumann designed a deterministic procedure to extract random bits from any von Neumann source.
The procedure takes outputs of neighboring random variables $X_{2m-1}$ and $X_{2m}$ and compares them.
If they are equal, they are both discarded, otherwise $X_{2m}$ is added to the output string.
Formally, von Neumann extractor $Ext$ is defined as follows:
$$
Ext(X_1,\dots,X_n) = Y_1,\dots Y_k,
$$
where $Y_i = X_{2m_i}$ and $m_1<m_2<\dots<m_k$ are all indices $m<n$ such that $X_{2m-1}\neq X_{2m}$
(see Fig.~\ref{fig:vonNeumann}).

The von Neumann extractor has all the expected properties of randomness extractors.
First of all, the output bits are uniformly distributed for any bias $\varepsilon$.
This is quite straightforward to see, because
$\Pr[X_i = 1 \wedge X_{i+1} = 0] = \varepsilon(1-\varepsilon) = \Pr[X_i = 0 \wedge X_{i+1} = 1]$.
Secondly, independence of the output bits is implied by the independence of input bits.

One of the most important drawbacks of this simple procedure is that it discards non-negligible portion of randomness.
The procedure was later improved by Peres \cite{Peres-IteratingVonNeumann's-1992}, who showed how to efficiently extract
the amount of random bits that is close to the entropy of the source by exploiting the discarded bits.

The von Neumann source is rarely regarded in the literature nowadays. The reason for this is that the
assumption of independence of bits $X_i$ is considered to be very strong and unrealistic.
However the von Neumann source is worth mentioning, because it conveys one
important message:
The set of probability distributions in the von Neumann source is uncountable
($\varepsilon$ is a real parameter), yet the source is extractable.
This hints on the fact that in the question of extractability the size of the source is less
relevant than it's structure and the structure of von Neumann source is too strong in this sense.

%%%%%%%%%%%%%%%%%%%
\subsubsection{Santha-Vazirani sources}\label{sec:SVS}
%%%%%%%%%%%%%%%%%%%
More general notion of weak sources randomness are so called Santha-Vazirani sources
(\emph{SV-sources}) \cite{SanthaVazirani-Generatingquasi-randomsequences-1986}.

\begin{definition}[Santha-Vazirani source]\label{def:SVsource}
A Santha--Vazirani source with a parameter $0\leq\varepsilon\leq\frac{1}{2}$ is defined as a sequence of
binary random variables $X_1,X_2,\dots$, such that
\begin{align*}
\forall i\in\mathbb{N}, \forall x_1,\dots,x_{i-1}\in\{0,1\},
\quad  \\
\Pr\left[X_i = 1|X_{i-1}=x_{i-1}\dots X_{1}=x_1\right]\in \left\langle\frac{1}{2}-
\varepsilon,\frac{1}{2}+\varepsilon\right\rangle.
\end{align*}
\end{definition}

\begin{figure}[tb]
\centering
\includegraphics[width=11cm,clip]{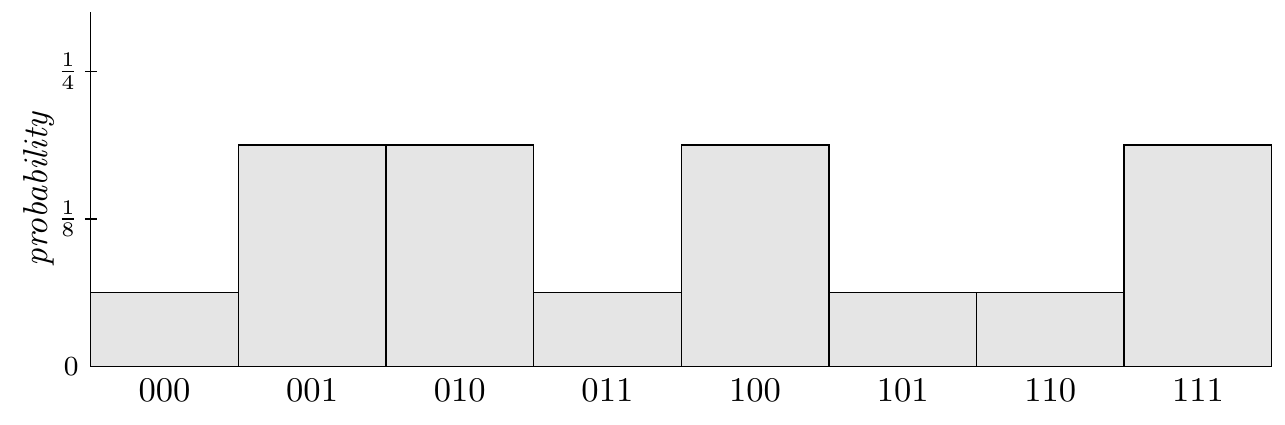}
	\caption{\it\small An example of a distribution of a SV-source with $n = 3$ and $\varepsilon = \frac{1}{4}$.
This example maximizes the probability of outputting
a string with odd parity. Note that each string appears with non-zero probability.}
\label{fig:SVS}
\end{figure}

Note that in this model the bias can change for each bit $X_i$ to some extent, and what is more,
the bias can depend on previously generated bits.
Informally, we suppose that each bit contains some amount of randomness (bounded from below) even
conditioned on the previous ones.
Here again, for fixed $n$ and $\varepsilon$, SV-source can be interpreted as a set of probability distributions over
$n$-bit strings (see Fig. \ref{fig:SVS}). However, the restriction on the allowed distributions is much less stringent
comparing to the von Neumann source
due to the allowed correlations between bits. For example, for $\varepsilon = \frac{1}{2}$ every probability distribution
over $n$-bit strings is in the set of allowed distributions $\mathcal{S}$. On the other hand, SV-source
with $\varepsilon = 0$ contains only a single distribution -- the uniform distribution $U_n$.
Another important property that we will use in
the Section \ref{sec:RandAmplif} is that for SV-sources with $\varepsilon < \frac{1}{2}$ every $n$-bit string appears with
non-zero probability.

Because the SV-sources are less restricted than von Neumann sources, they are
more suitable for modeling real world random number generators.
The price to pay are several negative results that have been proven for Santha-Vazirani soures.
Santha and Vazirani \cite{SanthaVazirani-Generatingquasi-randomsequences-1986} themselves
have shown that it is impossible to extract even a single unbiased bit
from an SV-source. More precisely, any compression of bits from SV-source with $\varepsilon$, in a form
of a boolean function $f:\{0,1\}^m\mapsto\{0,1\}$ cannot  produce another, improved SV source
with $\varepsilon' < \varepsilon$.
More importantly it has been shown by various authors
\cite{McInnes,DodisOngPrabhakaranEtAl-(im)possibilityofcryptography-2004}
that even slightly biased SV-sources, \textit{i.e.} sources with low $\varepsilon$, are not suitable for many
cryptographic purposes.
On the other hand Vazirani and Vazirani \cite{VaziraniVazirani-RandomPolynomialTime-1985} have shown how to
simulate a class of bounded error
randomized algorithms with a single SV-source.

However, even though the deterministic extraction fails, there is a possibility to post-process SV-sources with
the help of additional
randomness. An example of such additional resource is another independent SV-source.
In this setting Vazirani \cite{Vazirani} has shown that for any $\varepsilon$ and two independent sources with SV parameter
$\varepsilon$ there exists an efficient procedure to extract a single almost perfect bit.
More precisely, if $X = (X_1,\dots,X_n)$ is the outcome of the first source and $Y = (Y_1,\dots,Y_n)$
is the outcome of the second source,  post-processing function $Ext$ is defined as

 \begin{equation}
Ext(X,Y) = (X_1\cdot Y_1) \oplus (X_2\cdot Y_2) \oplus \dots \oplus (X_n\cdot Y_n),
\end{equation}
where $\oplus$ denotes sum modulo $2$. In other words the function $Ext$ is a scalar product
between the two $n$-bit strings $X$ and $Y$.
This function is very useful in other randomness  extractor constructions as we will see in the remainder
of this section and in the Section \ref{sec:RandAmplif}.
Extraction for SV-sources is sometimes called \emph{randomness amplification}, as it can be interpreted as transforming two
$\varepsilon$ SV-sources into another $\varepsilon'$ SV-source with $\varepsilon'<\varepsilon$, at the
rate of $\frac{1}{m}$ (\textit{i.e.} $m$ bits of the original sources are transformed into a single bit of the new,
improved source). Note that in general compressing more bits will result in a lower $\varepsilon'$ of the resulting SV-source.

This extraction function for SV-sources nicely demonstrates some important  concepts in
the area of randomness extraction.
First of all, in order to be able to extract, we need an additional, independent source of randomness,
be it another weak source or a short random seed.
Second, the quality of the output depends on both the length and quality of the input.
We will discuss these concepts in more detail in Subsection \ref{sec:Extractors}.

%%%%%%%%%%%%%%%%%%%
\subsubsection{Min-entropy sources}\label{sec:MESources}
%%%%%%%%%%%%%%%%%%%

Santha-Vazirani sources require that each produced bit contains some amount of randomness even
conditioned on the previous ones. In order to generalize this definition
Chor and Goldreich \cite{Chor} introduced sources, where the randomness is not guaranteed
in every single bit, but instead it is guaranteed in each $n$-bit block.
The randomness in blocks is guaranteed by it's
\emph{min-entropy} defined as
\begin{definition}[Min-entropy]\label{def:ME}
A \emph{min-entropy} $H_\infty(X)$ of a $n$-bit random variable $X$ is defined as:
$$H_\infty(X) = \min_{x\in\{0,1\}}\left(-\log_2(\Pr[X=x])\right).$$
\label{ME}
\end{definition}

Informally min-entropy is minus logarithm of the probability of the most probable element.
The randomness in blocks is therefore guaranteed by the restriction of the most probable
$n$-bit string appearing as the outcome of the block.
The  most probable element of a distribution is of a special interest as it also constitutes the
best strategy in trying to guess the outcome of the variable -- simply guessing 
the most probable element.
This leads us to the following formal definition:

\begin{definition}[Block source]
A \emph{$(n,k)$-block source} is modeled by a sequence of
$n$-bit random variables $X_1,X_2,\dots$, such that
$$
 \forall i\in\mathbb{N},\forall x_{1},\dots,x_{i-1}\in\{0,1\}^{n},\quad
  H_{\infty}(X_{i}|X_{i-1}=x_{i-1},\dots,X_{1}=x_{1})\geq k.\nonumber
$$
\label{def:min_entropy}\\
\end{definition}
In order to recover the view of such a randomness source as a set of probability distributions
over the same domain, we simply need to consider finite number of blocks.

It is also easy to see that SV-sources are recovered with $n=1$ and  $\varepsilon = 2^{-H_{\infty}(X)}-\frac{1}{2}$.
In order to show that the block sources are a strict generalization of SV-sources we must invoke
an argument suggesting that there are no deterministic extractors that can extract even a single
bit from the block sources.
A deterministic single-bit extractor is again considered to be a boolean function $f: \{0,1\}^n\mapsto \{0,1\}$.
We will proceed to show that for every $f$, there exists a probability distribution on the inputs
with min-entropy $n-1$, such that the output of $f$ is constant.

Consider an arbitrary $f$ and let us split the set of $n$-bit strings
into two subsets $D_0 = \{x\vert f(x) = 0\}$ and $D_1 = \{x\vert f(x) = 1\}$. Without the loss of generality
assume that $\vert D_0\vert \geq 2^{n-1}$.
Now consider a random variable $X$, such that $\Pr[X=x] = \frac{1}{\vert D_0\vert}$ for
$x\in D_0$ and $\Pr[X=x] = 0$ otherwise.
Random variable $X$ has min-entropy at least $n-1$,
but $f(X)$ outputs a constant bit $0$. This indicates that deterministic extraction is impossible even
for sources with very high min-entropy.
Moreover, this also implies that a block source cannot be transformed into an SV-source with
$\varepsilon<\frac{1}{2}$ and therefore min-entropy block sources are indeed a strict generalization of the
Santha-Vazirani sources.

Of course one might want to employ deterministic functions that try to
extract from more than one block of the source. In fact, such strategy doesn't provide any advantage,
as $m$ blocks can be seen as a single block of a source with block size of $mn$.

\begin{figure}[tb]
\centering
\includegraphics[width=11cm,clip]{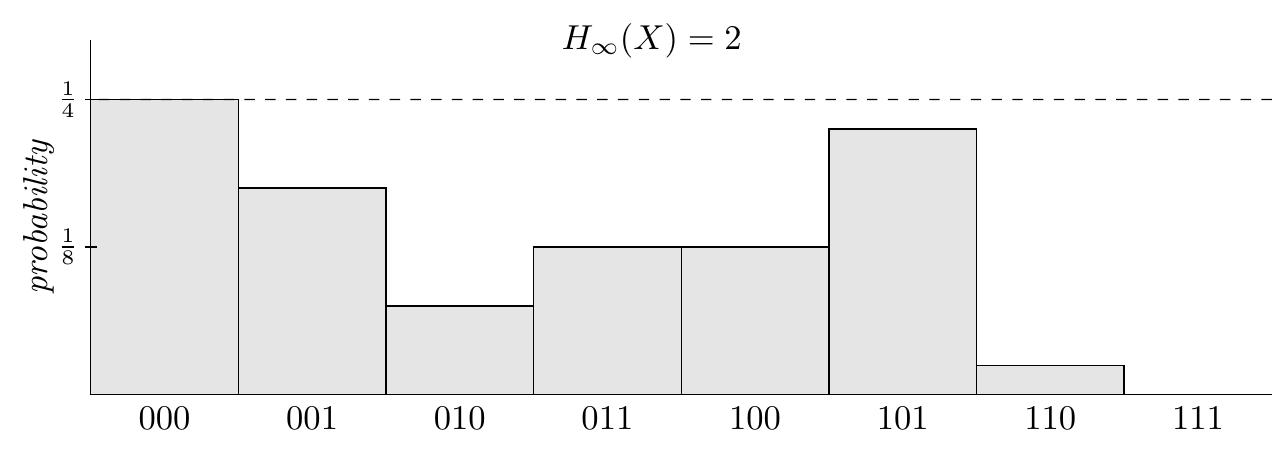}
	\caption{\it\small A possible probability distribution of a (3,2)--source. Notice that this model allows strings
               which have zero probability to appear.}
	\label{fig:MES}
\end{figure}

Up to this point we defined randomness sources as infinite streams of bits with different
requirements on their structure. This view is very useful when considering random number generators,
which supposedly can produce any number of random bits on demand.
However in most applications we use only a \emph{finite} number of random bits. Random
inputs into these applications can be treated as sources of randomness of finite size. This view
suggests yet another generalization of the block sources -- sources of randomness of finite output size,
where no internal structure, such as guaranteed entropy in every bit (SV-sources) or every block of certain size
(block sources)  is assumed. The only guarantee of randomness is it's overall min-entropy (see Fig. \ref{fig:MES}).

\begin{definition}[Min-entropy source.]
An $n$-bit random variable $X$, such that $$H_\infty(X)\geq k$$ is called an $(n,k)$-source.
\end{definition}

Note that in this view, a min-entropy source is a set of distributions with an upper bound on the probability
of the most probable element imposed by min-entropy. This is incidentally the probability of success of the best
strategy to guess the outcome of the variable.
Such sources were introduced by Zuckerman \cite{Zuckerman-SimulatingBPPusing-1996} and nowadays are the
most studied type of weak sources.
Also note that randomness extractors for this type of sources (discussed
in Subsection \ref{sec:Extractors}) can easily be used for block sources as well by simply applying them
block-wise.

\begin{figure}[tb]
\centering
\includegraphics[width=11cm,clip]{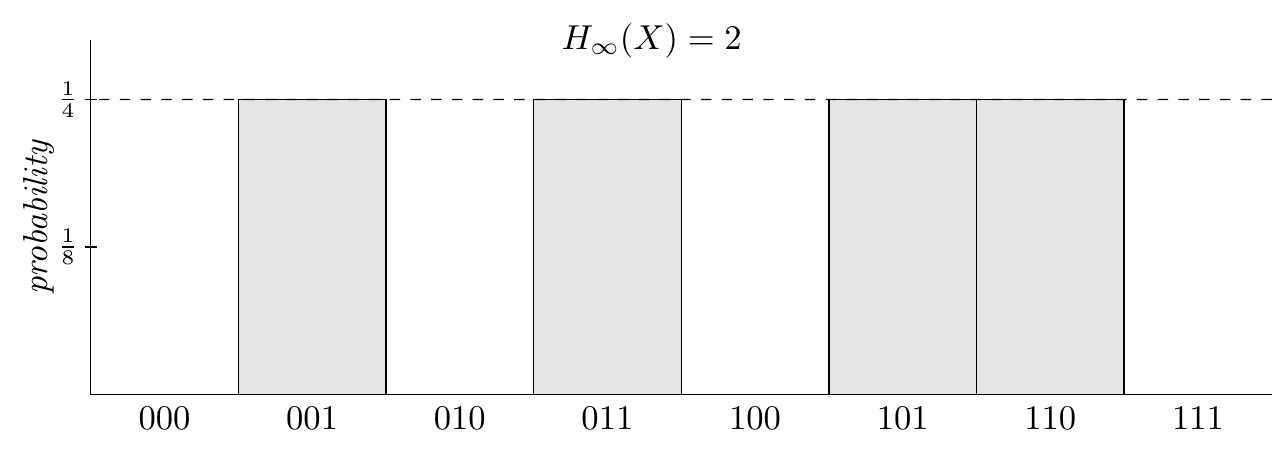}
	\caption{\it\small A possible distribution of a flat source  with $n = 3$ and $k = 2$.}
\label{fig:flatSource}
\end{figure}

Since we will discuss min-entropy sources in this paper as well, we need several more definitions.
For any $n$-bit random variable $X$ with $H_\infty(X) \geq k$,
let us denote it's \emph{min-entropy loss} as $c = n-k$
and it's \emph{min-entropy rate} as $\frac{k}{n}$.
The last important definition is that of the \emph{flat sources}.
\begin{definition}[Flat source]\label{def:flatSource}
Let $S\subset \{0,1\}^n$.
A random $n$-bit variable $X$ is  \emph{flat} on $S$, if for all $x \in S$,
$$\Pr[X = x] = \frac{1}{\vert S\vert}.$$
By extension, for all $y\notin S$, $\Pr[X = y] = 0$.
If $\vert S\vert \geq 2^k$, it is a $(n,k)$- flat source.
 \end{definition}
In other words, the flat sources with min-entropy $k$
are distributed according to a probability distribution that is uniform
on sufficiently large subset of possible outcomes (see Fig. \ref{fig:flatSource}).
More importantly, any source with min-entropy $k$ is a convex combination of flat sources with min-entropy
$k$ and it can be shown
that in many applications the flat sources manifest the worst case behavior. That is why the analysis is
often carried out on the flat sources only.

%Randomness extraction from $(n,k)$-sources is a very interesting and well developed field
%and it's results will be used throughout the rest of this thesis. In order to properly introduce this topic
%we devote the whole next subsection to it.

%%%%%%%%%%%%%%%%%%%%%%%%%%%%
\subsection{Randomness extractors}\label{sec:Extractors}
%%%%%%%%%%%%%%%%%%%%%%%%%%%%%

As mentioned previously, it is a very common scenario to post-process weak randomness,
with the use of \emph{randomness extractors}. These algorithms produce nearly perfect randomness,
 which can later be used in other applications.
The aim of randomness extraction from $(n,k)$-sources is to turn a bit string distributed according to
arbitrary probability distribution with min-entropy at least $k$ into a possibly shorter
bit string that is \emph{close} to being perfectly random. Concept of closeness can be summed by the
following definition.

\begin{definition}[$\varepsilon$-closeness]
Random variables $X$ and $Y$ over the same domain $D$ are $\varepsilon$-close, if:
$$
\Delta (X,Y) =\frac{1}{2}\sum_{x\in D}\left\vert \Pr[X=x] - \Pr[Y=x]\right\vert \leq\varepsilon
$$
\end{definition}
The usefulness of this definition perhaps becomes more evident when we point out that $\Delta (X,Y)$
can be equivalently defined as $\vert p_X(A) - p_Y(A)\vert\leq \varepsilon$ for every event $A\subseteq D$,
therefore variables $X$ and $Y$ are almost indistinguishable.

It can be shown that the min-entropy of a variable $X$ gives the upper bound on the number of extractable
almost uniformly distributed bits \cite{Shaltiel1}.
In other words, if $k$ bits can be extracted from a random variable $X$, then $X$ has min-entropy at least $k$.
Informally we say that  a $(n,k)$-distributed variable $X$ contains $k$ bits of randomness.
It is for this reason that min-entropy sources have become the most widely considered model of weak randomness
in the literature.

%%%%%%%%%%%%%%%%%
\subsubsection{Seeded extractors}
%%%%%%%%%%%%%%%%%

As we mentioned earlier, deterministic extraction is impossible for min-entropy sources.
Nevertheless, as we have learned in the case of SV-sources, extraction might be possible
with additional resources.
The most widely studied constructions are \emph{seeded extractors},
in which the extra resource is an additional short, uniformly distributed random string,
called \emph{the seed}.

\begin{definition}[Seeded extractor]\label{def:SeedExt}
A function
$
Ext:\{0,1\}^n\times\{0,1\}^s\mapsto\{0,1\}^m
$ is a \emph{seeded $(k,\varepsilon)$-extractor} if for every $(n,k)$-distributed random
variable $X$,
$$\Delta(Ext(X,U_s),U_m) \leq \varepsilon.$$
\end{definition}

Sometimes a stronger definition of extractor is needed and the output is required to be random even
to an entity that has seen the value of the seed. This can be formally written as

\begin{definition}[Strong extractor.]
A function
$
Ext:\{0,1\}^n\times\{0,1\}^s\mapsto\{0,1\}^m
$ is a \emph{strong seeded $(k,\varepsilon)$-extractor} if for every $(n,k)$-distributed random
variable $X$,
$$\Delta( U_s\circ Ext(X,U_s),U_s\circ U_m) \leq \varepsilon,$$
where $\circ$ is concatenation and two copies of $U_s$ denote the same random variable.
\end{definition}

Strong extractors can be seen as a set of deterministic extractors $\{Ext(\cdot,s)\vert s\in\{0,1\}\}$ with a
following property:
For any given $(n,k)$-source $X$ most of the extractors in the set constitute a good extractor for $X$.
This property will be used in Subsection \ref{MinEntropyAmp} as a basic building block for a randomness amplification
protocol.

There are several parameters against which the quality of the extractor can be evaluated.
First of all we want the seed $s$ to be as small as possible, because, as we have argued, (nearly) perfect
randomness is a scarce resource.
Second of all, we want the extractor to successfully extract randomness
even from sources with low min-entropy $k$. In fact, it is easier to extract randomness from
sources with higher min-entropy and the required min-entropy often depends on the length of
the source $n$.
This requirement becomes more clear when one realizes that the overall quality of the source is more accurately
expressed by the min-entropy rate $\frac{k}{n}$
then the total min-entropy $k$. Intuitively it should be clear that a $(100,2)$-source
is much worse than a $(3,2)$ source.

Naturally, we want $m>s$ to actually gain some randomness, ideally we want to achieve the maximum
possible size of the output $m = s+k$.
We also require the statistical distance $\varepsilon$ of the output from an uniform random variable $U_m$ to be
as small as possible, typically $\varepsilon$ is a function of all $n,k$ and $s$.
Last but not the least, we want the extractor to be efficient and efficiently constructible, meaning that given
parameters $n$ and $k$ the function $Ext$ must be constructible in polynomial time in both
of these parameters and it's evaluation should also be possible in time polynomial in both $n$ and $k$.

The optimal parameters of extractors obtained by probabilistic methods
\cite{RadhakrishnanTa-Shma-BoundsDispersersExtractors-2000} are seed length
of $s = \log n + O(1)$, output length of $m = k+s-O(1)$ and such optimal extractor is
able to extract from a source with any min-entropy $k$, regardless of it's length $n$. Note however,
that this extractor is non-explicit, and therefore not efficiently constructible.
Even though the construction of an optimal extractor is not known, there are known extractor constructions
that obtain optimal values for any pair of these three parameters.
The best recent constructions according to the output length and the seed length are introduced in
\cite{LuReingoldVadhanEtAl-Extractors:optimalup-2003,GuruswamiUmansVadhan-Unbalancedexpandersand-2009}.

As an example of a min-entropy extractor we will introduce a construction based on universal hashing
\cite{CarterWegman-Universalclassesof-1979}.
\begin{definition}[Universal hashing]
A set $\mathcal{H}$ of hash functions $H:\{0,1\}^n\mapsto \{0,1\}^\ell$ is a
\emph{universal family of hash functions} if for any $w_1\neq w_2 \in \{0,1\}^n$, and for any
$x_1,x_2\in \{0,1\}^\ell$,
$$
\Pr[h(w_1) = x_1 \wedge h(w_2) = x_2] = 2^{-2\ell},
$$
where the probability is taken over uniform choice of hash function $h\in\mathcal{H}$.
\end{definition}
There are known construction for such families of hash functions of size $2^{2n}$ for every $\ell< n$.
Let us parametrize such set as $\mathcal{H} = \{h_s\vert s \in \{0,1\}^{2n}\}$.

\begin{definition}[Universal hashing extractor]\label{def:UHE}
A function $Ext:\{0,1\}^n\times\{0,1\}^{2n}\mapsto \{0,1\}^m$ defined as
$$
Ext:(x,s) = s\circ h_s(x)
$$
is a $(m-s+2\log(1/\varepsilon),\varepsilon)$ extractor for every $\varepsilon$.
\end{definition}

This extractor was first introduced in \cite{SrinivasanZuckerman-Computingwithvery-1994} and it  has optimal
output length and can
extract from sources with any $k$ --
as long as $m = k+s -2\log(1/\varepsilon)-O(1)$, the output is at most $\varepsilon$ far from
uniform distribution.
For more details and history of seeded extraction consult any of the excellent surveys of this topic
\cite{Nisan,Shaltiel1} and references therein.

%%%%%%%%%%%%%%%%%%%%
\subsubsection{Extraction from several independent sources}\label{sec:2SExtractors}
%%%%%%%%%%%%%%%%%%%%

The disadvantage of the seeded extraction is that it requires uniformly distributed seed, which, as we argued before,
is difficult to obtain.
This fact leads to another direction in designing randomness extractors, which is to consider extracting randomness from
\emph{several independent} min-entropy sources.
We will define such extractor for two independent sources, while generalization to several sources is straightforward.

\begin{definition}[Two source extractor]\label{def:TwoSourceExt}
A function $\text{Ext} : \{0,1\}^n\times\{0,1\}^n\mapsto \{0,1\}^\ell$ is a $(k_X,k_Y,\varepsilon)$-extractor
if for every two independent $(n,k_X)$ source $X$ and $(n,k_Y)$ source $Y$
$$
\|\text{Ext}(X,Y) - U_\ell\|_1\leq \varepsilon.
$$
\end{definition}
A pair of sources was already considered by
Chor and Goldreich \cite{Chor}. Multi-source extractors were subsequently studied by many other
authors \cite{Dodis,Dodis2,Raz}
and the best recent constructions can be found in \cite{Barak2}.

One longstanding problem of multi-source extractors is to find explicit constructions
for the sources with low \emph{min-entropy rate}. Probability argument suggests that the lower bound on extractable
min-entropy is in the order $O(\log_2(n))$,
where $n$ is the length of the input strings. Nevertheless, explicit constructions existed only for sources with
min-entropy rate greater than $\frac{n}{2}$.
Only recently Bourgain \cite{Bourgain} has broken the $\frac{n}{2}$ barrier and shown how to construct extractors
for sources with min-entropy below $\frac{n}{2}$.

In order to show an example of a two source extractor,
let us first revisit the scalar product function. It turns out scalar product is essential in building two
source extractors.
In this context it is sometimes also called the \emph{Ha\-da\-mard extractor},
and is often used as a primitive to extract a single bit.

\begin{definition}[Hadamard extractor]\label{def:Hadamard}
A function $\text{Had}: \{0,1\}^n\times \{0,1\}^n \mapsto \{0,1\}$ defined as
$$\text{Had}(x,y) = \left(\sum_{i=1}^{n}  x_i \cdot y_i\right) \mod 2,$$ where $x = (x_1,\dots, x_n)$ and $y = (y_1,\dots,y_n)$
is a $(k_X,k_Y,2^{\frac{n-k_X-k_Y-1}{2}})$ extractor.
\end{definition}

In order to show how to expand Hadamard extractor in order to extract more than a single bit, we present
a construction introduced by Dodis \textit{et.~al.} \cite{Dodis}.

\begin{definition}[DEOR extractor]\label{def:DEOR}
For all $n>0$ there exists a set of $n\times n$ matrices $\{A_1,\dots,A_n\}$ over $GF(2)$
such that for any non-empty set $S\subseteq \{1,\dots,n\}, A_S = \sum_{i\in S} A_i$ has full rank.
Let $n\geq m>0$. A Function
$Ext_D: \{0,1\}^n\times\{0,1\}^n \mapsto \{0,1\}^m$ defined as
$$
Ext_D(x,y) = Had(A_1x, y), Had(A_2x, y),\dots, Had(A_mx, y)
$$
is a $(k_X,k_Y,2^{\frac{n+m-k_X-k_Y-1}{2}})$ extractor. Here $GF(2)$ is a field of addition and multiplication
modulo $2$.
\end{definition}

We have shown in \cite{BoudaPivoluskaPlesch-ImprovingHadamardextractor-2012} that Hadamard
extractor can be improved, especially for sources with high min-entropy.
We focused on one of the weaknesses of the Hadamard extractor -- if one of the randomness sources, say $X$,
happens to be uniform, the extractor fails to
produce unbiased bit. In our paper we proposed a function which produces bits with constantly better bias
for $(n,k)$-sources with $k<n-1$.
What is more, the distance $\varepsilon$ of the produced bit from a uniform bit approaches $0$ as $k$
approaches $n$.
Our construction is presented in the next definition.

\begin{definition}[BPP extractor]
A function $\text{Had}_{\oplus}: \{0,1\}^n\times \{0,1\}^n \mapsto \{0,1\}$ defined as
$$\text{H}_\oplus(x,y) = \left(x_1 + y_1 + \sum_{i=2}^{n}  x_i \cdot y_i\right) \mod 2,$$ where
$x = (x_1,\dots, x_n)$ and $y = (y_1,\dots,y_n)$
is a $(k_X,k_Y,2^{\frac{n-k_X-k_Y-3}{2}})$ extractor.
\end{definition}

Our extractor therefore obtains $\sqrt{2}$ times smaller bias than the Hadamard extractor.
Moreover, contrary to the Hadamard extractor, the bias of the proposed extractor is $0$ if at
least one of the input sources is uniform.
Another interesting property is that our construction can
beat the $\frac{n}{2}$ bound in some cases.

Our extractor can be plugged into the DEOR (see Def. \ref{def:DEOR}) construction, in which case we obtain
a $(k_X,k_Y,\frac{\sqrt{3}}{2}2^{\frac{n+m-k_X-k_Y-1}{2}})$  extractor,
which is slightly better than the original construction.
Surprisingly, this advantage in the extraction without the strongness property is
not retained in the strong extraction scenario. In fact, our extractor is $\sqrt{\frac{3}{2}}$
worse than the extractor based on the Hadamard construction.

%%%%%%%%%%%%%%%%%%%%%%
\subsection{Weak sources with quantum side information}\label{sec:QWS}
%%%%%%%%%%%%%%%%%%%%%%

So far, we have considered only classical weak sources, \textit{i.e.} sources described
by classical random variables with unknown probability distribution or, equivalently,
classical random variables with adversaries holding some \emph{classical}
information about their outcome. In quantum information, however, this is not
the most general model of weak randomness sources -- a potential adversary might
obtain side information
about the source in form of quantum states.

In order to formalize this approach we need to introduce some new notation.
\begin{definition}[cq-state]
Let $Z$ be a classical random variable and $\{\rho_z\}_{z\in \mathcal{Z}}$
a set of density matrices.
Then we denote the \emph{cq-state}:
$$
Z\rho_Z = \sum_{z\in \mathcal{Z}} \Pr[Z=z]\ket{z}\bra{z}\otimes\rho_z.
$$
\end{definition}

Supposing the adversary holds the quantum part of a cq-state and the device
produ\-cing randomness holds the classical part,
the weak source is obtained by measuring the classical part of the state
in the computational basis -- quantum mechanics guarantees that the measurement
outcomes will be distributed according to the distribution $p_Z$ of the random
variable $Z$. Moreover according to the Copenhagen interpretation of quantum
mechanics, this randomness is objective, therefore weak randomness is
modeled in the spirit of our second definition -- the adversary obtains
information about the source via a side channel: in this case, for each outcome
$z$ the adversary obtains a state $\rho_z$.
Generally the adversary holds a mixed state $\rho_Z = \sum_{z\in\mathcal{Z}}\Pr[Z = z]\rho_z$ and
tries to infer as much information as possible
about the corresponding classical outcome.

In order to clearly  formulate the capabilities of the adversary, the following scenario
is typically considered. The adversary obtains the outcome $z$ of the random variable
$Z$ and stores quantum information about it in a quantum memory, by applying
a quantum operation $U_z$ to it, resulting in a state $\rho_z$.
If all the states in $\{\rho_z\}_{z\in Z}$
are distinguishable, the adversary can obtain all the outcomes with probability $1$,
simply by using the distinguishing measurement. Therefore there remains a question
how to bound the adversary's knowledge in a meaningful way.

There are
two ways to do this, both inspired by the classical min-entropy sources.
First of all, the dimension of states $\rho_z$ can be restricted. In such scenario,
the adversary is given the outcome of the variable, but is allowed to store
only $b$ qubits about it. If one allows the adversary only classical memory,
min-entropy sources are recovered -- storage of $b$ classical bits about
a source that is uniformly distributed ensures that it's resulting min-entropy
conditioned on the adversary's memory is at least $n-b$.

Recall that min-entropy of a classical source can also be interpreted as
the upper bound on the probability of correct guess of it's outcome.
In the same spirit we can restrict the amount of information the reduced
state $\rho_Z$ contains about the random variable $Z$. Formally,
we will restrict adversary's probability to guess the outcome of $Z$ correctly
by restricting it's \emph{guessing entropy} given $\rho_Z$.
As adversary's strategy consists of measuring $\rho_Z$, we need to optimize
over all possible POVMs.

\begin{definition}[Guessing entropy]
Let $X\rho_X$ be an arbitrary $cq$-state. The guessing entropy of $X$ given $\rho_X$ is
$$
H_g(X\leftarrow \rho_X) = -\log\max_M \mathbb{E}_{x\leftarrow X}[\text{Tr}(M_x\rho_x)],
$$
where the maximum is taken over all $\vert\mathcal{X}\vert$ outcome POVMs
$M = \{M_x| x\in \mathcal{X}\}$.
\end{definition}
Perhaps, more intuitive form is
$$
H_g(X\leftarrow \rho_X) = -\log\max_M \Pr[M(\rho_X) = X],
$$
where $M(\rho_X)$ is a random variable we obtain by measuring the state $\rho_X$
by a POVM $M$.  An alternative concept, often used in literature is called
\emph{conditional min-entropy} \cite{Renner2005a}, but it has been shown that the two definitions
are equivalent \cite{KonigRennerSchaffner-OperationalMeaningof-2009}.

The definition of the weak source is now straightforward.
A source $X$ is a $(n,k)$-source against quantum memory $\rho_X$,
if $H_g(X\leftarrow \rho_X)\geq k$.

Randomness extraction is studied for both sources with bounded storage
(see \textit{i.e.} \cite{KonigMaurerRenner-powerofquantum-2005,DeVidick-Near-optimalExtractorsAgainst-2010}
and the references therein)
and sources with guaranteed guessing entropy \cite{DePortmannVidickEtAl-Trevisan'sExtractorin-2012}.
One of the most prominent results is that
quantum side information can in some cases offer a significant advantage to the adversary
in extraction scenario, when compared to it's
classical counterpart. This was proven by Gavinsky \textit{et.~al.}
\cite{GavinskyKempeKerenidisEtAl-ExponentialSeparationsOne-way-2007}
who constructed a strong randomness extractor which is secure against classical adversaries, but
fails to produce almost perfect randomness against adversaries holding quantum
information.
More precisely, their construction outputs almost perfectly random bits against an
adversary with $o(\sqrt{n})$ bits of storage, while it fails against an adversary
with $O(\log(n))$ quantum bits used for storage.

On the positive side, some of the existing extractors for classical weak sources
have been proven to be secure against both types of quantum side information,
among them the construction based on universal hashing
\cite{TomamichelRennerSchaffnerEtAl-LeftoverHashingagainst-2010} (see Def. \ref{def:UHE}).
The recently  best construction against adversaries with bounded storage can be found in
\cite{DeVidick-Near-optimalExtractorsAgainst-2010}
 and against guessing entropy adversaries in
\cite{DePortmannVidickEtAl-Trevisan'sExtractorin-2012}.

Extraction without the access to uniform seed is even more complicated in case of
sources with quantum side information.
We will consider the case of extraction with two weak sources,
which can easily be generalized to multiple source extractors.

In this scenario we assume that there are two non-communicating
adversaries, one for each weak source. After the sources produce their
outcomes, the two adversaries meet and try to guess the outcome of the
extractor. In order to see that this level of abstraction is necessary, assume
only a single adversary. In the side information formalism introduced earlier, he
first receives the outcomes $x$ and $y$
of random variables $X$ and $Y$ representing the weak sources, and tries to
save some restricted
information about them in a quantum memory. Because he can see both
random inputs of the extractor, he can simply calculate it's outcome $Ext(x,y)$
and store it. The outcome $Ext(x,y)$ by definition contains almost no information about the
individual inputs, thus fulfilling the restricted information requirements.
In other words in this model the outputs of variables $X$ and $Y$ are easily made
correlated via adversary's memory that is why we need to assume two
non-communicating adversaries.

To add more complexity to the problem, we can allow the adversaries' memories
to be entangled. This leads to several different models: bounded memory adversaries
with/without entanglement and guessing entropy adversaries with/without entanglement.
Kasher and Kempe \cite{KasherKempe-Two-SourceExtractorsSecure-2010}
 studied the DEOR construction introduced in the previous
section (see Def. \ref{def:DEOR}) in various settings of this type and were able to
 obtain
positive results for both entangled and non-entangled bounded memory adversaries
and non-entangled guessing entropy adversaries. The remaining scenario
with entangled guessing entropy adversaries proved to be too strong and
the DEOR construction does not provide any security in this case.
Moreover, it is not clear if such strong constructions of extractors are even possible.

This concludes the short introduction into the wide field of weak randomness
and we are ready to present the main topic of this paper.

\setcounter{equation}{0} \setcounter{figure}{0} \setcounter{table}{0}\newpage
%%%%%%%%%%%%%%%%%%%%
%%%%%%%%%%%%%%%%%%%%%%%%%%%%%%%%%%%
\section{Bell inequalities and randomness}\label{sec:Bell}%%%%%%%%%
%%%%%%%%%%%%%%%%%%%%%%%%%%%%%%%%%%%%%%%%
%%%%%%%%%%%%%%%%%%%%%%
In this section we will explain in detail
what Bell type inequalities are and
how their violation guarantees that the outcomes of certain measurements
are fundamentally undetermined. This fact is exploited in the construction
of randomness generation protocols.

%%%%
\subsection{The local set and determinism}
%%%%

\begin{figure}[!b]
\begin{center}
\includegraphics[scale = 1]{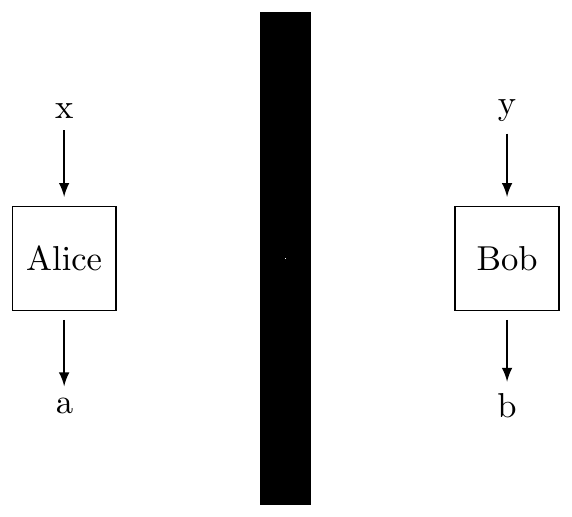}
\caption{\it\small Two party Bell type experiment. Alice's measurement setting is $x$ and Bob's $y$.
Their respective outcomes are $a$ and $b$. The black barrier represents their inability to communicate, \textit{e.g.}
by separating the experiments by a space-time interval.}
\label{scenario}
\end{center}
\end{figure}

The simplest scenario of a Bell type non-locality test consists of two spatially separated
observers, Alice and Bob, who measure a bipartite system produced by a common source.
Both Alice and Bob can choose one of several possible measurement settings.
After the measurement both of them record the outcome.
Let us label Alice's choice of measurement setting $x\in \mathcal{X} = \{1,\dots, M_A\}$
and Bob's choice $y\in \mathcal{Y} = \{1,\dots, M_B\}$ and their
outcomes $a\in \mathcal{A} = \{1,\dots m_A\}$ and $b\in \mathcal{B} = \{1,\dots,m_B\}$ (see Fig. \ref{scenario}).
If this procedure is repeated many times,
Alice and Bob can communicate their measurement settings and outcomes to each other
and
estimate probabilities $p_{AB|XY}(ab\vert xy) = \Pr[AB = ab\vert XY = xy]$, where $X,Y$ are the random variables governing
the inputs and $A,B$ the random variables governing the outputs of the boxes.
We say the outcomes are correlated, if for some $x,y,a,b$
%%%%%%%%%%%%%%% eq. 1 %%%%%%%%%%%%%%%%%
\begin{equation}
p_{AB|XY}(ab\vert xy) \neq p_{A|X}(a\vert x)p_{B|Y}(b\vert y).
\end{equation}
Existence of correlations isn't anything surprising. In fact correlations
are very natural and can be classically explained
by some common cause of the observed statistics.
%In particular finding out that outcomes of the measurements of distant parts of a bipartite system are correlated does
%not imply any super-luminal influence of one particle upon another.
Formally, one can model the cause of these correlations
by a set of random variables $\Lambda$,
which have causal influence on both measurement outcomes,
but are inaccessible to the observers. In a
\emph{local hidden variable} model, taking into account all the possible
causes $\Lambda$, outcomes of the experiments are fully independent, \textit{i.e.},
for all $a,b,x,y,\lambda$
%%%%%%%%%%%%% eq.2 %%%%%%%%%%%%%%%%%
\begin{equation}
p_{AB|XY\Lambda}(ab\vert xy,\lambda) = p_{A|X\Lambda}(a\vert x,\lambda)p_{B|Y\Lambda}(b\vert y,\lambda).
\end{equation}
This in fact represents an explanation of the correlations according to which
Alice's outcome depends only on her local measurement setting $x$ and some common cause for the correlations $\lambda$
and not on distant Bob's measurement setting and outcome, and analogously Bob's outcome doesn't
depend on anything that Alice does.
This is in fact a crucial assumption required by the theory of relativity, which forbids non-local causal influence
for spatially separated entities.
To complete the picture we must take into account the
probability distribution of $\Lambda$ -- $p_\Lambda$. This gives rise to a \emph{local hidden variable} condition:
%%%%%%%%%%%%% eq. 3 %%%%%%%%%%%%%%%%%%%%%
\begin{equation}\label{LocCon}
p_{AB|XY}(ab\vert xy) = \int_\Lambda \mathrm{d}\lambda p_\Lambda(\lambda)
p_{A|X}(a\vert x,\lambda)p_{B|Y}(b\vert y,\lambda).
\end{equation}
This characterization contains an implicit assumption -- the measurement settings $x$ and $y$ can be chosen
independently of $\lambda$. Formally,
\begin{equation}\label{MeasurementIndependence}
p_{\Lambda|XY}(\lambda\vert x,y) = p_\Lambda(\lambda).
\end{equation}

Notice that so far we haven't assumed anything about the determinism of measurements in the local model,
as condition (\ref{LocCon}) states only that the outcomes are probabilistically determined.
In \emph{deterministic local hidden variables} model, which is a special
case of the above, each outcome is uniquely determined by $\lambda$ and the corresponding input $x$, \textit{i.e.}
for each outcome $a$, input $x$ and hidden cause $\lambda$, $p_{A|X\Lambda}(a\vert x,\lambda)$ is equal to either $1$ or $0$,
and similarly for $b,y$ and $\lambda$.

In fact, \emph{local hidden variables} are fully equivalent to \emph{deterministic local hidden variables}, as first proven by Fine
\cite{Fine-HiddenVariablesJoint-1982}.
The reason why both definitions are equivalent stems from the fact that all randomness present in the probability functions
$p{A|X\Lambda}(a\vert x,\lambda)$ and $p_{B|Y\Lambda}(b\vert y,\lambda)$, can always be incorporated into the shared 
random variable.
To show this, let us introduce two continuous variables $\mu_A,\mu_B\in[0,1]$ and introduce a new common variable
$\Lambda' = (\Lambda,\mu_A,\mu_B)$. Let
%%%%%%%%%%%% eq. 4 %%%%%%%%%%%%%%%%%%
\begin{equation}
 p_{A|X\Lambda'}(a\vert x,\lambda') = \left\{
  \begin{array}{l l}
  1 & \quad \textrm{if } F(a-1\vert x,\lambda)\leq \mu_A < F(a \vert x, \lambda) \\
  0 & \quad \textrm{otherwise},
  \end{array} \right.
\end{equation}
where $F(a\vert x, \lambda) = \sum_{a'\leq a} p_{A|X\Lambda}(a'\vert x, \lambda)$, be new deterministic
function governing outcomes of Alice and define analogous function
for Bob. If we choose both $\mu_A$ and $\mu_B$ with uniform distribution,
we will recover the prediction of the general model.

So far, we have shown that observed correlations that admit decomposition as in Eq. (\ref{LocCon})
can be explained by a fully deterministic model -- outcomes of the measurement are
completely predetermined by the measurement settings and some hidden variables.

In the next subsection we will show how to certify that there is no local hidden variable
model for the observed correlations, as in the case for certain measurements of
quantum systems.

%%%%%%%%%%%%%%
\subsection{Bell inequalities}\label{BellSection}
%%%%%%%%%%%%%%%%
In the previous subsection we have shown the equivalence between
deterministic and general local hidden variable models that can both explain correlations observed.
This result has in fact one more corollary -- we need to consider only finite number of hidden
variables. Indeed, in a deterministic model each variable in $\Lambda$ specifies an outcome
for one concrete input. The general model is a probabilistic mixture of these assignments from
outputs to inputs. Since the total number of inputs and outputs is finite, so is the total number
of different assignments and thus there is a finite number of hidden variables.

More formally, we can equivalently write down the model in (\ref{LocCon}) as follows:
Let us define the values of hidden variables $\Lambda$ as $\lambda = (a_1,\dots,a_{M_A},b_1,\dots,b_{M_B})$.
For each value of $\lambda$ we can construct the corresponding
deterministic input to output assignment $d^\lambda$ as:
%%%%%%%%%% eq. 5 %%%%%%%%%%%%%%%%%
\begin{equation}
 d^\lambda(ab\vert xy) = \left\{
  \begin{array}{l l}
  1 & \quad \textrm{if } a = a_x \textrm{ and } b = b_y \\
  0 & \quad \textrm{otherwise}.
  \end{array} \right.
\end{equation}
There are $m_a^{M_A}m_b^{M_B}$ possible values of $\lambda$ and thus
$m_a^{M_A}m_b^{M_B}$ deterministic measurement outcome assignments.
Observed probability $p_{AB|XY}(ab\vert xy)$
can be explained by local hidden variables, if it can be written as a convex combination of such
deterministic local points:
%%%%%%%%%%%% eq. 6 %%%%%%%%%%%%%%%%%
\begin{equation}\label{VerticesPolytope}
p_{AB|XY}(xy|ab) = \sum_\lambda p_\Lambda(\lambda) d^\lambda(ab|xy),
\end{equation}
where $p_\Lambda(\lambda) \geq 0, \sum_\lambda p_\Lambda(\lambda) = 1$, \emph{i.e.} $p_\Lambda$
is the probability distribution of the deterministic points $d^\lambda$.

The set of possible hidden variable models is a convex hull
of a finite number of deterministic points $d^\lambda$, and therefore
in terms of geometry it is a polytope.
Any linear inequality defining a half-space in which the whole \emph{local polytope} $\mathcal{L}$ resides
can be used as a witness that a distribution violating this inequality is non-local.
Inequalities of this type are called \emph{Bell inequalities}.
To identify the optimal set of Bell inequalities, it suffices to recall basic results
in the theory of polytopes --
a polytope can be represented not only by all it's vertices, as in Eq. (\ref{VerticesPolytope}),
but equivalently it can be represented by a finite number of half-spaces -- the facets of the polytope.
Each such facet can be expressed by a linear inequality of the form
$
\sum_{a,b,x,y} c_{abxy} p_{AB|XY}(ab|xy) \leq S_l,
$
where $c_{abxy}$ are some linear coefficients defining the Bell inequality and $S_l$ is the
maximum value attainable by local probabilities 
\begin{equation}
P = \left\{ p_{AB|XY}(ab|xy)\vert a\in\mathcal{A},b\in\mathcal{B},x\in\mathcal{X},y\in\mathcal{Y}\right\},
\end{equation}
belonging to the local polytope $\mathcal{L}$.
Such set of probabilities $P$ is also called \emph{behavior}.
Hence an observed behavior $P$ lies in the local polytope $\mathcal{L}$,
if and only if:

\begin{equation}\label{eq:BellInequality}
\sum_{a,b,x,y} c_{abxy} p_{AB|XY}(ab|xy)\leq S_l^i \quad\forall i\in I,
\end{equation}
where $I$ is a finite index set of linear inequalities corresponding to the facets of the local polytope.
Conversely, if behavior $P$ is non-local,
it necessarily violates at least one of these Bell inequalities.
Note that some facets are trivial and correspond to positivity conditions ($p_{AB|XY}(ab\vert xy)\geq 0$).
These are obviously never violated by any physical behavior. All other facets are violated by some
non-local behaviours, some of them even by quantum behaviours as we will show next.
It is important to note
that although there exist algorithms for obtaining all the polytope facets, given it's vertices, they
become extremely time-consuming as the number of inputs, outputs or parties grow. This
is the reason why the study of Bell inequalities is a fruitful research area up to these days.

As an example we introduce here one of the most studied Bell inequalities.
Consider the simplest scenario where both Alice and Bob choose one of two measurements
$x,y\in\{0,1\}=\mathcal{B}$ and obtain one of two measurement outcomes, which we label $a,b\in\{-1,1\}=\mathcal{A}$.
In this case the local polytope $\mathcal{L}$ has been fully characterized \cite{Fine-HiddenVariablesJoint-1982}.
The only non-trivial
facet inequality is the CHSH inequality introduced in \cite{ClauserHorneShimonyEtAl-ProposedExperimentto-1969}.
Let $\langle a_xb_y\rangle = \sum_{ab} ab\cdot p_{AB|XY}(ab\vert xy)$
be an expectation value of the product $ab$ after measuring $x$ and $y$. The CHSH inequality then reads:
\begin{equation}\label{CHSH}
I_{CHSH} = \langle a_0b_0\rangle + \langle a_0b_1\rangle + \langle a_1b_0\rangle - \langle a_1b_1\rangle \leq 2.
\end{equation}
Let us now analyze classical strategies. In order to maximize the CHSH expression $I_{CHSH}$,
we simultaneously want to
achieve the highest possible value for $\langle a_0b_0\rangle , \langle a_0b_1\rangle $ and
$ \langle a_1b_0\rangle$ and the lowest possible value of $\langle a_1b_1\rangle$.
It is easy to see that with deterministic assignments, we can achieve the best value for three
out of the four expressions. As an example consider a strategy such that for any question both
Alice and Bob answer $1$. Then $p_{AB|XY}(11|00) = p_{AB|XY}(11|10) = p_{AB|XY}(11|01) = 1$
which maximizes the first three expectation values, but also requires $p_{AB|XY}(11|11) = 1$.
As argued before, all the other local strategies can be seen as convex combinations
of such deterministic assignments and therefore the inequality holds.

Quantum strategy that violates this Bell inequality involves measuring the state
\begin{equation}
\ket{\Psi^+} = \frac{1}{\sqrt{2}} \left(\ket{00} + \ket{11}\right).
\end{equation}
The corresponding
measurements can be expressed by the following observables.
Alice's observables are
\begin{align}
A_0 = \left(\begin{matrix}
1&0\\0&-1
\end{matrix}\right) \quad  \text{and} \quad
A_1 = \left(\begin{matrix}
0&1\\1&0
\end{matrix}\right),
\end{align}
while Bob's observables are
\begin{align}
B_0 = \frac{1}{\sqrt{2}}\left(\begin{matrix}
1&1\\1&-1
\end{matrix}\right) \quad  \text{and} \quad
B_1 = \frac{1}{\sqrt{2}}\left(\begin{matrix}
1&-1\\-1&-1
\end{matrix}\right).
\end{align}
By the laws of quantum mechanics we have
\begin{equation}
\langle a_xb_y\rangle = \bra{\Psi^+}A_x\otimes B_y\ket{\Psi^+}.
\end{equation}
After doing the calculations
we can see that
\begin{equation}
\langle a_0b_0\rangle =
\langle a_0b_1\rangle =
\langle a_1b_0\rangle =
- \langle a_1b_1\rangle = \frac{1}{\sqrt{2}},
\end{equation}
yielding the value of the term
\begin{equation}
I_{CHSH} = {2\sqrt{2}},
\end{equation}
which is certainly larger than $2$.

Another well known Bell inequality we will extensively use in this paper is called
GHZ inequality \cite{GreenbergerHorneShimonyEtAl-Bellstheoremwithout-1990}.
We will introduce it in an alternative formalism used to describe quantum non-locality
inspired by game theory. This formalism is very useful and every Bell inequality can be expressed
as a game, including the CHSH game introduced previously as we will see in Subsection \ref{sec:ExpExp}.

\begin{figure}[tb]
\begin{center}
%\begin{framed}
\fbox{\includegraphics[width=6cm,clip]{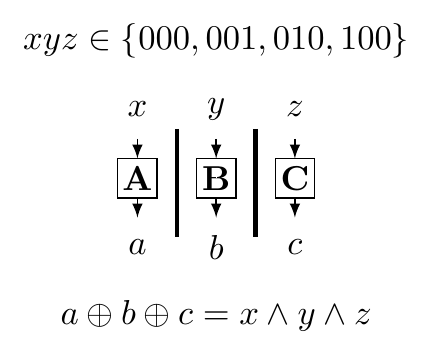}}
%\end{framed}
\end{center}
\caption{\it\small A GHZ test scenario. Three non-communicating boxes $\mathbf{A},\mathbf{B},\mathbf{C}$
each take a single bit input and each produce a single bit output. The test is successful if the condition
$a\oplus b\oplus c=x\wedge y\wedge z$ holds. The best classical strategy achieves $75\%$ success probability,
the best quantum strategy achieves success probability $1$ and produces two random bits -- random variables
$A$ and $B$ describing the outcomes of boxes $\mathbf{A}$ and $\mathbf{B}$
are fully random and independent. Random variable $C$ is correlated with $A$ and $B$ via the inputs.}
\label{fig:GHZscenario}
\end{figure}

The GHZ inequality requires three non-communicating parties and
it can be defined in terms of a three party game (see Fig.~\ref{fig:GHZscenario}).
Each of the three non-communicating boxes receives a single input bit  and produces a single output bit.
Let us denote the input bits of the respective boxes by
$x,$ $y$ and $z$ and the corresponding output bits by $a$, $b$ and $c$.
For the valid input combinations holds that $x\wedge y\wedge z=1$, \textit{i.e.} we consider only inputs
$xyz\in\{111,100,010,001\}$ simultaneously passed to all boxes. The value $v$
of the GHZ term is a function of the $4$ conditional probabilities $p_{ABC|XYZ}$ and the
joint probability distribution $p_{XYZ}$ of the inputs:
\begin{align}
v= &  \sum_{a\oplus b\oplus c = 1}p_{ABC|XYZ}(abc|111)p_{XYZ}(111)+\nonumber\\
+  &  \sum_{a\oplus b\oplus c = 0}p_{ABC|XYZ}(abc|100)p_{XYZ}(100)+\nonumber\\
+  &  \sum_{a\oplus b\oplus c = 0}p_{ABC|XYZ}(abc|010)p_{XYZ}(010)+\nonumber\\
+  &  \sum_{a\oplus b\oplus c = 0}p_{ABC|XYZ}(abc|001)p_{XYZ}(001). \label{eq:mermin}%
\end{align}
In particular, for the uniform input distribution we set
$p_{XYZ}(111)=p_{XYZ}(100)=p_{XYZ}(010)=p_{XYZ}(001)=\frac{1}{4}$ and denote the GHZ term by
$v_{u}$.

Assuming the uniform distribution on all four inputs, the maximal value of
$v_{u}$ achievable by classical device \cite{GreenbergerHorneShimonyEtAl-Bellstheoremwithout-1990}
is $\frac{3}{4}$
(thus the GHZ inequality reads $v_{u}\leq\frac{3}{4}$) and there exists a
classical device that can make any $3$ conditional probabilities
simultaneously equal to $1$. In the quantum world we can achieve $v_{u}=1$ and
satisfy perfectly all $4$ conditional probabilities using the tripartite GHZ
state $\frac{1}{\sqrt{2}}({|000\rangle}+{|111\rangle})$ and measuring
$\sigma_{X}$ ($\sigma_{Y}$) when receiving $0$ ($1$) on input.

The beautiful property of the GHZ inequality is that the violation $v$
gives us directly the probability that the device passes a test
\begin{equation}\label{eq:GHZWinCon}
a\oplus b\oplus c=x\wedge y\wedge z.
\end{equation}
The probability of failing
this test reads $1-v$. This property will be extensively used
in construction of device independent randomness amplification protocols of Section
\ref{sec:RandAmplif}.

The discussions about quantum non-locality have been part of quantum theory from the beginning and
many have considered it controversial \cite{EinsteinPodolskyRosen-CanQuantum-MechanicalDescription-1935}.
However, quantum violations of Bell inequalities have now been convincingly
verified in many experiments (see for example \cite{AspectGrangierRoger-ExperimentalRealizationof-1982}).

Now we are finally ready to formulate the main message of this Section --
\emph{violation of a Bell inequality guarantees at least some amount of randomness in outcomes
of the experiments}. This fact can be intuitively understood by the following
argumentation: Local model (\ref{LocCon}) is equivalent to a deterministic model, where to each setting $x,y$ and 
hidden variable $\lambda$
the outcomes $a$ and $b$ are deterministically assigned.
However, such model is excluded by the violation of a Bell inequality.
The observed correlations thus cannot be explained by deterministic assignments and therefore the measurement outcomes
are fundamentally undetermined.

This intuition certainly requires more clarification. In fact just as every
local explanation is equivalent to a deterministic local explanation, it can be shown that every non-local behavior
can be explained by a model that deterministically assigns outputs $a$ and $b$ depending on \emph{both}
measurement settings $x$ and $y$. However every such explanation is necessarily signaling -- if such explanation
were true, Alice would be able to infer some information about the outcome of Bob's spatially separated measurement
setting $y$ only by looking at her outcome $a$ and input $x$. Such interaction could be exploited to send signals faster
than the speed of light, which is deemed impossible by the theory of relativity, leaving us with the original explanation
-- outcomes of the measurements are fundamentally undetermined. What is more, this type of randomness can be certified --
any observed correlations violating some Bell inequality guarantee presence of randomness.

It is important to stress that to certify the randomness of outcomes we didn't have to assume anything
about the inner working procedures of the measurement devices or the source of measured particles --
violation of a Bell inequality itself
is sufficient. This is especially interesting for cryptography, in which it allows for reduction of the assumptions regarding the
security.
Quantum protocols which do not require the specification of the devices are called \emph{Device Independent (DI)}.
A variety of protocols with this property have been devised ranging from self testing \cite{MayersYao-Selftestingquantum-2004}
to quantum key distribution \cite{Ekert1991}
and random number generation, which is the main focus of this paper. For more thorough introduction to
Bell inequalities and device independent outlook on quantum physics see
excellent surveys~\cite{BrunnerCavalcantiPironioEtAl-Bellnonlocality-2014,Scarani-device-independentoutlookquantum-2012}.

%%%%%%%%%%%%%%%%%
\subsection{The set of quantum behaviors and the no-signaling set}\label{sec:QuantumSet}
%%%%%%%%%%%%%%%%%%%%%

In this subsection we will define the set of behaviors achievable by generalized
measurement of a bipartite quantum state,
which then can be easily generalized to more parties.
Let us define a bipartite quantum behavior as a vector of probability
distributions $p_{AB|XY}(ab\vert xy)$, which can be obtained by a measurement
of a bipartite quantum system:
\begin{equation}\label{QuantDist}
p_{AB|XY}(ab|xy) = Tr(\rho M^x_a \otimes M^y_b),
\end{equation}
where $\rho\in \mathcal{H}_A \otimes \mathcal{H}_B$ is a bipartite quantum state of arbitrary dimension,
and $M^x_a$ and $M^y_b$
are elements of Alice's POVM $M^x = \{M^x_a\vert a\in\mathcal{A}\}$ and Bob's POVM $M^y = \{M^y_b\vert b\in\mathcal{B}\}$ respectively.

It is interesting to study the set $\mathcal Q$ of quantum behaviours in the context of Bell inequalities. For example it is very
interesting to ask what is the maximal violation of Bell inequalities with quantum resources. In fact, the
state and measurements presented in the previous subsection achieves the maximum violation of
the CHSH inequality as shown in \cite{Cirel'son-Quantumgeneralizationsof-1980}.
Generally, unlike the set of local correlations $\mathcal L$,
the set of quantum behaviors $\mathcal Q$ is not a polytope and is quite difficult to characterize.
In a seminal paper of Navascu\'es~\textit{et.~al} \cite{Navascues2008} the authors
introduced an infinite hierarchy of semi-definite conditions $C_i$, $i= 1,2,\dots$, which are
necessarily satisfied by all probabilities of the form (\ref{QuantDist}). The number
of conditions rises with the index $i$, however the higher in the hierarchy the conditions are,
the more precise is the characterization of the set $p_{AB|XY}(ab\vert xy) = Tr(\rho M_a^x\otimes M_b^y)$.
This formulation allows solving optimization problems over the set of quantum behaviors
to arbitrary precision using the technique called semi-definite programming.

\begin{figure}[!tb]
\vspace*{-0.3cm}
\begin{center}
\includegraphics[width=8.6cm,clip]{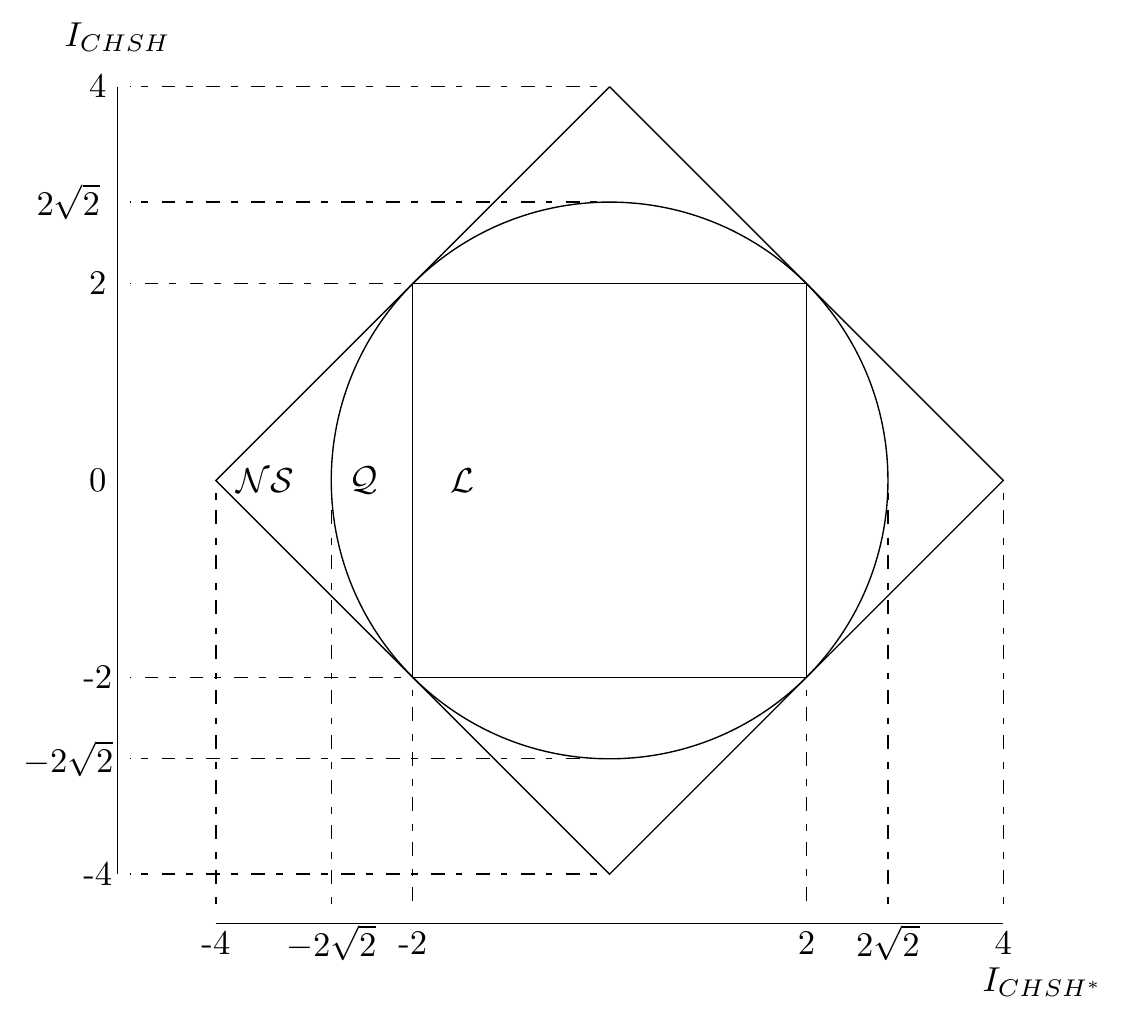}
\vspace*{-0.2cm}
\caption{\small\it A two-dimensional cut of the no-signaling polytope,
where $\mathcal L$ is the local polytope, $\mathcal Q$ is the set of quantum behaviors and $\mathcal{NS}$
is the non-signaling polytope. The values for two Bell-type expressions are
plotted in the figure, $I_{CHSH}$ is the value of the CHSH expression \ref{CHSH}
and $I_{CHSH^*}$ is it's symmetric version with labels $0$ and $1$ interchanged. Notice that a no-signaling
distribution can achieve the CHSH value of 4, which is it's algebraic maximum.}
\label{LQNS}
\end{center}
\vspace*{-0.4cm}
\end{figure}

The last set of correlations that is often studied in the literature is the set of 
\emph{no-signaling probability distributions} denoted $\mathcal{NS}$.
The no-signaling set of distributions requires only that $p(ab\vert xy)$ are proper probability distributions,
\textit{i.e.} $\forall a,b,x,y: p_{AB|XY}(ab\vert$ $xy) \geq 0$, 
$\forall x,y: \sum_{a,b} p_{AB|XY}(ab\vert xy) = 1$, and the no-signaling condition:
\begin{align}
\forall a,x,y,y' & &\sum_{b = 1}^{M_B} p_{AB|XY}(ab\vert xy) = \sum_{b = 1}^{M_B} p_{AB|XY}(ab\vert xy')\nonumber\\
\forall b,x,x',y & &\sum_{a = 1}^{M_A} p_{AB|XY}(ab\vert xy) = \sum_{a = 1}^{M_A} p_{AB|XY}(ab\vert x'y).
\end{align}
This condition expresses the inability to utilize these correlations to send signals. The set of
all no-signaling distributions is again a polytope, and a strict hierarchy can be shown (see Fig.~\ref{LQNS}):
\begin{equation}
\mathcal L\subset \mathcal Q\subset \mathcal {NS}.
\end{equation}

The advantages of considering the no-signaling set are twofold. First of all, because of the fact that the set forms a polytope
it might make the analysis easier. Moreover, for certain impossibility theorems for quantum set, it is sufficient
to prove the theorems for a larger set -- the no-signaling set. The second advantage is that theorems proven
for no-signaling set will hold even against possible post-quantum theories, which might allow stronger than
quantum correlations. This is especially interesting in the field of cryptography, where it is desirable to
construct cryptosystems secure also in the presence of  possible future theories.

\setcounter{equation}{0} \setcounter{figure}{0} \setcounter{table}{0}\newpage
%%%%%%%%%%%%%%%%%%%
%%%%%%%%%%%%%%%%%%%%%%%%%%%%%%%%%%%%
\section{Randomness expansion}\label{sec:RandExp}
%%%%%%%%%%%%%%%%%%%%%%%%%%%
%%%%%%%%%%%%%%%%%%%%
In this section we will discuss protocols for generating random numbers in a device independent way.
The crucial requirement for the  randomness expansion protocols to work properly is the existence of
a short random seed.
Part of this seed is used to randomly choose settings in a Bell experiment and
after verification that obtained outcomes violate a Bell inequality. If the violation is detected we know
that the outcomes of the experiment contain some amount of entropy which we need to estimate.
Then the rest of the preexisting randomness
is used for classical post-processing of the outcomes via seeded randomness extractors. The result is a random
string, which is longer than the initial seed, hence the name -- \emph{randomness expansion}.

%%%%%%%%%%%%%%
\subsection{Quadratic expansion}
%%%%%%%%%%%%%%%

These protocols were first suggested by Colbeck \cite{ColbeckKent-Privaterandomnessexpansion-2011},
but we use the protocol of
Pironio~\textit{et.~al.} \cite{PironioAc'inMassarEtAl-Randomnumberscertified-2010}
for introduction.
The protocol is based on a CHSH experiment.
Intuitively, the greater the violation $I_{CHSH}$ of the CHSH inequality (\ref{CHSH}) is, the more randomness has been
produced in the outcomes of the experiment.
Although this intuition is not entirely correct as later 
shown in \cite{Ac'inMassarPironio-RandomnessversusNonlocality-2012}, where
the maximum production of randomness has been achieved by a non-maximal CHSH violation.

At first suppose the black-boxes performing the CHSH test were the same in each of $n$ rounds  of the protocol
(a round is a single instance of the CHSH experiment)
and for simplicity assume, a Bell violation $I_{CHSH}$ of the underlying measurement process is given.
Let $A$ and $B$ be the random variables describing the outputs of a single run of the experiment.
Measure used to quantify the amount of randomness present in $A$ and $B$ is the
min-entropy  $H_\infty(AB\vert xy) = -\log_2 \max_{ab} p(ab\vert xy)$, conditioned on the inputs
$x$ and $y$.
Recall that the amount of nearly perfectly random bits obtainable
from a partially random source is roughly equal to it's min-entropy (see Def. \ref{def:ME}).
The aim is therefore to obtain a lower bound $f(I_{CHSH})$ on min-entropy, if the process achieves violation $I_{CHSH}$
of the CHSH inequality:
\begin{equation}
\forall x,y \quad H_\infty(AB\vert xy) \geq f(I_{CHSH}).
\end{equation}
Having such lower bound for a single run, the min-entropy of all the outputs is at least $nf(I_{CHSH})$ and
a randomness extractor can be used to transform this randomness into $O(nf(I_{CHSH}))$ bits that are close
to being uniformly distributed and uncorrelated to any information the adversary may hold.

In the following we will show how to obtain the lower bound $f(I_{CHSH})$.
Let us label $p^*(ab\vert xy)$ the maximum value of $p(ab\vert xy)$, where $x,y$ are given and the
maximum is taken over all possible values of $a$ and $b$ and all possible
quantum probability distributions that achieve
CHSH violation of value $I_{CHSH}$. Such maximization problem can be written in a form
\begin{align}
\label{max}
p^{*}(ab\vert xy) = &\quad\max & p(ab\vert xy)\\
\label{CHSHcondition}
& \quad\textrm{subject to} & \langle a_0b_0\rangle + \langle a_0b_1\rangle +
\langle a_1b_0\rangle - \langle a_1b_1\rangle = I_{CHSH}\\
\label{quant}
& & p(ab\vert xy) = Tr(\rho M_a^x\otimes M_b^y).
\end{align}
Recall that Eq. (\ref{CHSHcondition}) is a linear constraint and can be written as a linear combination
of probabilities of the form
$\sum c_{abxy} p(ab\vert xy) = I_{CHSH}$, where $c_{abxy}$ are constants.
The only remaining complication is to show that it is possible to
express Eq. (\ref{quant}) in a more useful way.
This can be done by the techniques of Navascues~\textit{et.~al.} \cite{Navascues2008} introduced in previous section.
We can characterize the approximation of the quantum set by semi-definite conditions
$C_i$.
By doing so, we can obtain
relaxations of the original problem $R_i$ in the form of semi-definite programs (SDP).
Moreover, the higher in the hierarchy $C_i$ is, the more precise upper bound on $p^*(ab\vert xy)$
we can obtain, which in turn gives us better lower bounds on the min-entropy $H_\infty(AB\vert xy)$.
Expressing optimization problems in forms of SDP guarantees that we can find the global maximum
with arbitrary precision.
In order to obtain a lower
bound that does not depend on the inputs, we need to calculate the maximum for all combinations of
inputs $x$ and $y$.
The authors of \cite{PironioAc'inMassarEtAl-Randomnumberscertified-2010} used these techniques to
derive the following lower bound:
\begin{equation}
\forall x,y \quad H_\infty(AB\vert xy) \geq f(I_{CHSH}) = 1 - \log_2\left(1+\sqrt{2 - \frac{I_{CHSH}^2}{4}} \right).
\end{equation}

The scenario given above is however only an idealization for the case of known CHSH violation $I_{CHSH}$.
In a real world protocol, to obtain the CHSH violation we would need infinite number of rounds,
and moreover, the measurements and the measured state in round $i$ can in principle depend
on the data from previous rounds. This needs to be dealt with using a statistical approach
which takes into account such memory effects.
What we do is that we repeat the test $n$ times and estimate the observed violation $I_{CHSH}^{obs}$.
It is given by
\begin{equation}\label{CHSHestimate}
I_{CHSH}^{obs} =  [a_0b_0] + [a_0b_1] + [a_1b_0] - [a_1b_1],
\end{equation}
where $[a_xb_y] = \sum_{ab} ab \frac{N(ab\vert xy)}{n p(xy)}$
and $N(ab\vert xy)$ is the number of times outcomes $a$ and $b$ have been
observed after measuring $x$ and $y$ and $p(xy)$ is the probability of a pair $x,y$ appearing as an input.
However, in finite number of rounds such violation can be obtained with a positive probability
even with a fully deterministic strategy. Authors of \cite{PironioAc'inMassarEtAl-Randomnumberscertified-2010}
were able to upper bound, by $\delta$,
the probability that the value of $I_{CHSH}^{obs}$ deviates from the real value of Bell
violation $I_{CHSH}$  by more than  $\varepsilon$:
\begin{equation}\label{securityParameter}
\delta = \exp\left(-\frac{n\varepsilon^2}{2(1/q + I_q)^2}\right),
\end{equation}
where $I_q$ is the maximum obtainable quantum violation of a Bell inequality and $q = \max_{xy} p(xy)$
is the probability of the most probable measurement setting. Combining all the previous results,
the min-entropy of the produced string $R = (a_1, b_1; a_2, b_2;\dots;  a_n,b_n)$ given all
the inputs $D = (x_1, y_1;x_2, y_2; \dots; x_n, y_n)$
can be bounded from below by
\begin{equation}\label{MinEntropyLowerBound}
H_\infty(R\vert D)\geq nf(I_{CHSH}^{obs}-\varepsilon),
\end{equation}
with probability more than $1-\delta$.
What is more, the proof holds even for
the case of different measurements and measured states in each round.
The whole protocol is summarized in Fig.~\ref{fig.QuadraticExpansion}:

\begin{figure}[tb]
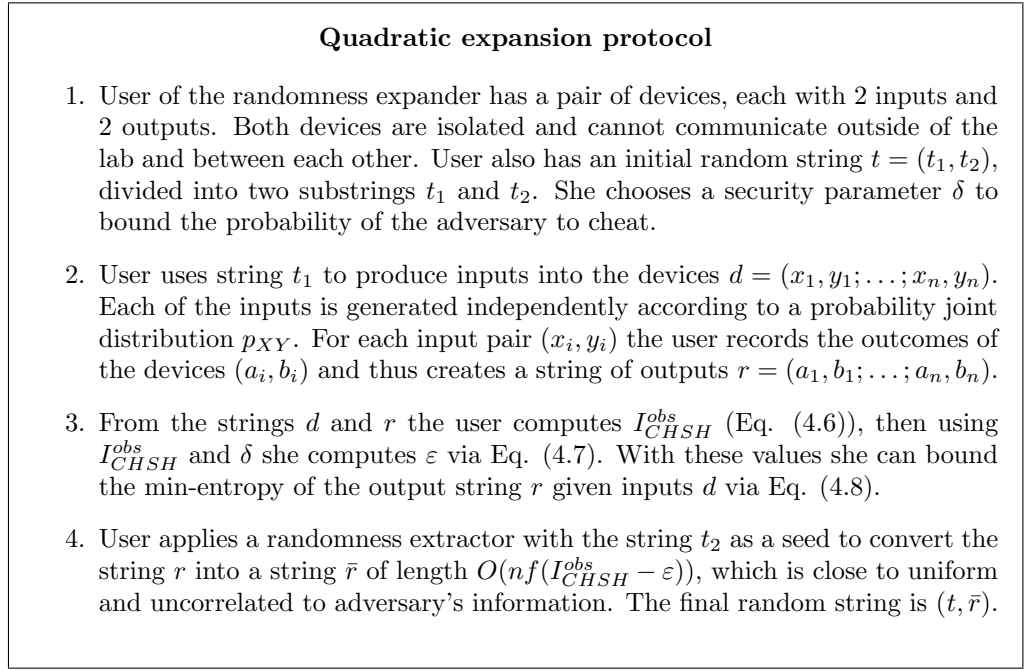

\begin{framed}
\begin{center}
\textbf{Quadratic expansion protocol}
\end{center}
\begin{enumerate}
\item User of the randomness expander has a pair of devices,
each with 2 inputs and 2 outputs. Both devices are isolated and cannot
communicate outside of the lab and between each other.
User also has an initial random string $t = (t_1,t_2)$,
divided into two substrings $t_1$ and $t_2$. She chooses a security parameter $\delta$ to bound the probability
of the adversary to cheat.
\item User uses string $t_1$ to produce inputs into the devices $d = (x_1,y_1;\dots;x_n, y_n)$. Each of the
inputs is generated independently according to a probability joint distribution $p_{XY}$. For each input pair $(x_i,y_i)$
the user records the outcomes of the devices $(a_i,b_i)$ and thus creates a string of outputs
$r = (a_1,b_1;\dots;a_n,b_n)$.
\item From the strings $d$ and $r$ the user computes $I_{CHSH}^{obs}$ (Eq. (\ref{CHSHestimate})), then using
$I_{CHSH}^{obs}$ and $\delta$
she computes $\varepsilon$ via Eq. (\ref{securityParameter}). With these values she can bound the min-entropy of the output
string $r$ given inputs $d$ via Eq. (\ref{MinEntropyLowerBound}).
\item User applies a randomness extractor with the string $t_2$ as a seed
to convert the string $r$ into a string $\bar{r}$
of length $O(nf(I_{CHSH}^{obs}-\varepsilon))$, which is close to uniform and uncorrelated to adversary's information.
The final random string is $(t,\bar{r})$.
\end{enumerate}
\end{framed}
\caption{\it\small Quadratic expansion protocol of Pironio~\textit{et.~al.} \cite{PironioAc'inMassarEtAl-Randomnumberscertified-2010}.}
\label{fig.QuadraticExpansion}
\end{figure}

In order to asses the efficiency of the scheme, we need to examine the length of the final random string.
It can be shown that if $n$ is large enough, it is possible to start with a short random seed of the length
$O(\sqrt{n}\log_2\sqrt{n})$ to produce a string of length $O(n)$.
It turns out that to achieve such quadratic expansion it is crucial not to choose measurement settings
with uniform probability. Indeed, if the user chooses one of the possible inputs with probability $1-3q$ and
the other three with probability $q$, with $q$ small, the randomness required to generate the inputs is then
equal to $nO(-q\log_2 q)$. Choosing $q = 1/\sqrt{n}$ thus requires $O(\sqrt{n} \log_2\sqrt{n})$ random
bits. On the other hand, amount of randomness in the string $r$ is given by (\ref{MinEntropyLowerBound}),
with $\varepsilon$ equal to $O(1)$. Hence, for a constant Bell violation the string $r$ contains $O(n)$ bits of randomness.
The protocol thus achieves quadratic expansion.

Protocol introduced by Pironio~\textit{et.~al.} \cite{PironioAc'inMassarEtAl-Randomnumberscertified-2010} was the first protocol with rigorous proof of security and what is more,
in order to show the concept of device independent randomness expansion is viable with current technology,
the authors also implemented their protocol and were able to obtain 42 new random bits with 99\% confidence.
However, the protocol still had some weaknesses to be fixed. First of all the protocol is not proven to be
\emph{universally composable} against a full quantum adversary -- the bound (\ref{MinEntropyLowerBound}) is
derived against adversaries that measure their quantum systems prior to the randomness extraction.
Recall that full quantum adversaries might store the side-information in a quantum memory and measure their systems
later, perhaps after a part of the extracted string has been revealed (as a part of another cryptographic
protocol, \textit{e.g.} privacy amplification). Secondly, this protocol rises a question,
whether another, more efficient protocol can be designed -- a protocol that achieves exponential, or even unbounded
expansion. Both of these questions have been resolved in subsequent work and we will examine them in the rest of
this section.

%%%%%%%%%%%%%%%%%%%%%%
\subsection{Exponential expansion}\label{sec:ExpExp}
%%%%%%%%%%%%%%%%%%%%%%%

We will present the protocol of Vidick and Vazirani \cite{vazirani2012certifiable}. In order to do so we will first reformulate
the CHSH inequality in a language of game theory. Alice and Bob are now a non-communicating players
that receive an input $x,y\in\{0,1\}$ and reply with bits $a,b\in\{0,1\}$. They win if and only if
\begin{equation}\label{CHSHalt}
a\oplus b = x \wedge y,
\end{equation}
where $\oplus$ is sum modulo $2$ and $\wedge$ is the logical AND.
This is actually only a differently expressed CHSH inequality (\ref{CHSH}).
All  classical strategies (\textit{i.e.} any strategy Alice and Bob could agree on
before they start playing) achieve at most $75\%$ probability to win the game. On the other hand,
if Alice and Bob share an entangled state, they can achieve better probability of winning -- about $85\%$.
These properties of quantum and classical strategies are a simple corollary of the CHSH inequality
presented in the previous section. The protocol for randomness expansion now consists of playing
multiple rounds of the game with two devices with random inputs and determining the probability
of winning from the outcomes. If the estimated probability
to win is more than $75\%$, the devices have used a quantum strategy,
which implies that some randomness was produced.
The difficult part is again to make these statements quantitative.

First of all let us discuss a protocol that can achieve exponential expansion against an adversary without
quantum memory. The first crucial idea to design
such protocol was already hinted in the quadratic expansion protocol of previous subsection --
do not choose inputs into devices with uniform distribution.
In the Vazirani-Vidick protocol authors use default inputs, \textit{i.e.} $x = y = 0$, into the devices in a significant portion of
the rounds and use random choices of $x$ and $y$ only occasionally, in order to check if the devices
use quantum strategy. This can obviously help to use shorter initial random seed, however makes
testing of the CHSH condition more cumbersome. For example consider a protocol in which only a small
fraction of the inputs is chosen randomly, while the rest of the inputs is fixed to $x=0$ and $y=0$.
If the devices only output $a=0$ and $b=0$ in every round, the CHSH condition would be
satisfied with probability almost $1$, on average over the whole protocol. This demonstrates the
need for a more sophisticated way
to check the CHSH condition, as we will see in the following protocol.

Let $n$ be the length of a random string to be generated and $\delta$ a security parameter.
Divide all the runs in the protocol into $m = C\lceil n\log (1/\delta)\rceil$ blocks of the length
$k = 10\lceil\log^2 n\rceil$, where $C$ is a large constant. Inputs in a given block consist of a fixed
pair $(x,y)$ repeated in the whole block. For most of the blocks the input is $(0,0)$ and
only for randomly chosen blocks, called Bell blocks, the inputs $(x,y)$ are chosen with uniform probability.
The number of Bell blocks is approximately $\Delta = 1000\lceil\log (1/\delta)\rceil$. Authors introduced the
blocks of input in order to check the CHSH condition (\ref{CHSHalt}) in a more sophisticated way -- the condition
needs to be fulfilled by at least $84\%$ of outputs in \emph{each block} in order for the protocol
to pass the CHSH test.

The formal statement guarantees the existence of a constant $C$, such that the following holds.
%Let the security parameter $\delta>0$ and the intended
%number of generated bits $n$ be given.
%Then there exists a constant $C$, such that  if we set
%$\Delta = 1000\lceil\log(1/\delta)\rceil$ and $\ell = Cn$ as the parameters of the protocol depicted
%in Fig. \ref{fig:ExponentialExpansion}, the following statement holds.
Let $A$ and $B$ be random variables describing the output of the two devices used for randomness generation
and $P_{CHSH}$ an event in which the protocol passes the CHSH test.
For all large enough $n$ at least one of the following holds:
\begin{align}
&\textrm{Either}& H_\infty(B\vert P_{CHSH}) \geq n\\
&\textrm{Or} &\Pr[P_{CHSH}]\leq\delta
\end{align}
We will omit the proof of this statement, and direct the reader to the original paper \cite{vazirani2012certifiable}.
With this qualitative statement we now can present the protocol, which is described in Fig. \ref{fig:ExponentialExpansion}.

\begin{figure}[tb]
\begin{framed}
\begin{center}
\textbf{Exponential expansion}
\end{center}
\begin{enumerate}
\item Compute $\ell = Cn$ and $\Delta =1000\lceil\log(1/\delta)\rceil$ from inputs $n$ and $\delta$.
Set $k = \lceil 10 \log^2 \ell \rceil$ and $m = \Delta\ell$.
\item Choose the Bell blocks $T\subseteq \{1,\dots,m\}$ by randomly selecting each block with probability $1/\ell$.
Repeat, for $i = 1,\dots, m$
	\begin{enumerate}
	\item If $i\notin T$, then
		\begin{enumerate}
			\item Set $(x,y) = (0,0)$ as inputs for $k$ consecutive rounds of CHSH test and collect the outputs $(a,b)$.
			\item If $a\oplus b$ has more than $\lceil0.16k\rceil$ 1's then reject and abort the protocol, otherwise continue.
		\end{enumerate}
	\item If $i\in T$, then
		\begin{enumerate}
			\item Choose $(x,y)\in\{0,1\}^n$ uniformly at random and use it as input for $k$ rounds of CHSH test. Collect the
			outputs $(a,b)$.
			\item If $a\oplus b$ differs from $x\wedge y$ in more than $\lceil0.16k\rceil$ rounds, reject and abort the 					protocol. Otherwise continue.
		\end{enumerate}
	\end{enumerate}
\item If all steps accepted, then accept.
\end{enumerate}
\end{framed}
\caption{\it\small Exponential expansion protocol of  Vidick and Vazirani \cite{vazirani2012certifiable}.}
\label{fig:ExponentialExpansion}
\end{figure}

Notice that only $O(\Delta \log\ell)$ bits of the initial randomness were used.
In step (2.) $O(\Delta\log\ell)$ bits were used to choose the Bell blocks
and in step (2.b.i.) 2 bits per Bell block of randomness were used, \textit{i.e.} $O(2\Delta\log\ell)$ bits altogether.
Taking into account $O(\log n)$ bits needed for randomness extraction,
we only need $O(\log n)$ bits of initial randomness to produce $O(n)$ bits of perfect randomness.
This protocol however is again secure only against an adversary without quantum memory.

%%%%%%%%%%%%%%%%%%%%%%%%%%%%%%%%%%%%%%%%%%%%
\subsection{Exponential expansion against full quantum adversaries}\label{sec:QSExp}
%%%%%%%%%%%%%%%%%%%%%%%%%%%%%%%%%%%%%%%%%%%

In order to achieve full composable security against quantum adversaries we need two ingredients.
First of all we need to
grant the adversary a quantum system $\rho_E$, possibly entangled
to the devices used for the Bell test and a protocol that can guarantee that the outputs of the Bell test
contain some entropy even conditioned on this quantum system.
This is not a trivial task, especially in the composable security setting, where part of
the produced string can later be revealed. It been shown that after revealing part of the generated random string,
as in the case in some cryptographic protocols -- privacy amplification in quantum key distribution being the prime
example --
the measurement of the system $\rho_E$ enables the prediction of the rest of the string with
inverse polynomial probability \cite{vazirani2012certifiable}. This is much higher than the inverse exponential 
probability that is available by measuring $\rho_E$ without any advice bits.

The second ingredient is a randomness extractor
which is secure even in the presence of quantum side information. We have already discussed
existence of such extractors in Subsection \ref{sec:QWS}.

The authors of \cite{vazirani2012certifiable} were able to construct a protocol with desired properties,
using a modified CHSH game. In the game each of the players receives one of three possible inputs,
which correspond to three measurement settings used in the CHSH game. They are labeled
$(A,0),(A,1),(B,0)$, where $(A,0)$ denotes the measurement Alice would perform
in a CHSH test, if her input was 0 and similarly for the others. The honest strategy is to perform
measurements and use the state as in honest CHSH inequality as defined in Subsection \ref{BellSection}.
Important properties are that whenever $x = y$, the expected results are $a=b$, whenever
$x = (A,1)$ and $y = (A,0)$, the outcomes are expected to be equal with probability $\frac{1}{2}$
and whenever $x = (A,0)$ and $y = (B,0)$, then according to the CHSH game, about $85\%$
of the outcomes are expected to be the equal. Any other combination of settings does not appear
in the protocol.

\begin{figure}[tb]
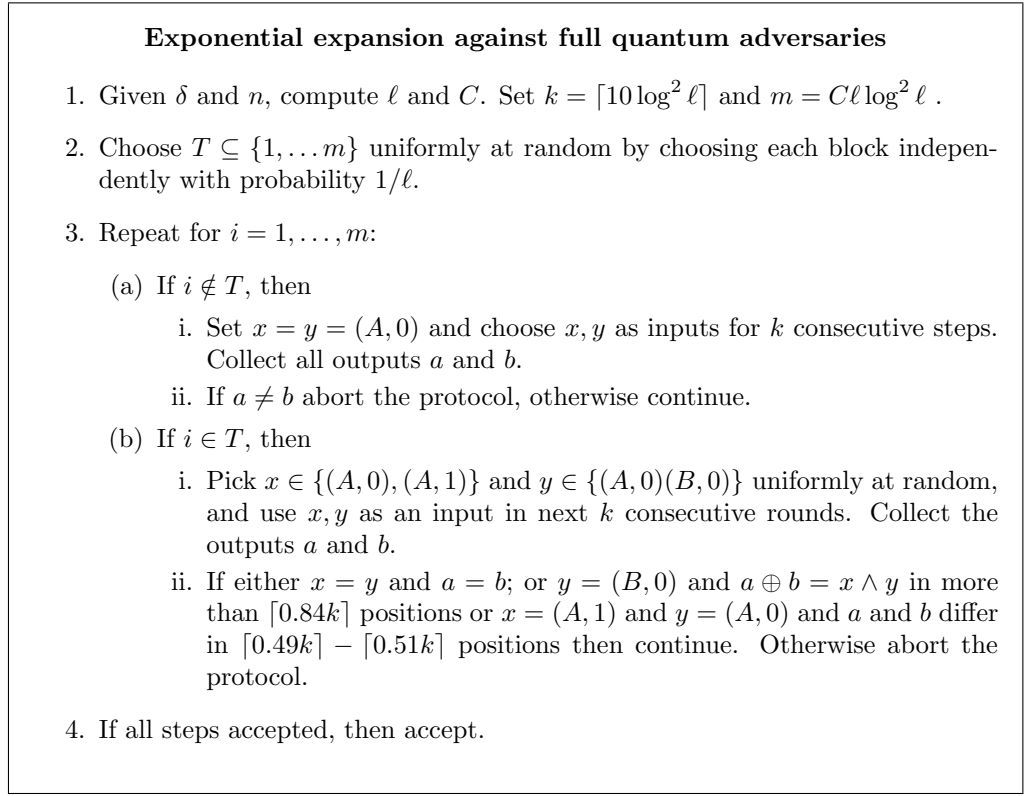

\begin{framed}
\begin{center}
\textbf{Exponential expansion against full quantum adversaries}
\end{center}
\begin{enumerate}
\item Given $\delta$ and $n$, compute $\ell$ and $C$.
Set $k = \lceil 10 \log^2\ell\rceil$ and $m = C\ell\log^2\ell$ .
\item Choose $T\subseteq \{1,\dots m\}$ uniformly at random by choosing each block independently with probability
$1/\ell$.
\item Repeat for $i = 1,\dots,m$:
	\begin{enumerate}
	\item If $i\notin T$, then
		\begin{enumerate}
		\item Set $x = y = (A,0)$ and choose $x,y$ as inputs for $k$ consecutive steps. Collect
		all outputs $a$ and $b$.
		\item If $a\neq b$ abort the protocol, otherwise continue.
		\end{enumerate}
	\item If $i\in T$, then
		\begin{enumerate}
		\item Pick $x\in\{(A,0),(A,1)\}$ and $y\in\{(A,0)(B,0)\}$ uniformly at random, and use $x,y$ as an
		input in next $k$ consecutive rounds. Collect the outputs $a$ and $b$.
		\item If either $x = y$ and $a=b$; or $y = (B,0)$ and $a\oplus b = x\wedge y$ in more than $\lceil 0.84k \rceil$
		positions or $x = (A,1)$ and $y=(A,0)$ and $a$ and $b$ differ in $\lceil 0.49k\rceil - \lceil 0.51k \rceil$ positions then
		continue. Otherwise abort the protocol.
		\end{enumerate}
	\end{enumerate}
\item If all steps accepted, then accept.
\end{enumerate}
\end{framed}
\caption{\it\small Exponential expansion protocol of  Vidick and Vazirani \cite{vazirani2012certifiable} secure against full quantum adversaries. }
\label{fig:ExponentialExpansionQuantum}
\end{figure}

The protocol is again performed in $m$ blocks of size $k$.
Most of the blocks have constant input $(A,0)$ and a randomly chosen subset $T\subseteq \{1,\dots m\}$
of Bell blocks has inputs chosen uniformly at random -- Alice gets either $(A,0)$ or $(A,1)$ and Bob gets either
$(A,0)$ or $(B,0)$.
The protocol takes as an input $n$ -- the number of bits to be produced, and the security parameter $\delta$.
The quantitative claim uses additional constants $\alpha, \gamma,\ell$ and $C$,
which can be directly calculated from $n$ and $\delta$, in order to calculate the number and length of the blocks.
Let $\alpha,\gamma > 0$, $\alpha = -\log\delta$ and $\gamma \leq 1/(10+8\alpha)$. Set $C = \lceil 100\alpha\rceil$,
and $\ell = n^{1/\gamma}$. Let $P_{CHSH}$ be an event in which the protocol output is accepted and
$B'$ a random variable
describing Bob's output bits conditioned on $P_{CHSH}$. Let $\rho_E$ be an arbitrary quantum system, possibly
entangled with devices executing the protocol. Then for large enough $n$ at least one of the following holds
\begin{align}
&\textrm{Either}& H_\infty(B'\vert \rho_E) \geq n\\
&\textrm{Or} &\Pr[P_{CHSH}]\leq\delta.
\end{align}
The full protocol is described in Fig. \ref{fig:ExponentialExpansionQuantum}.

Let us analyze the length of a random seed needed in the protocol.
To choose the set $T$, we need $O(\log^3\ell)$ bits. In each of the Bell blocks, the number of bits needed
to choose the inputs is $4$. All in all we need $O(\log^3\ell)$ bits of randomness to produce $O(\ell^\gamma)$
bits, where $\gamma$ is a constant depending directly on the security parameter $\delta$.

The protocols we just presented managed to improve the original proposal \cite{PironioAc'inMassarEtAl-Randomnumberscertified-2010} in both
providing super-polynomial expansion and security against full quantum adversaries.
However, their drawback is that they are not \emph{robust}. Notice that in both protocols the tolerated
deviation from the full quantum strategy is very low --  by using  honest quantum
strategy the devices can fulfill the CHSH condition (\ref{CHSHalt}) in about $85\%$ of the rounds
and the accepted success rate in the protocol is only $84\%$. This deems the protocols to be
unpractical.
This issue was later addressed  by Miller and Shi \cite{MillerShi-RobustProtocolsSecurely-2014}, who designed a
\emph{robust} protocol with exponential expansion secure against quantum adversaries, which, as we will see in the
next section, is suitable for protocol concatenation.

%%%%%%%%%%%%%%%%%%%%%%
\subsection{Concatenation of protocols and unbounded expansion}\label{subsec:concat}
%%%%%%%%%%%%%%%%%%%%%%%

After having designed a randomness expansion protocol, one of the most natural questions
is if several expansion devices $D_0,D_1,\dots, D_n$ can be chained together such that output of device $D_i$ --
almost perfectly random string -- is used as input into device $D_{i+1}$.
In this way the resulting output  could be much longer with multiple devices,
ultimately leading to unbounded expansion.

The concatenation idea was present in the work on randomness expansion
from the beginning \cite{ColbeckKent-Privaterandomnessexpansion-2011,PironioAc'inMassarEtAl-Randomnumberscertified-2010},
 but was seriously analyzed
for the first time by Fehr~\textit{et.~al.} \cite{FehrGellesSchaffner-Securityandcomposability-2013}. They used concatenation
with quadratically expanding protocol secure against adversary holding classical information
in order to obtain polynomially expanding protocol. We will examine their
protocol in more detail as it nicely demonstrates the difficulties of expansion
protocol concatenation.

First of all, let us split the expansion protocol into two components -- expansion and extraction.
The expanding component uses black-box devices and if successful, produces
a string with high min-entropy towards both the adversary and the input, which is not
necessarily uniformly distributed.
The extraction component takes this string with high entropy and transforms it
into the outcome of the protocol -- a string that is almost uniform towards the
adversary. Note that both components require private seed $S$ -- a string uniform towards
both the device and the adversary.

\begin{figure}[tb]
\begin{center}
\includegraphics[width=6cm,clip]{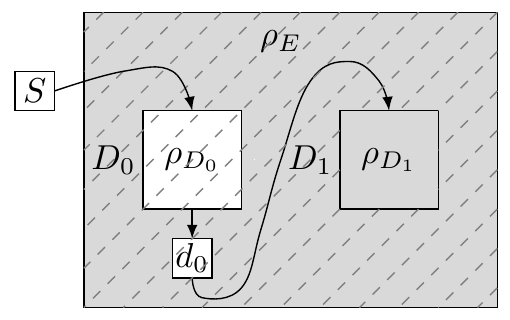}
\end{center}
\vspace*{-0.5cm}
\caption{\small \it The adversarial system of $D_0$ is colored gray in the picture and the adversarial system of $D_1$ is dashed.
Original seed $S$ is secure against adversarial system of $D_0$. Input into $D_1$ is $d_0$, which is not secure against
 the whole adversarial system of $D_1$, namely it is not secure against $D_0$. If it can still be used, then expansion protocol
$D_1$ is input secure.}
\label{fig:protocolConcat}
\end{figure}

Let us now again
consider randomness expansion devices $D_0$ and $D_1$ and suppose that the whole
seed $S$ was used as an input into device $D_0$ to maximize the length of it's
output $d_0$. The adversarial system for the device $D_0$ consists of the system $\rho_E$
(the system the adversary holds) and system $\rho_{D_1}$ (internal system of $D_1$).
Since the original seed $S$
is secure against both of these systems as well as internal system $\rho_{D_0}$ of $D_0$,
the output $d_0$ is also secure against $\rho_{D_1}$ and $\rho_E$.
Now we would like to use $d_0$ as an input into device $D_1$. The adversarial system
for the device $D_1$ consists of $\rho_E$ and $\rho_{D_0}$. Note that device $D_0$
can hold a whole copy of $d_0$ in it's memory, therefore $d_0$ is secure against $\rho_E$ and
$\rho_{D_1}$, but not $\rho_{D_0}$ (see Fig. \ref{fig:protocolConcat}). Can we still use it as a seed for the device $D_1$?
If the answer is \emph{yes}, we call the protocol \emph{input secure}.

\begin{figure}[tb]
\begin{center}
\includegraphics[width=7cm,clip]{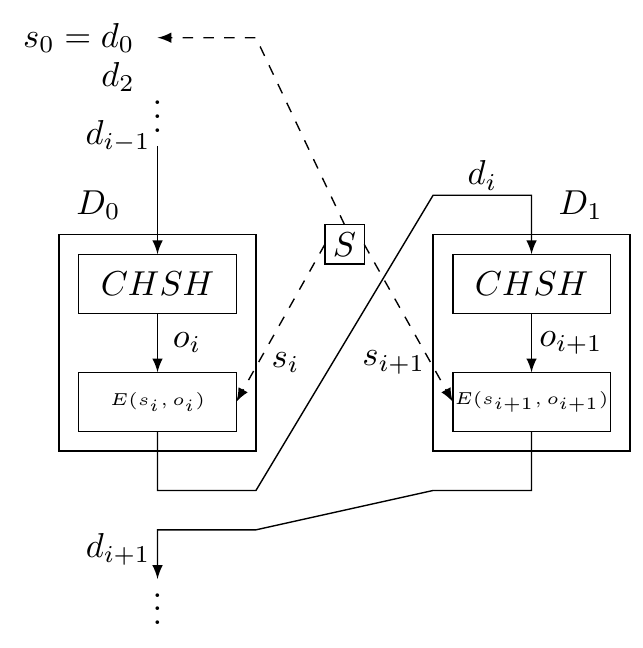}
\end{center}
\vspace*{-0.8cm}
\caption{\small\it A schematic drawing of concatenation protocol of Fehr~\textit{et.~al.}
\cite{FehrGellesSchaffner-Securityandcomposability-2013}. The protocol first inputs the first part of the seed $s_0$
into the device $D_0$. Then the devices are alternated such that the outcome $o_i$ of the expansion part in $i^{th}$
round is extracted with the use of a fresh seed $s_i$, before being used as input into $i+1^{st}$ round.
After $k$ rounds the output
of the whole protocol is obtained.}
\label{fig:protocolFehr}
\end{figure}

Fehr~\textit{et.~al.} \cite{FehrGellesSchaffner-Securityandcomposability-2013} 
designed a protocol, for which they have proven input security of the expansion part 
under the assumption of an classical adversary. Unfortunately they haven't been able to 
show input security for the extraction part. 
However, even in this setting they could use the protocol concatenation 
considering only two alternating
independent devices in order to achieve their goal of polynomial expansion.
Let us label the two devices $D_0$ and
$D_1$. Considering the model with the classical adversary we
assume that the devices $D_0$
and $D_1$ are not entangled together. Part of initial seed $S$, which is independent of the adversary and
both devices, is used with device $D_0$.
Since $D_1$ holds only classical information about $D_0$, the output of $D_0$ is guaranteed
to be almost random to $D_1$. As stated earlier, expansion component of $D_1$ will
output raw string with high min-entropy towards both the adversary and $D_0$, even though
it's input wasn't secure against $D_0$. However it is necessary to take the seed for the extraction
part of $D_1$ from the original seed $S$. In this way the two devices can alternate until fresh seeds
for the extractors are available and thus jointly create an output much longer than the
original protocol (see Fig. \ref{fig:protocolFehr}).

For a long time it was an open question
whether a full input secure expansion protocol exists.

\begin{figure}[tb]
\begin{center}
\includegraphics[width=5cm,clip]{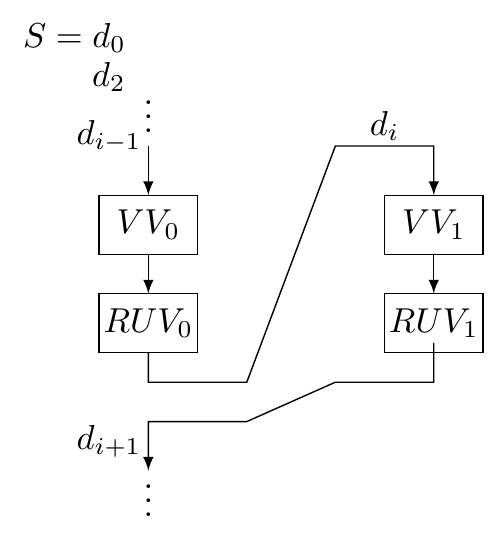}
\end{center}
\vspace*{-0.8cm}
\caption{\small\it Concatenation protocol of Coudron and Yuen \cite{CoudronYuen-InfiniteRandomnessExpansion-2014}.
The protocol uses two sub-protocols. The first one, denoted $VV$ is the Vazirani and Vidick protocol introduced in Subsection
 \ref{sec:QSExp} and the second one, denoted $RUV$ is an input secure protocol of Reichardt\textit{~et.~al.}
 \cite{ReichardtUngerVazirani-ClassicalLeashQuantum-2013}. The concatenation protocol is initialized with a random seed
$S$ and then the devices alternate in a way depicted in the scheme, until the desired number of rounds is reached.}
\label{fig:protocolCoudron}
\end{figure}

This question was concurrently solved by both Coudron and Yuen \cite{CoudronYuen-InfiniteRandomnessExpansion-2014}
 and Miller and Shi \cite{MillerShi-RobustProtocolsSecurely-2014}.
Coudron and Yuen proposed a protocol conceptually similar to that of Fehr.~\textit{et.~al.}
\cite{FehrGellesSchaffner-Securityandcomposability-2013}.
The protocol is alternating between two expansion devices $D_0$ and $D_1$ running
the quantum proof protocol of Vazirani and Vidick \cite{vazirani2012certifiable}. What makes the whole protocol
input secure is the fact that each output of an expansion protocol is first decoupled  (see Fig. \ref{fig:protocolCoudron})
from both devices $D_0$ and $D_1$ by another protocol taken from the work of
Reichardt~\textit{et.~al.} \cite{ReichardtUngerVazirani-ClassicalLeashQuantum-2013} (refered to as RUV protocol).
The RUV protocol is input secure
and therefore produces a random string even if it's seed is secure only against internal
state of the device running the protocol. The disadvantage of the RUV protocol is that it's output is actually
shorter than it's input. Nevertheless, the shrinking factor is only polynomial, therefore
coupling it with exponentially expanding Vazirani and Vidick protocol provides the
desired unbounded expansion with only four devices.
%To summarize
%the protocol alternates between devices $D_0$ and $D_1$ running
%the expansion protocol. The output of device $D_0$ is decoupled by device
%$D_2$ running the RUV protocol and similarly output of $D_1$ is decoupled
%by device $D_3$.

\begin{figure}[tb]
\begin{center}
\includegraphics[width=5cm,clip]{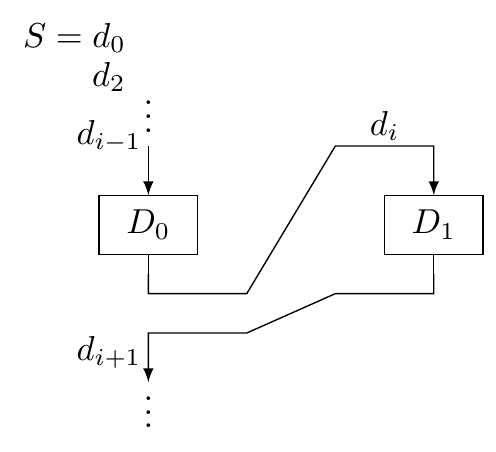}
\end{center}
\vspace*{-0.8cm}
\caption{\it\small Miller and Shi \cite{MillerShi-RobustProtocolsSecurely-2014} concatenation of expansion protocols.
They have shown that any device obtaining security with globally secure seed remains secure if used with seeds secure
against the device only. This fact allows for very simple alternation of two independent devices, resulting
in unbounded expansion.}
\label{fig:protocolShi}
\end{figure}

The solution of Miller and Shi \cite{MillerShi-RobustProtocolsSecurely-2014} heavily leans on a result by
Chung~\textit{et.~al.} \cite{ChungShiWu-PhysicalRandomnessExtractors-2014},
mainly their equivalence lemma. The equivalence lemma is very powerful and
somewhat surprisingly dodges the problem of input secure extractor components.
It treats the expansion protocol as a whole and shows that any expansion
protocol with globally secure seed retains the same parameters if used
with only a device secure seed. This lemma therefore automatically allows them
to use any expansion protocol in an alternating way without any other assumptions (see Figure \ref{fig:protocolShi}).

\subsection{Experimentally feasible adversaries}
To finish the review of the randomness expansion protocols we mention a set of papers
that deal with a slightly different view on the topic of device independence.

In this view the device independence is motivated by the fragility of quantum devices that might
easily lead to imperfect functioning rather then by an adversary in the system. Thus here we relax our assumption
from the all powerful adversary (within given limitations) to a model imperfect devices. Instead, the devices are expected
to be designed and constructed in an honest way, but might malfunction. But this malfunctioning is limited to carrying
out the expected tasks in a wrong way (or not carrying them out at all) rather than performing completely new tasks.
Thus, if there is \textit{e.g.} no quantum memory in the design of the protocol, the malfunctioning device will not 
be able to use it, as it is technically not possible.

Another motivation for examining this type of protocols comes from the fact that it is arguably very difficult to guarantee that the adversarially constructed devices do not contain classical transmitters of any sort. An dishonest provider would probably just install a device that would broadcast the final key in a classical way and not bother to break the security on the quantum level. Thus it has more sense to trust the provider to be honest, but possibly might be slovenly in the production process.

Let us first mention work of Pironio and Massar \cite{PironioMassar-Securityofpractical-2013} and
Fehr~\textit{et.~al.} \cite{FehrGellesSchaffner-Securityandcomposability-2013}.
Both papers used tools from \cite{PironioAc'inMassarEtAl-Randomnumberscertified-2010}
and improved their analysis of security against classical side information.
Moreover Fehr~\textit{et.~al.} \cite{FehrGellesSchaffner-Securityandcomposability-2013} developed the concatenation idea
described in the previous section and Pironio and Massar  \cite{PironioMassar-Securityofpractical-2013} argued
why classical security is sufficient in real world expansion protocols.

Even though their protocols aren't secure in full generality they still
provide a number of very useful properties. Traditional random number generators
suffer from several problems, which device independent randomness expansion
can successfully overcome. One of the most severe problems is monitoring the quality
of the output.
Deterioration of the output quality of random number generators is difficult to detect
and should be monitored constantly. As we have seen, this is an implicit property
of the randomness expansion protocols -- all the produced data had to pass a Bell test.
Another advantage is the estimation of entropy, which does not rely on any statistical
test -- violating a Bell inequality is the proof of randomness as such.

The work with similar adversarial setting was later continued in work by Bancal~\textit{et.~al.}
\cite{BancalSheridanScarani-Morerandomnessfrom-2014,BancalScarani-MoreRandomnessFrom-2014} and
Nieto-Silleras~\textit{et.~al.} \cite{Nieto-SillerasPironioSilman-Usingcompletemeasurement-2014}
who suggest techniques to certify more randomness from
the correlations observed during a non-locality test.

 In  \cite{SilmanPironioMassar-Device-IndependentRandomnessGeneration-2013} a different relaxation of the
assumptions was considered and authors constructed a protocol, in which the devices
used are allowed a small amount of communication and authors of
\cite{CoudronVidickYuen-RobustRandomnessAmplifiers:-2013} gave an upper and lower bound on expansion rates for different types
of protocols.

\setcounter{equation}{0} \setcounter{figure}{0} \setcounter{table}{0}\newpage
%%%%%%%%%%%%%%%
%%%%%%%%%%%%%%%%%%%%%%%%%%%%%%%
\section{Randomness amplification}\label{sec:RandAmplif}
%%%%%%%%%%%%%%%%%%%%%%%%%%%%%%%
%%%%%%%%%%%%%%%
In this section, we will focus on a task closely related to randomness expansion.
Recall that one of the crucial assumptions in randomness expansion protocols is the
existence of an independent random seed, which is used to choose the measurement settings
in a Bell test and as a seed for randomness extractor during the  classical post-processing.
The random and independent measurement choice corresponds to the assumption (\ref{MeasurementIndependence}), 
which states that the preparation of the measured state does not depend on the measurement settings.
Relaxation of this assumption can be modeled by granting
the adversary some information about the random seed.
In different context, this can be seen as a limitation on the ``free will'' of the experimentalist;
a line of research is devoted to this topic, see \textit{e.g.} \cite{KoflerPaterekBrukner-Experimenter'sfreedomin-2006,
PawlowskiHorodeckiHorodeckiEtAl-QuantumCryptographyand-2010,
LawsonLindenPopescu-Biasednonlocalquantum-2010} 

The first difficulty stemming from partial information about the choice of measurement settings
is that the programming of the
devices can depend on them.
Therefore the adversary can prepare the devices to expect some inputs more often than the others, hence
their internal state is not fully independent on the inputs.
The second difficulty is the post processing.
In the case of randomness expanders, the initial random seed was split into two \emph{independent}
parts, one used for the measurement choices, the other for post-processing via randomness extractors,
which require independent random seed to function properly.
This is no longer possible in amplification scenario -- we cannot split the initial weakly random
seed into two independent parts. Because post-processing is more difficult, most of the existing
amplification protocols produce only single independent random bit per run. Only recently, thanks
to the development of input-secure unbounded expansion, protocols that produce more bits
per run have been discovered.

In this section we review amplification protocols for Santha-Vazirani weak sources and then protocols for
min-entropy sources.

%%%%%%%%%%%%%%
\subsection{Santha-Vazirani amplification with many devices.}
%%%%%%%%%%%%%%%%%%%

The first randomness amplification protocol of Colbeck and Renner \cite{ColbeckRenner-Freerandomnesscan-2012}
 that appeared in the literature was able
to amplify any SV-source (see Def. \ref{def:SVsource}) with $\varepsilon < (\sqrt{2} - 1)^2/2 \approx 0.086$, under
the assumption of quantum adversaries.
Under the stronger assumption of no signaling adversaries, the protocol can amplify SV-sources with $\varepsilon< 0.058$.
Their protocol uses a two party chained Bell inequality \cite{BraunsteinCaves-ChainedBellInequalities-1989}.
In Grudka~\textit{et.~al.} \cite{GrudkaHorodeckiHorodeckiEtAl-Freerandomnessamplification-2014}  analysis of two party
protocol against no-signaling adversary was improved and it was shown that any SV source
with $\varepsilon < 0.0961$ can be amplified.

The reason why the first protocols weren't able to amplify SV-sources for arbitrary $\varepsilon>\frac{1}{2}$
is that this task requires a very specific Bell inequality.

First of all, to be able to design a protocol for full randomness amplification, it is
crucial to use Bell inequalities, which can be maximally violated by quantum mechanics.
To see this consider a particular cheating strategy. The adversary programs the devices
with an optimal classical strategy. Using such strategy some inputs will be compatible
with the optimal violations and some will not. However, the adversary 
is assumed to be able to influence the input randomness up to the Santha-Vazirani parameter $\varepsilon$
 and might set the classical strategy in each run in such a way
that the outcomes will be incompatible with the set of least probable inputs only.
It can easily be seen
that the worse the randomness is (the higher the $\varepsilon$), the more successful this strategy becomes, 
as the ``bad'' inputs become less probable.
If the parameter
of the SV source $\varepsilon$ increases above some threshold, the probability to successfully provide compatible
outcomes with this classical strategy reaches the quantum limit.
This attack is no longer possible when the maximal attainable violation is observed, because
the adversary is forced to provide correlations attaining the maximum violation in every round
of the protocol, which is impossible with classical strategy.
An example of Bell inequality with this property is the  GHZ inequality introduced in Subsection \ref{BellSection}.

The second property of Bell tests required to design full amplification protocol
is relevant if we want to obtain security against non-signaling adversaries.
What we are looking for is a function that can post-process the outcomes of measurement into a single
\emph{non-deterministic bit}. It turns out this is not elementary and even inequalities fulfilling the
first property of maximum quantum violation don't have to have this property.
For example, it can be shown
that for every function post-processing the outcomes of the GHZ inequality, there
is a non-signaling distribution, which fully violates the inequality, but fixes the outcomes
of such function.

%These two
%The breakthrough came with the work of Galego et.~al. \cite{GallegoMasanesEtAl-Fullrandomnessfrom-2013},
%where the authors showed
%how to amplify an \emph{arbitrary} SV-source with $\varepsilon<1/2$ with the use of the following
%5-party Mermin inequality.
This is the reason why a breakthrough article
\cite{GallegoMasanesEtAl-Fullrandomnessfrom-2013} introducing the first amplification
protocol able to amplify arbitrary SV-source with $\varepsilon<1/2$  consider the following 5-party inequality coming
from a family of inequalities generalizing the GHZ inequality -- so called Mermin inequalities \cite{Mermin1990}.

Let us denote the inputs for the five parties
$\vec{x} = (x_1,\dots, x_5)$ and the outputs $\vec{a} = (a_1,\dots, a_5)$, with $x_i,a_i\in\{0,1\}$.
We will express the inequality as a linear combination of the non--signaling probabilities $p(\vec{a}\vert \vec{x})$,
governing the outputs, given the inputs. The inequality than reads:
\begin{equation}\label{5GHZ}
\sum_{\vec{a},\vec{x}} c_{\vec{a},\vec x}\cdot p(\vec{a}\vert \vec{x}) \geq 6,
\end{equation}
with linear coefficients
\begin{equation}
c_{\vec{a},\vec{x}} = (a_1\oplus a_2\oplus a_3\oplus a_4\oplus a_5)\delta_{\vec{x}\in X_0} +
(a_1\oplus a_2\oplus a_3\oplus a_4\oplus a_5 \oplus 1)\delta_{\vec{x}\in X_1}
\end{equation}
where
$$
\delta_{\vec{x}\in X_i} = \left\{
  \begin{array}{l l}
  1 & \quad \textrm{if}\quad \vec{x}\in X_i\\
  0 & \quad \textrm{if}\quad \vec{x}\notin X_i,
  \end{array} \right.
$$
and
\begin{align*}
X_0 = \{&(10000), (01000), (00100), (00010), (00001), (11111)\}\\
X_1 = \{&(00111), (01011), (01101), (01110), (10011), (10101), \\
&(10110), (11001), (11010), (11100)\}.
\end{align*}
Note that only half of the possible inputs appear in the inequality.
The maximum non-signaling violation of this inequality corresponds to the situation with left hand side of Eq. \ref{5GHZ} equal to
zero and it can be achieved by a quantum strategy of measuring state
$\vert GHZ\rangle_5 = \frac{1}{\sqrt{2}}(\vert 00000\rangle + \vert 11111\rangle)$ with measurements
$\sigma_x$ and $\sigma_y$.

The post-processing function required to obtain non-deterministic bits from the outcomes of the measurement
here is the majority function $maj(\vec{a})$ of the
the first three parties involved in the protocol. It can be shown that for every non-signaling
distribution fully violating the inequality, the predictability of the majority function is at most $3/4$.
The property of the majority function can be interpreted as an amplification procedure
of an arbitrary SV source with $\varepsilon<\frac{1}{2}$ into an SV-source with $\varepsilon = \frac{1}{4}$.
To finish the protocol it needs to be equipped with two additional components.

The first of them
is an estimation procedure to make sure that the untrusted devices indeed do yield the required
Bell violation and the other thing is a procedure $f$ that can transform sufficiently many bits
with bias $\varepsilon = \frac{1}{4}$ generated in the Bell experiment into a random bit with bias $\varepsilon$
arbitrary close $0$. The second task looks to be in direct contradiction with the result of
Santha and Vazirani \cite{SanthaVazirani-Generatingquasi-randomsequences-1986},
which claims it is impossible to classically extract bits from such a source.
 The reason
it is possible here is that the bits are produced in a quantum process and the Santha-Vazirani
classification (the value of $\varepsilon$) does not sufficiently describe all the properties of these bits.
In other words, the bits do contain an additional structure.
Using the techniques of \cite{Masanes-Universallycomposableprivacy-2009}, the authors of the protocol were able to
show the existence of such function $f$. Unfortunately, even though the existence of such function is proven,
the precise construction is yet unknown.

The protocol uses as a resources an SV-source $X$ with arbitrary $\varepsilon < \frac{1}{2}$ and $N$
$5$-partite GHZ states (see Fig. \ref{fig:SVamplifMany}).

\begin{figure}[tb]
\begin{framed}
\begin{center}
\textbf{Santha-Vazirani amplification protocol with many devices}
\end{center}
\begin{enumerate}
 \item Use $X$ to generate $N$ quintuplets of bits $\vec{x}_1,\dots,\vec{x}_n$, which are used as inputs into $5N$
devices. The devices outputs are labeled $\vec{a}_1,\dots,\vec{a}_n$.
\item Quintuplets that are not valid inputs for 5-party Mermin inequality are discarded. If less than $N/3$ quintuplets
remain, abort.
\item The rest of the quintuplets are organized into $N_b$ blocks each having $N_d$ quintuplets.
One of the blocks is chosen, randomly according to the random source $X$, to be the distillation block.
\item Check, if all the non-distilling blocks maximally violate 5-party Mermin inequality. If not, abort the protocol.
\item Produce the bit $k = f({maj}(\vec{a}_1),\dots,{maj}(\vec{a}_{N_d}))$, where
$f$ is the distillation function, $(\vec{a}_1,\dots,\vec{a}_{N_d})$ are outputs in the distillation block and
{$maj(\vec{a})$} is the majority function of the first three outputs of quintuplet $a$.
\end{enumerate}
\end{framed}
\caption{\it\small Santha-Vazirani amplification protocol of Galego~\textit{et.~al.} \cite{GallegoMasanesEtAl-Fullrandomnessfrom-2013}.}
\label{fig:SVamplifMany}
\end{figure}

The authors have shown in a rather complicated proof that the probability $p_{guess}$ of the adversary
to guess the output bit $k$ correctly, using all the available information, can be upper bounded as:
\begin{equation}
p_{guess} \leq \frac{1}{2} +
\frac{3\sqrt{N_d}}{2}\left[\alpha^{N_d}+2N_{b}^{\log_2(1/2+\varepsilon)}
(32\beta(1/2-\varepsilon)^{-5})^{N_d}\right],
\end{equation}
where $\alpha$ and $\beta$ are real numbers such that $0<\alpha<1<\beta$.

Note that $\varepsilon_f = \frac{3\sqrt{N_d}}{2}\left[\alpha^{N_d}+2N_{b}^{\log_2(1/2+\varepsilon)}
(32\beta(1/2-\varepsilon)^{-5})^{N_d}\right]$ is the bias of the newly produced bit,
which can be made arbitrary close to zero
($p_{guess}$ can be made arbitrarily close to $\frac{1}{2}$) by setting
$N_b = (32\beta(1/2-\varepsilon)^{-5})^{2N_d/\vert\log_2(1/2+\varepsilon)\vert}$ and increasing $N_d$, so that
$N_dN_b\geq N/3$.
Notice that this protocol works only in a noiseless setting and the number of steps and devices increases
as the $\varepsilon_f$ of the produced bit approaches 0.

\subsection{Santha-Vazirani amplification with eight devices.}

The work of Gallego~\textit{et.~al.}  \cite{GallegoMasanesEtAl-Fullrandomnessfrom-2013} was followed by several results. 
At first in \cite{MironowiczPawlowski-Amplificationofarbitrarily-2013}, a tripartite
amplification protocol was presented, which could amplify SV source with arbitrary $\varepsilon<1/2$,
in a noisy setting in finite time, secure against quantum adversaries.
Subsequently, in a breakthrough paper \cite{BrandaoRamanathanGrudkaEtAl-RobustDevice-IndependentRandomness-2013}  the authors
showed an amplification protocol secure against non-signaling adversaries for any SV-source with
$\varepsilon<1/2$, which uses only a finite number of devices and tolerates a constant rate of error.
We will review their protocol in this subsection.

The protocol of Brand\~ao~\textit{et.~al.} \cite{BrandaoRamanathanGrudkaEtAl-RobustDevice-IndependentRandomness-2013} uses two independent non-communicating devices composed of four
non-signaling components to test a 4-partite Bell type inequality with inputs chosen according to a given
Santha-Vazirani source with $\varepsilon < \frac{1}{2}$.
The runs of both devices are divided into blocks of equal lengths.
After the necessary number of rounds two blocks are chosen according to the bits
drawn from the given Santha-Vazirani source -- one block
from the first device and the second block from the second device.
Subsequently, the violation of the used Bell inequality is calculated from the inputs
and outputs of the devices.
If the violation is high enough, a 2-source extractor is applied to the block outputs
in order to create outputs of the protocol.
In order to establish the correctness of the protocol authors have proven three crucial claims.

The first claim is that one can use as little as two devices with four components each. The
main problem here is that in order to justify the use of a two source
extractor, it's inputs have to be independent. Although at the end of the protocol, the inputs for
the extractor are chosen from two non-communicating devices, they are not trivially independent.
Correlations between the blocks can be caused by several factors, for example pre-shared
randomness or the fact that their inputs are chosen according to an SV-source which are generally
correlated.

The crucial technique here is the use of a variant of quantum de Finetti
theorem \cite{BrandaoHarrow-QuantumDeFinetti-2013}. The quantum de Finetti theorems
essentially show us that if a $k$-partite quantum state is permutation-symmetric, the reduced state of its
subsystems of size $l \ll k$ is close to a convex combination of $l$-partite identical separable quantum states.
Moreover, after conditioning on the measurement outcomes of the rest of the states (the states that are not
part of the small $l$-partite subsystem in question), the state of the
chosen subsystem collapses into a state that is close to factorized.
The permutation symmetry can be obtained by simple uniform random choice of the subsystems.
Then by the de Finetti
theorem the subsystems are factorized and therefore their measurement outcomes are uncorrelated.
The main problem in this context is to show that similar result holds even if the subsystems are chosen
according to a (arbitrary weak) Santha-Vazirani source instead of the uniform one.

The second claim is that the violation of the used Bell inequality can certify randomness of the measurement
outcomes even if the measurement settings are not chosen uniformly, but according to an SV-source.
Moreover, it is also necessary to quantify the amount of min-entropy in the outcomes that passed the Bell test
in order to use the appropriate two source extractors.

The third claim is concerned with the use of an appropriate two source extractor at the end of the protocol.
It is important to stress that the authors use a non-explicit extractor, which is known to exist, but its
construction is not generally known yet (\textit{i.e.} it is not efficiently constructible -- see Subsection \ref{sec:Extractors}). 
The reason for this is that recent two source extractor constructions
require both input sources to have relatively high-min entropy rate as discussed in Subsection \ref{sec:Extractors},
which is not the case of the measurement
outcomes in the chosen blocks. However, the protocol can be extended to a case with multiple
devices with four components. In such a protocol a multi-source extractor is used at the end and there are
known extractor constructions which require much lower entropy rate. The price to pay is the increase on
the number of non-communicating devices, which nevertheless stays constant.

In order to introduce the protocol more formally, let us fix the notation.
The first device will be used for $N_1 = N_dN_b^1$ runs and the second device for $N_2 = N_dN_b^2$ runs.
Each run is divided into $N_b^1$ and $N_b^2$ blocks of size $N_d$ respectively
The $N_1$ input quadruples for the first device are denoted ${\vec x_1}$, similarly $N_2$ input
quadruples for the second device are denoted ${\vec u_2}$. Output quadruples are denoted
${\vec a_1}$ and ${\vec a_2}$ respectively.

The Bell inequality used in the protocol has the following form. Each of the four devices
receives one bit input and produces one bit output, therefore the inputs are labeled
${\vec x} = \{x_1,x_2,x_3,x_4\}$ and outcomes ${\vec a} = \{ a_1,a_2,a_3,a_4\}$.
The measurement settings that appear in the Bell term are divided into two sets
\begin{equation}
  X_0 = \{0111,1011,1101,1110\} \quad\text{and}\quad X_1 = \{0001,0010,0100,1000\}
\end{equation}
The inequality then reads:

%\begin{equation}
%\sum_{\vec x, \vec a}\left( \mathtt{I}_{\oplus_{i=1}^4 a_i = 0} \mathtt{I}_{\vec x \in X_0}
%+ \mathtt{I}_{\oplus_{i=1}^4 a_i = 1} \mathtt{I}_{\vec x \in X_1} \right)
%p(\vec a\vert\vec x)\geq 2,
%\end{equation}

\begin{equation}
\sum_{\vec a, \vec x}c_{\vec a, \vec x}\cdotp(\vec a\vert\vec x)\geq 2,
\end{equation}
with linear coefficients  
\begin{equation}
c_{\vec{a},\vec{x}} = (a_1\oplus a_2\oplus a_3\oplus a_4)\delta_{\vec{x}\in X_0} +
(a_1\oplus a_2\oplus a_3\oplus a_4 \oplus 1)\delta_{\vec{x}\in X_1}
\end{equation}
where
$$
\delta_{\vec{x}\in X_i} = \left\{
  \begin{array}{l l}
  1 & \quad \textrm{if}\quad \vec{x}\in X_i\\
  0 & \quad \textrm{if}\quad \vec{x}\notin X_i,
  \end{array} \right.
$$
 Local models
can achieve minimal value of $2$, while there exists a quantum strategy
that achieves the algebraic minimum of 0. Moreover, the authors have shown
that there doesn't exist any no-signaling distribution that can simultaneously
achieve low values of the Bell term and deterministic outcomes.

\begin{figure}[tb]
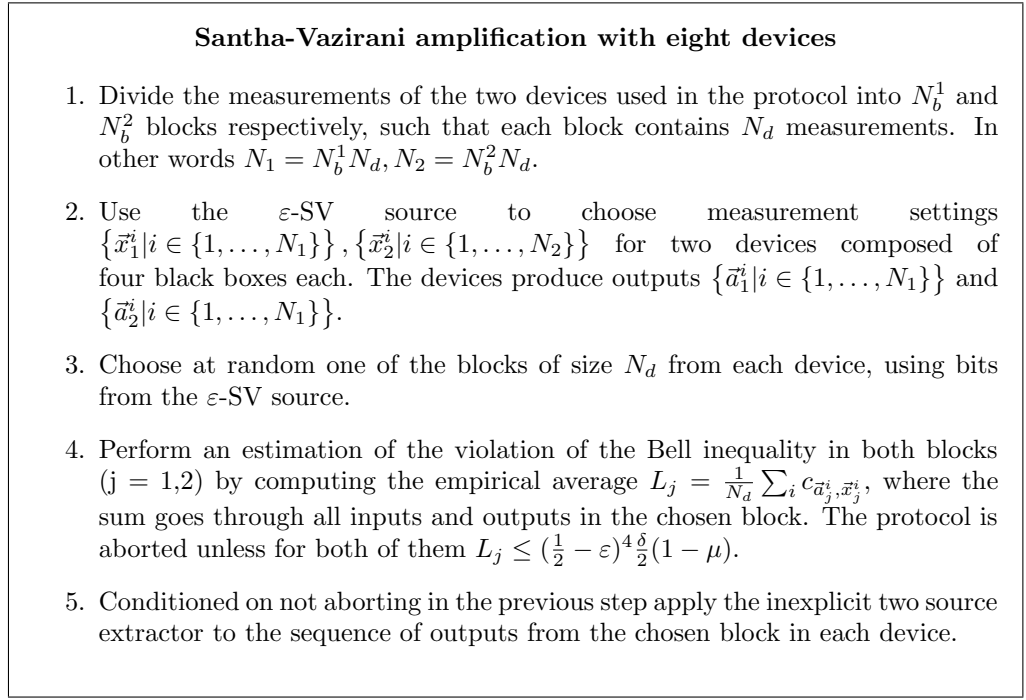

\begin{framed}
\begin{center}
\textbf{Santha-Vazirani amplification with eight devices}
\end{center}
\begin{enumerate}
\item Divide the measurements of the two devices used in the protocol into $N_b^1$ and $N_b^2$ blocks respectively, such that
each block contains $N_d$ measurements. In other words $N_1 = N_b^1N_d, N_2 = N_b^2N_d$.
\item Use the $\varepsilon$-SV source to choose measurement settings ${\left\{\vec x_1^i\vert i\in\{1,\dots,N_1\}\right\},
\left\{\vec x_2^i\vert i\in\{1,\dots,N_2\}\right\}}$ for two
devices composed of four black boxes each. The devices produce outputs ${\left\{\vec a_1^i\vert i\in\{1,\dots,N_1\}\right\}}$ and 
${\left\{\vec a_2^i\vert i\in\{1,\dots,N_1\}\right\}}$.
\item Choose at random one of the blocks of size $N_d$ from each device, using bits
from the $\varepsilon$-SV source.
\item Perform an estimation of the violation of the Bell inequality in both blocks (j = 1,2)
by computing the empirical average $L_j = \frac{1}{N_d}\sum_{i} c_{\vec{a}_j^i,\vec{x}_j^i}$,
where the sum goes through all inputs and outputs in the chosen block.
The protocol is aborted unless for both of them $L_j \leq (\frac{1}{2}-\varepsilon)^4\frac{\delta}{2}(1-\mu).$
\item Conditioned on not aborting in the previous step apply the inexplicit two source extractor to the
sequence of outputs from the chosen block in each device.
\end{enumerate}
\end{framed}
\caption{\small\it Protocol for Santha-Vazirani source randomness amplification with 8 devices by Brand\~ao~\textit{et.~al.} \cite{BrandaoRamanathanGrudkaEtAl-RobustDevice-IndependentRandomness-2013}.}
\label{8device}
\end{figure}

In the protocol we will be using empirical average of the Bell term over $N_d$ runs.
This serves the same role as before. We would need infinite time to
check the Bell inequality precisely, so we are using average obtained in each of the blocks instead.
Here it is
defined as:
\begin{equation}
L = \frac{1}{N_d}\sum_{i=1}^{N_d} c_{\vec{a},\vec{x}}.
\end{equation}
If this empirical average is lower than $(\frac{1}{2}-\varepsilon)^4\frac{\delta}{2}(1-\mu)$ with fixed
constants $\delta,\mu>0$, the outputs of the $N_d$ runs  have min-entropy linear in $N_d$.
The whole protocol is shown in Fig. \ref{8device}.

Note that because this protocol can output more than one bit, it can be concatenated with one of the
input secure expansion protocols in order to obtain unbounded amount of almost random bits.

%%%%%%%%%%%%%%%%%%%%%%%%
\subsection{Min-entropy amplification with many devices}\label{MinEntropyAmp}
%%%%%%%%%%%%%%%%%%%%%%%%%%%%%

In order to illustrate the complications in the analysis of min-entropy amplification,
let us first recall the crucial difference between Santha-Vazirani sources and min-entropy sources.
The main difference is that any string from a SV-source with $\varepsilon<\frac{1}{2}$
has non-zero probability to appear.
As we have discussed in Subsection \ref{sec:MESources}, this is not the case of min-entropy sources.
%Consider for example an $n$-bit random string with the first bit fixed to $0$.
%Although the min-entropy of such string is almost ($n-1$), half of the possible strings
%-- those starting with bit $1$ -- have zero probability to appear.
This fact creates many difficulties.
As an example consider the case of GHZ scenario (see Subsection \ref{BellSection}).
If one of the four possible measurement settings
is known by the adversary to have zero probability to be used, the classical deterministic strategy
to violate the GHZ inequality exists. Note that this is not only the case of the GHZ inequality,
but similar property is inherent in every Bell-type scenario as shown by Le~\textit{et.~al.}
\cite{ThinhSheridanScarani-Belltestswith-2013}.
This is perhaps the main reason why min-entropy amplification took longer time to develop than
it's Santha-Vazirani counterpart.

Let us start by a protocol for block-min entropy sources (defined in Subsection \ref{sec:MESources})
by Bouda, Paw\l{}owski, Pivoluska and Plesch \cite{BoudaPawlowskiPivoluskaEtAl-Device-independentrandomnessextraction-2014}.
The protocol is based on GHZ game introduced in Subsection \ref{BellSection}.

First of all note that in principle we need only two bits to generate an input into a GHZ test,
as only four out of eight input bit combinations appear in the inequality.
The simplest approach would be for the protocol to simply take a block of size $n$ from the weak source,
divide it into $\frac{n}{2}$ two bit substrings and use these two bit substrings to create inputs
for $\frac{n}{2}$ rounds of the GHZ test.
Unfortunately this approach does not work. Recall that in order to use
deterministic strategy in a single run of a GHZ game, it is sufficient, if one of the inputs
have $0$ probability to be used.
Thus the adversary can choose a probability distribution for the block source
in such a way that only three out of four possible values appear in each two bit substring.
Such source would allow the adversary to use a deterministic strategy to win the GHZ game
in every round.
The the min-entropy rate $R$ of this distribution is $\frac{\log_2 (3)}{2}$ and therefore
with this strategy amplification is impossible for sources with $R \leq \frac{\log_2 (3)}{2}$.

\begin{figure}[tb]
\begin{center}
\includegraphics[width=6.5cm,clip]{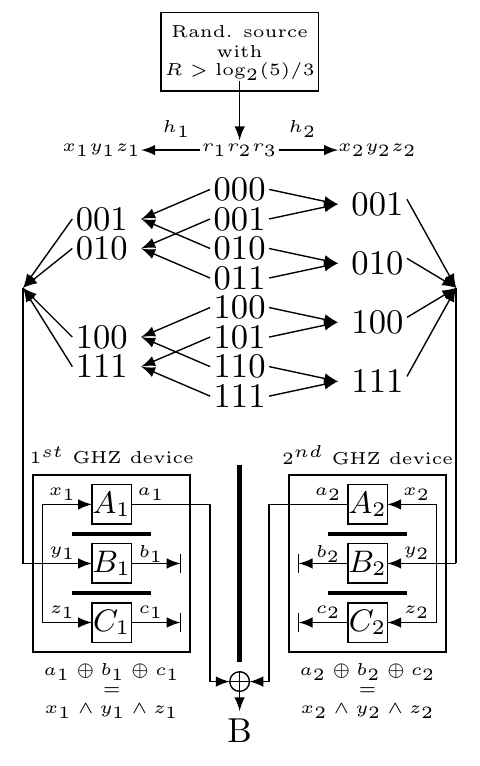}
\end{center}
\vspace*{-0.7cm}
\caption{\small\it An illustration of the ``hash trick''
for blocks of length $n = 3$.
We use two independent devices with two hash functions. At least $3$
out of $8$ source outputs have to be fixed in order to win both GHZ games  with deterministic strategy with probability $1$,
which is mote than $2$ out of $8$ required with the use of a single device.}
\label{2hash}
\end{figure}

The main idea of the protocol construction that allows us to go around this difficulty is to first use several
specific hash functions depending on each bit of the block in order to generate inputs into multiple
independent devices performing the GHZ test.
The reason why this trick works is that if the inputs are chosen in this way for
multiple independent devices with the use of  the suitable set of hash functions, it is much more demanding
to achieve probability $0$ of some input for \emph{all the devices  simultaneously}.
To illustrate how hashing can improve the performance see Fig. \ref{2hash}.

It might seem that to lower the required min-entropy, using simply more hash functions to create the
inputs into more independent GHZ testing devices will suffice.
However, it is only partially true and this method has it's limits. Whenever less than four out of all possible strings
appear, each hash function can output only less than four of it's outputs with non-zero
probability. Therefore, the only requirement we have is that each block contains at least two bits
of entropy, \textit{i.e.} it is a $(n,2)$ block source. Note that as $n$ goes to infinity, rate of such source
goes to $0$.

Therefore the first step is to find a set of hash functions $\mathcal H$
that ensures that every quadruple of the possible block outputs of size $n$ will be hashed to four
different  two bit strings by at least one hash function $h\in \mathcal H$.
Such construction ensures that at least one of the many independent GHZ devices paired with
the hash functions has
a distribution on the inputs that prevents it from winning the GHZ game with probability $1$
with deterministic strategy.

\begin{figure}[tb]
\begin{center}
\includegraphics[width=8cm,clip]{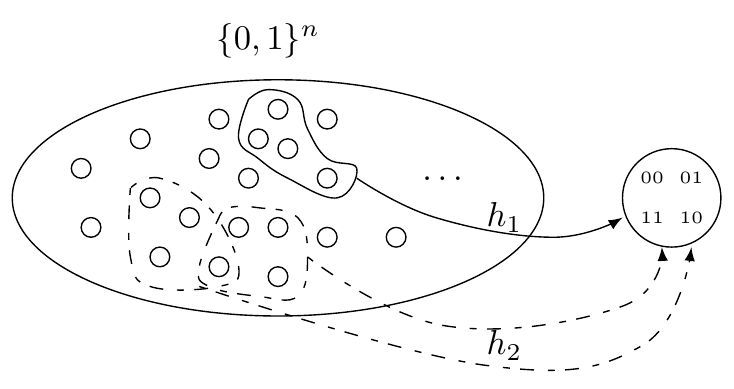}
\end{center}
\vspace*{-0.7cm}
\caption{\small\it For each quadruple of $n$ bit strings we need a hash function to map it to four different
outputs. Several quadruples can share a hash function as demonstrated by function $h_2$ (dashed) and 
thus one does not need the full set of all hash functions.}
\label{hashExample}
\end{figure}

This can be easily achieved by considering the set of all hash functions $h: \{0,1\}^n \mapsto \{0,1\}^2$,
however the size of such set $\vert \mathcal H\vert = m$ is impractically large ($m = 4^n$ in this case), as a single
function in fact ``covers'' several quadruples (see Fig. \ref{hashExample}).
On the other hand for large $n$ one hash
function covers as many as $9\%$ of all four-tuples, independently of $n$. So
the size of an optimal set of hash functions might not depend on $n$ at all.

In what follows
we show a construction of $\mathcal H$ with $m$ polynomially large in $n$.

Let us consider a sequence of random variables $Z=\left(  Z_{0},\dots,Z_{N-1}\right)$
such that $\mathcal{Z}_{i}\in\{0,1,2,3\}$. The outcomes of such a
random experiment are $N$-position sequences from the set $\{0,1,2,3\}^{N}$.
It is easy to see that each such sequence specifies uniquely a particular
function $h:\{0,...,N-1\}\rightarrow\{0,1,2,3\}$, and vice versa. Since now on
we will use them interchangeably.

Let us assume that random variables $Z$ satisfy the condition that for every
$4$--tuple of positions $j_{0},j_{1},j_{2},j_{3}$ and every $4$-element string
$z_{0}z_{1}z_{2}z_{3}\in\{0,1,2,3\}^{4}$ it holds that
\begin{equation}\label{Eq1}%
\Pr\bigl[Z_{j_{0}}=z_{0}\wedge Z_{j_{1}}=z_{1}\wedge Z_{j_{2}}=z_{2}\wedge
Z_{j_{3}}=z_{3}\bigr]>0.
\end{equation}
Note that for our purposes even a weaker assumption on $Z$ is sufficient: It
is enough if for every $4$--tuple of positions $j_{0},j_{1},j_{2},j_{3}$ there
exists at least one $4$-element string $z_{0}z_{1}z_{2}z_{3}\in\{0,1,2,3\}^{4}%
$ with all $z_{0},z_{1},z_{2},z_{3}$ begin mutually different and satisfying
(\ref{Eq1}). However, the stronger condition will make it easier to find a
suitable set.

Let us denote $\mathcal H=\{z\in\{0,1,2,3\}^{N}\text{ such that }P[Z=z]>0\}$. Using the
probabilistic method we see that for each $4$-tuple of positions
$j_{0},j_{1},j_{2},j_{3}$ and every $4$-element string $z_{0}z_{1}z_{2}%
z_{3}\in\{0,1,2,3\}^{4}$ there exists a function $h\in \mathcal H$ such that
\begin{equation}
h(j_{0})=z_{0}\wedge h(j_{1})=z_{1}\wedge h(j_{2})=z_{2}\wedge h(j_{3}%
)=z_{3}.
\end{equation}
The number of functions in $\mathcal H$ is the same as the number of (nonzero
probability) sample space elements of $Z$. It remains to construct $Z$ with a
sample space as small as possible. In order to do so, we will need the following
definition and theorem.

\begin{definition}[$s$-wise $\delta$-dependence]
Binary random variables $\left(  Z_{0},\dots,Z_{N-1}\right)$ are \textbf{$s$-wise $\delta$-dependent} iff for all
subsets $S\subseteq\{0,\dots,N-1\},\left\vert S\right\vert \leq s$
\begin{equation}
||U_{\vert S\vert}-D_{ S }||_1 \leq\delta,
\end{equation}
where $U_{\vert S\vert}$ is a uniform distribution over $|S|$-bit strings and $D_S$ is a
marginal distribution over subset of variables specified by $S$.
\end{definition}

\begin{theorem}[\cite{NaorNaor-Small-BiasProbabilitySpaces:-1993}]
The logarithm of the cardinality of the sample space needed to
construct $N$ $s$-wise $\delta$-dependent random variables is $O\left(
s+\log\log N+\log\frac{1}{\delta}\right) .$

\end{theorem}
In our case we are interested in $s=4$.
Let us consider two sequences $X_{0},\dots,X_{N-1}$ and $Y_{0},\dots,Y_{N-1}$
of binary $4$-wise $\delta$-dependent random variables, both sequences being
mutually independent. Let $Z_{i}=2X_{i}+Y_{i}$.

As both $X$ and $Y$ are $\delta$-dependent, their distance from the uniform
distribution for every subset of size at most $4$ is at most $\delta$. Assuming
there is a zero probability for at least one binary string out of
$\{0,1\}^{4}$ at positions $(0,1,2,3)$ we have that the distance of such a
distribution from the uniform distribution is at least $2\times2^{-4}=2^{-3}%
$.

Hence, assuring that $\delta<2^{-3}$ we find that for each $4$ positions
there is a nonzero probability of every $4$-bit sequence appearing. Hence, for
the sequence of random variables $Z$ it holds that in every $4$-tuple of
positions every string out of $\{0,1,2,3\}^{4}$ appears with non-zero probability.

In our case we need two independent sets of $N=2^{n}$ $4$-wise $1/8$-dependent
random variables, resulting in a sample space of $O(n^{c})$, bearing the
desired polynomial construction.

Another preliminary result we need to show is a form of rigidity theorem for the GHZ game.
Rigidity theorems for Bell-type experiments show that if the value of the experiment is close
to being optimal, so is the strategy that achieved it. In our case, the claim is that
if the GHZ game is won with probability $1$, then the optimal quantum strategy has been used,
which in turn guarantees random outcomes of the measurements. The exact version
of such rigidity for CHSH experiment has been known for some time \cite{MayersYao-Selftestingquantum-2004}.
It was later followed
by the more robust results for more general games (also the GHZ game), claiming that if the probability of
winning is \emph{close to} $1$,
then the strategy that achieved it is \emph{close to} the optimal one \cite{MillerShi-Optimalrobustquantum-2012,
McKague-Self-testinggraphstates-2010}.
 Such theorems show us that
the GHZ test behaves ``nicely'' in the sense that it is impossible to achieve a slightly suboptimal violation
with strategies that are dramatically different from the optimal one and therefore
even suboptimal violation is a witness of randomness being produced.

We used the following computational form of a rigidity theorem obtained by
semi-definite programing, which is concerned only about the produced randomness.
The formal statement is the following:

\begin{figure}[tb]
\center
\vspace*{-0.3cm}
\includegraphics[width=11cm,clip]{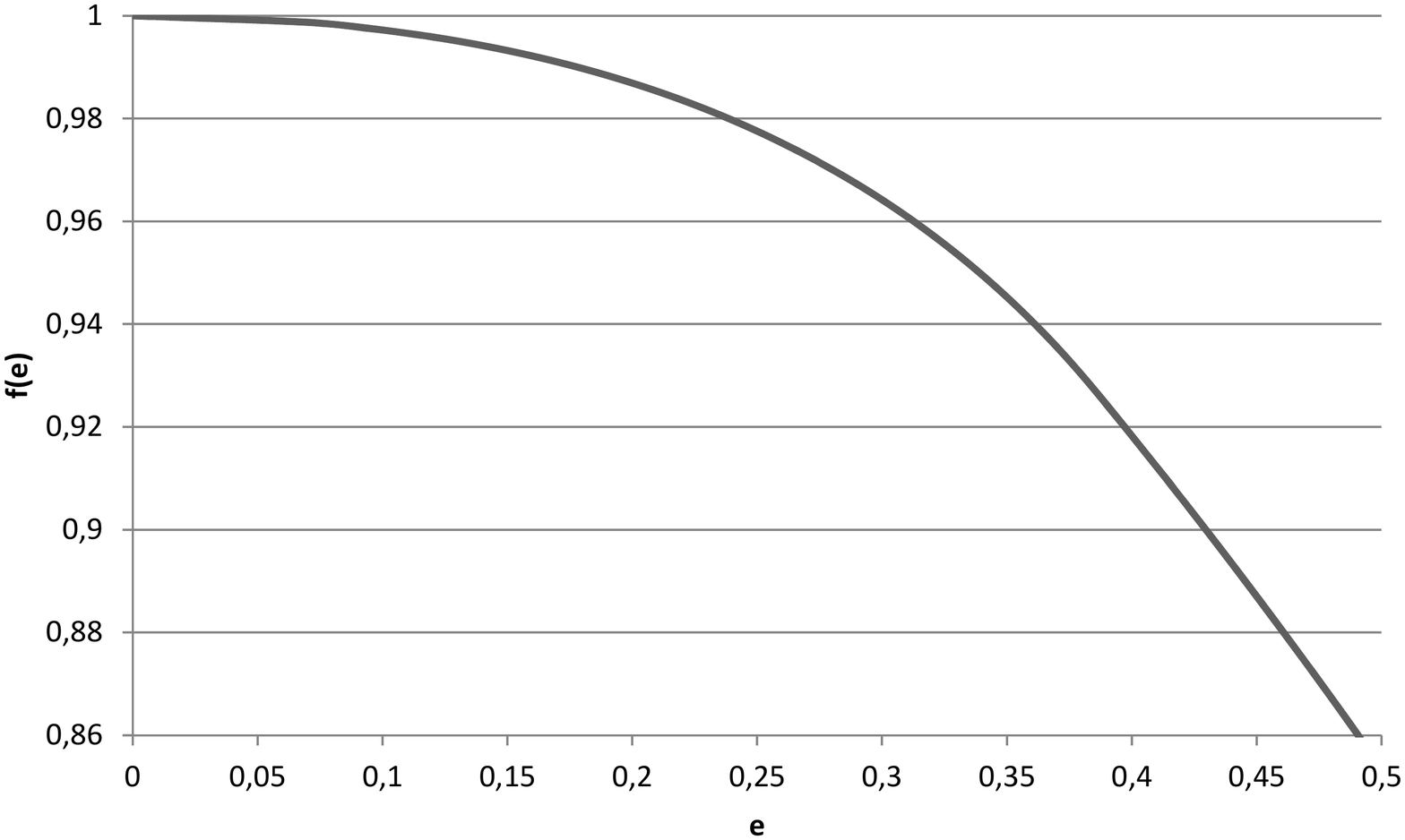}
\vspace*{-1.0cm}
\caption{\small\it Depicted is the value of GHZ
term $v=f(\varepsilon)$ needed to certify the bias of the output bit of the first device performing the GHZ test
is at most $\varepsilon$. This can be seen as a sort of rigidity term for the GHZ inequality and is
one of the main tools in analysis of the protocol by Bouda~\textit{et.~al.}\cite{BoudaPawlowskiPivoluskaEtAl-Device-independentrandomnessextraction-2014}} %
\label{FEpsilon}
\end{figure}

Take an arbitrarily long linearly ordered sequence of $t$ Mermin devices $D_{1}...D_{t}$
with uniform distribution on inputs, and each device
knows inputs and outputs of its predecessors,
but devices cannot signal to its predecessors. Let us assume that the inputs
of devices are described by random variables $XYZ_{1},\dots,XYZ_{t}$, and the
outputs by $ABC_{1},\dots,ABC_{t}$. Then there exists a function
$f(\varepsilon)$ such that if the value of the Mermin variable
(\ref{eq:mermin}) using uniform inputs is at least $v_{u}\geq f(\varepsilon)$, then the output bit
$A_{t}$ has a bias at most $\varepsilon$ conditioned on the input and output
of all its predecessors and the adversarial knowledge. This function can be
lower bounded by a semidefinite program (SDP) using any level of the
hierarchy discussed in Subsection \ref{sec:QuantumSet}. By using the second level of the
hierarchy one can obtain the bound on $f(\varepsilon)$ as a function of
$\varepsilon$ shown in Fig.~\ref{FEpsilon}.

We can set $t=1$ (having just a single device) and get the lower
bound on the detection probability of producing a bit biased by more than
$\varepsilon$, which is $1-f(\varepsilon)$. More independent
non-communicating devices can be ordered into any sequence and thus this limit
holds for any of these devices simultaneously.

\begin{figure}[tb]
\vspace*{-0.2cm}
\begin{center}
\includegraphics[width=6cm,clip]{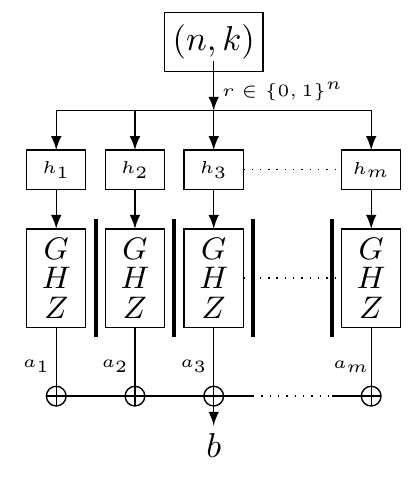}
\vspace*{-0.5cm}
\caption{\small\it The block source amplification protocol of Bouda~\textit{et.~al.}
\cite{BoudaPawlowskiPivoluskaEtAl-Device-independentrandomnessextraction-2014}.
The output $r$ of a weak source is mapped into inputs for independent GHZ tests with the use of hash functions
$h_i$ coming from a specially constructed set $\mathcal H$. The construction of $\mathcal H$ guarantees that at least
one of the GHZ tests obtains input distribution that cannot be won with deterministic strategy with probability $1$,
thus producing almost random outputs with high probability. Since the output $A_i$ of the good device is independent
of other outputs, the XOR of all the outputs has at most the same bias as $A_i$.
}
\label{pictureSingleRound}
\end{center}
\end{figure}

Now we are armed with all the tools to analyze a single round
of the protocol depicted in Fig. \ref{minEntProt} and Fig. \ref{pictureSingleRound}.

\begin{figure}[!tb]
\begin{framed}
\begin{center}
\textbf{Block source amplification protocol with many devices}
\end{center}
\vspace*{-0.7cm}
\begin{enumerate}
\item Obtain a (weakly) random $n$ bit string $r$ from the $(n,k)$ block source.

\item Input into each device $D_{i}$  the $3$ bit string $x_iy_iz_i$ chosen from set
$\{111,100,010,001\}$ --
each one corresponding to one of the possible outputs of $h_{i}(r)$
and obtain the outputs $a_{i}$, $b_{i}$ and $c_{i}$.

\item Verify whether for each device $D_{i}$ the condition $a_{i}\oplus
b_{i} \oplus c_{i}=x_{i}\cdot y_{i}\cdot z_{i}$ holds. If this is not true,
abort the protocol.

\item Output $b=\bigoplus_{i=1}^{m}%
a_{i}.$
\end{enumerate}
\end{framed}
\vspace*{-0.5cm}
\caption{\small\it Single round protocol for block source amplification by
Bouda~\textit{et.~al.} \cite{BoudaPawlowskiPivoluskaEtAl-Device-independentrandomnessextraction-2014}. The full protocol is a simple
repetition of the single round protocol, each time with fresh a block from the block source and new devices. The outcome
of the full protocol is an XOR of outcomes from all the rounds.}
\label{minEntProt}
\end{figure}

Let us now analyze how the single round protocol works. 
We will first analyze the single round protocol, if we restrict the adversary and allow her
to use only flat sources (see Def. \ref{def:flatSource}). In that case only $4$ strings appear with positive probability of $\frac{1}{4}$.
For such a weak source our construction of the
the set $\mathcal H$ of hash functions assures that there exists a function $h_{j}\in \mathcal H$ that
has its four outputs uniformly distributed. Thus, inputs for the corresponding device $D_{j}$
are uniform on this flat distribution.

Then by our rigidity claim we know that if the adversary wants to achieve bias $\varepsilon$ for the output
 bit $b$, she can do so only with probability $v_{u}\leq f(\varepsilon)$ -- otherwise the win condition
will not be satisfied in device $D_j$ and the protocol will abort.

More importantly, the set of all $(n,2)$ distributions is convex and the flat distributions are
exactly all the extremal points of this convex set.
Thus any $(n,2)$ distribution $d$ can be expressed as a convex combination of at most $N$ $(n,2)$ flat
distributions $d_{i}$ (Caratheodory theorem) as $d=\sum_{i=1}^{N}p_{i}d_{i}$
for some $p_{i}\geq0$, $\sum_{i=1}^{N}p_{i}=1$.

The probability that the win condition is fulfilled for a mixed distribution $d$ is then
upper bounded by the weighted sum of successful cheating probabilities of the flat distributions $d_i$.
As all these are upper bounded by $f(\varepsilon)$, the following statement holds:
\begin{equation}
v_{u}\leq\sum
_{i=1}^{N}p_{i}v_{u,i}\leq f(\varepsilon),
\end{equation}
where $v_{u,i}$ is the cheating probability of $i$-th flat distribution constituting the mixture.

To summarize this part, having any $(n,k)$ source with $k\geq2$, with a single
round of a protocol, we can produce a single bit that is biased at most by
$\varepsilon$ with the cheating probability of the adversary $f(\varepsilon)$.

However a good protocol should allow the user to choose both the target parameters of
the produced bit -- the bias $\varepsilon \ll 1$ and the cheating probability $\delta \ll 1$.
The single round protocol obviously doesn't fulfill this property, as the only pairs of allowed
$(\varepsilon,\delta)$ are $(\varepsilon,f(\varepsilon))$. This downside can be overcome
by repeating the protocol many times, each time with a new block from the weak source
and new set of devices. The outcome of the protocol is a simple XOR of all the single round
protocol outputs.

If we again consider flat weak sources, we know that in each round of the protocol there was at least one
device that received uniform inputs. All these devices were new, thus we can order them in any particular time sequence.
Using the rigidity result
we see that for each such a sequence the last bit will be biased by no more than $\varepsilon$,
unless the last round was cheated, which can be done with probability upper bounded by $f(\varepsilon)$.
To achieve the bias of the product bit
at least $\varepsilon$, all the rounds must be cheated, as any of them can be treated as the last one.
Probability of doing this is upper bounded by
$f(\varepsilon)^{l}$. Thus, choosing $l>\frac{\log\delta}{\log f(\varepsilon)}$
will guarantee the fulfillment of the conditions for the parameters $\varepsilon$ and $\delta$.
Using the Caratheodory theorem we can extend this results to non-flat sources as well.

Summing up, with an $(n,k)$ block source and $O\left(  \frac{\log\delta}{\log
f(\varepsilon)}Poly\left[  n\left\lceil \frac{2}{k}\right\rceil \right]
\right) $ Mermin devices we can produce a single random bit with bias smaller
than $\varepsilon$ with probability larger than $1-\delta$. For producing more
bits we simply repeat the whole procedure: all the bits produced will have bias
smaller than $\varepsilon$ conditioned on the bits produced so far, with
linear scaling of resources. Moreover, in the paper \cite{BoudaPawlowskiPivoluskaEtAl-Device-independentrandomnessextraction-2014}
 we have  shown that this protocol can
be extended to a robust protocol, which is able to tolerate certain amount of errors.

\begin{figure}[tb]
\begin{center}
\includegraphics[width=9cm,clip]{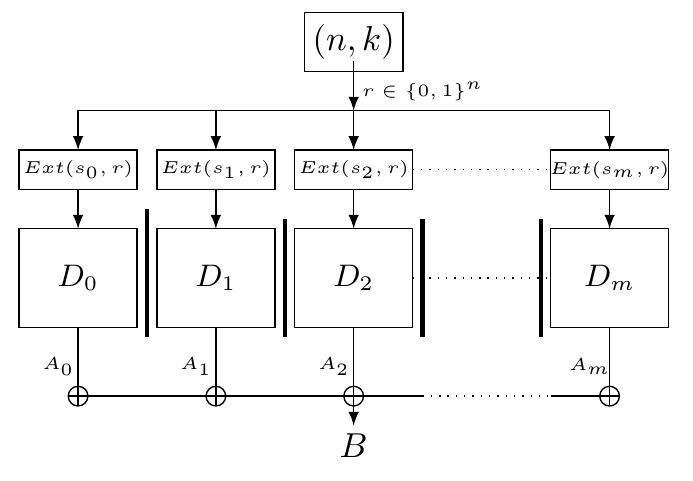}
\end{center}
\vspace*{-0.7cm}
\caption{\small\it Chung, Shi and Wu protocol for min-entropy amplification
\cite{ChungShiWu-PhysicalRandomnessExtractors-2014}.
Here $s_i$ is a binary $m = O(\log n)$ bit representation of $i$
used as a seed into extractor $Ext$.
For any $(n,k)$ distribution at least one seed $s_i$ constitutes a good extractor.
Output of these extractors is used as an input into expansion devices $D_i$. As all expansion
devices are input secure, their outputs $A_i$ are longer than the inputs and independent of each other, therefore
output string $B$ is almost perfectly distributed.}
\label{fig:protocolExt}
\end{figure}

To finish this subsections we will briefly mention a protocol for general min-entropy sources
proposed by Chung~\textit{et.~al.}~\cite{ChungShiWu-PhysicalRandomnessExtractors-2014}.
Their idea is very similar to the presented protocol. The first
difference is that they use a different set of hash functions. Namely they are using a
de-randomized strong seeded extractor. Strong extractor can be seen as a set of hash functions $\mathcal H$
with a property that for any input source $X$ with enough min-entropy most of the hash functions
$h_i\in \mathcal H$ produce almost random outputs. In practice therefore random choice over $\mathcal H$
constitutes a good extractor. However, if you use every function from the set, you can guarantee
that for each $X$ with $H_\infty(X)\geq k$, at least one of them outputs $k$ fully
random bits. These bits can be used as an input
into a randomness expansion protocol to produce many random bits. Of course similarly to the presented
protocol the user doesn't know which hash function produces random outputs and therefore all of them
has to be used with their own independent and non-communicating device for the expansion part and
the outcomes of the expansion protocols need to be summed together (see Fig. \ref{fig:protocolExt}).
Since there is no guarantee that the outputs of two different hash functions are independent,
their result heavily leans on the equivalence lemma discussed in Subsection \ref{subsec:concat},
because we need a guarantee that the expansion protocol used as a part of the amplification
protocol is input secure.

\begin{figure}[!t]
\begin{framed}
\begin{center}
\textbf{Single device protocol for min-entropy amplification}
\end{center}
\begin{enumerate}
\item Draw a string $X = (R_{1}^{1},R_{2}^{1};R_{1}^{2},R_{2}^{2};\dots;R_{1}^{n}%
,R_{2}^{n})$ from a $(2n,2Rn)$ min-entropy source.
\item Use a single GHZ device to test the GHZ game $n$ times, with input
$R_1^i,R_2^i,R_1^i\oplus R_2^i \oplus 1$ in the $i$th round.
\item Test if $a_i\oplus b_i\oplus c_i = a_i\wedge y_i \wedge z_i$ in every round. If this doesn't hold, abort the protocol.
\item Conditioned on not aborting the protocol, the output bit is $O = Ext(a,b)$.
\end{enumerate}
\end{framed}
\vspace*{-0.4cm}
\caption{\small\it Protocol for a single device min-entropy amplification by Plesch and Pivoluska \cite{PleschPivoluska-Device-independentrandomnessamplification-2014}.
In the protocol, $Ext$ is a specific two source extractor, $a = (a_1,a_2,\dots,a_n)$ are outputs of the
first box and $b = (b_1,b_2,\dots,b_n)$ are outputs of the second box used in the GHZ test.}
\label{fig:singleDeviceProtocol}
\end{figure}

%%%%%%%%%%%%%%%%%%%%%%%%%
\subsection{Min-entropy amplification with a single device}
%%%%%%%%%%%%%%%%%%%%%%%%%

Both presented protocols from the previous subsection require number of devices
that grows with the (block) size of the input source. It is still an open question if protocols
for min-entropy amplification
with constant number of devices exist. An important step towards an answer was made by
us in \cite{PleschPivoluska-Device-independentrandomnessamplification-2014}.
We examined a protocol that uses a single GHZ testing device (see Fig. \ref{fig:GHZscenario})
taking input from a single min-entropy $(2n,2Rn)$ source, with min-entropy rate $R$.
The protocol is depicted in Fig. \ref{fig:singleDeviceProtocol}.

In the round $i$
two bits $R_{1}^{i}R_{2}^{i}$ are used to choose one out of four possible
input combinations. Let $r = r_{1}^{1}r_{2}^{1},...,r_{1}^{n}r_{2}^{n}$ be a
concrete realization of $X$. Let define $E_{i}(r)$ as
\begin{equation}
E_{i}(r)=-\log_2\left(\max_{kl\in\{0,1\}^{2}} P(R_{1}^{i}R_{2}^{i} = kl \vert R_{1}^{1}R_{2}%
^{2},...,R_{1}^{i-1}R_{2}^{i-1} = r_{1}^{1}r_{2}^{2},...,r_{1}^{i-1}%
r_{2}^{i-1})\right).
\end{equation}
This is an important parameter characterizing the amount of entropy in the input of
the $i^th$ round. Recall that devices can have memory and thus know the history of previous inputs and outputs.
Therefore, if $E_i(r)$ is less than $\log_2 (3)$, then, conditioned on the previous inputs, only three out of four
two-bit strings might appear as an input into the devices in the $i^{th}$ round of the protocol and the round
-- consisting of a GHZ test (see Fig. \ref{fig:GHZscenario}) -- can be won with 
probability $1$ with deterministic strategy.

We analyzed the protocol in two specific adversarial scenarios. The first scenario
analyzes the bias of the output bit in a case where the adversary doesn't want to risk
aborting the protocol at all. The second scenario analyzes the probability of aborting the protocol
in case the adversary wants the protocol to produce a constant bit at all risk.

In the light of the definition of $E_i(r)$ we can divide the rounds of the protocol into two types.
\begin{enumerate}
\item It holds that $E_{i}(r)>\log_{2}(3)$; in this case
$P(R_{1}^{i}R_{2}^{i}=x_{i}y_{i}|R_{1}^{1}R_{2}^{1},...,R_{1}^{j}R_{2}^{j}=r_{1}^{1}r_{2}%
^{1},...,r_{1}^{j}r_{2}^{j})>0$ for all four possible values of $x_{i}%
,y_{i}\in\{00,01,10,11\}$ and the only strategy succeeding in the GHZ test
with probability $1$ is the honest strategy of measuring GHZ
states. As discussed before, in this case bits $a_{i}$ and $b_{i}$
are uniformly distributed and independent of each other as well as all the
other previous inputs $x_{j},y_{j},z_{j},\quad j\in\{1,\dots,i\}$ and outputs
$a_{j},b_{j},c_{j},\quad j\in\{1,\dots,i-1\}$. Probability of any other
strategy to fulfill the win condition is bounded away from $1$.

\item It holds that $E_{i}(r)\leq\log_{2}(3)$; there exists a probability
distribution $P$, such that $P(R_{1}^{i}R_{2}^{i}=x_{i}y_{i}|R_{1}^{1}%
R_{2}^{2},...,R_{1}^{i-1}R_{2}^{i-1}=r_{1}^{1}r_{2}^{2},...,r_{1}^{i-1}%
r_{2}^{i-1})=0$ for at least one possible value of $x_{i},y_{i}\in
\{00,01,10,11\}$. In this case there exists a classical strategy (which can be
encoded in the common information $\lambda$) that succeeds in
the GHZ test  with probability $1$.
\end{enumerate}

In the first scenario, the adversary must program the boxes to play an honest strategy during the
rounds of type 1 in order not to abort the protocol, however in the rounds of type two,
the test can be successful even with a deterministic strategy. Worse still, because the boxes have
memory, the deterministic strategy can depend on the outputs and inputs of the previous rounds, and thus
compromise the randomness produced in the honest rounds.

As an example consider a very general scenario where the resulting bit $O$ is computed
as a sum of partial results from individual rounds $o_i$. Let $o_i$ be a result of a round $i$ of type 1,
arbitrarily random. Let $j$ by a subsequent round of type 2. Devices and source can agree in advance that
in round $j$ they will output results obtained in the round $i$ independently on the inputs. In such case
$o_j = o_i$ and $o_j \oplus o_i =0$ and thus perfectly deterministic.
The price to pay is the fact that the source had to select a specific outcome in the round $j$,
which decreases its entropy.

Therefore the analysis of the first scenario boils down to finding out to what
extent can the outcomes of the rounds
of type 2 negate any randomness produced in the rounds of type 1, given a specific entropy of the source. Assume that
$k$ out of $n$ rounds are of type 1. Without the loss of generality we can
assume that all $k$ rounds of type 1 are realized before $n-k$ rounds of type
2. In fact, this order of rounds gives the adversary the best possible
situation to react in rounds of type 2 on the randomness already produced in
rounds of type 1.

In such ordering we have:
\begin{align}
A  &  =\left(  \vec{A}_{k},f_{\lambda}^{k+1}(\vec{A}_{k}),\dots,f_{\lambda
}^{n}(\vec{A}_{k})\right)  ,\nonumber\\
B  &  =\left(  \vec{B}_{k},g_{\lambda}^{k+1}(\vec{B}_{k}),\dots,g_{\lambda
}^{n}(\vec{B}_{k})\right)  ,
\end{align}
where $\vec{A}_{k}=(a_{1},\dots,a_{k}),\vec{B}_{k}=b_{1},\dots,b_{k})$ are
outcomes of the rounds of type 1. Functions $f_{\lambda}^{j}$ and $g_{\lambda
}^{j}$, $k+1\leq j\leq n$ are particular strategies in round $j$ of type 2
attempting to increase the bias of the final bit, depending on the outcomes of
the rounds of type 1 and common information $\lambda$. Recall that $\lambda$
is the common information between the devices, source and the adversary. All
these parties can be correlated only via this random variable. In a regime
where the adversary doesn't want to risk getting caught at all, this means
that although vectors $A$ and $B$ are generally not independent, they can only
be dependent via $\lambda$. Therefore given $\lambda$, $A$ and $B$ are
independent and their respective conditional min-entropies are $H_{\infty
}(A|\lambda)=H_{\infty}(B|\lambda)=k$. Thus we can use any two source
extractor $Ext$ to extract the entropy present in $A$ and $B$. Since $A$ and
$B$ are independent given $\lambda$, it holds that $\left(  Ext(A,B)|\lambda
\right)  $ will be distributed according to the properties of the particular
extractor (close to being uniformly distributed given the previously shared information
$\lambda$).

We used Hadamard extractor (see Def. \ref{def:Hadamard}) in our analysis.
Recall that the distance of the output of the extractor, $\left(  Had(A,B)|\lambda\right)$,
is guaranteed to be
$2^{(n-H_{\infty}(A|\lambda
)-H_{\infty}(B|\lambda)-2)/2}$-close to a uniformly distributed bit as long as
$H_{\infty}(A|\lambda)+H_{\infty}(B|\lambda)\geq\frac{n}{2}$. Therefore as
long as $k>\frac{n}{2}$, regardless of the strategy employed in rounds of type
2, the output bit is, at least to some extent, random. Note here that the
requirement on $k$ could in principle be made lower by using different two-source extractors (see Subsection \ref{sec:2SExtractors}).
For example Bourgain's extractor
\cite{Bourgain} produces non-deterministic bit
as long as the sum of the entropies of $A$ and $B$ is greater than
$2n(1/2-\alpha)$ for some universal constant $\alpha$ and non-explicit
extractors can go as low as $k=O(\log n)$ \cite{Chor}.

In the light of the previous analysis we can obtain the upper bound for the
min-entropy rate, for which full cheating (maximum bias with probability of
getting caught equal to $0$) is possible. In order to do so, let us represent
$2n$ bit strings that the biased source $X$ can output with non-zero
probability by a graph tree of depth $n$, where

\begin{itemize}
\item each vertex has at most $4$ children and each edge from parent
to child is labeled by one of $\{00,01,10,11\}$,

\item each vertex represents prefix of a concrete realization of $r$ with
$r_{1}^{1},r_{2}^{1},\dots,r_{1}^{i},r_{2}^{i}$ encoded in the edge labels on
the path from the root of the tree to the given vertex,

\item each leaf represents a concrete realization of $r$.
\end{itemize}

Clearly, each vertex has at least $\left\lceil 2^{E_{i}(r)}\right\rceil$ children. A vertex
with $2^{E_{i}(r)}>3$ will be called an honest vertex, as in this vertex an honest
quantum strategy must be used, whereas all other vertices will be called
dishonest vertices.

\begin{figure}[tb]
\begin{center}
\includegraphics[width=9cm,clip]{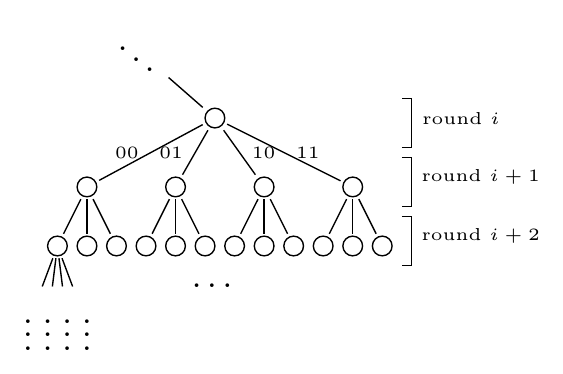}
\vspace*{-0.5cm}
\caption{\small\it An optimal tree
representation of the input weak source that potentially enables full cheating
in Pivoluska and Plesch protocol (see Fig. \ref{fig:singleDeviceProtocol}). Each node in the tree
level $i$ represents the probability distribution of the inputs for $i^{th}$ round.
If the node has $4$ children, all four inputs appear with positive probability
in $i^{th}$ round and the node is called honest.
To win this round with probability $1$ quantum strategy must be used.
Otherwise classical deterministic strategy exist and the node is called dishonest.
Optimal tree alternates between honest and dishonest vertices.}%
\label{randomtree}%
\end{center}
\end{figure}

To give an upper bound on the min-entropy for which the adversary can fully
cheat, we need to find a tree with a maximal number of leafs, such that for
each path from the root to the leaf the number of honest vertices is smaller
or equal to the number of dishonest vertices. Apparently such a tree can be
constructed by alternating between honest and dishonest vertices along
each path (see Fig. \ref{randomtree}); such tree has $\sqrt{12}^{n}$
leafs. Uniform distribution over $\sqrt{12}^{n}$ leaves maximizes the
min-entropy that can be used to realize such tree, yielding the min entropy
rate of $\mathbf{{R}}_{H}\mathbf{=}\log_{2}\mathbf{(}12\mathbf{)/}4$. For any
higher min-entropy rate, there exists a leaf such that the number of honest
vertices on the path from the root to the leaf is higher than the number of
dishonest vertices, therefore the adversary cannot know the outcome of the
protocol with probability $1$ without risking to be caught.

If the actual min-entropy rate of the source used is expressed as ${R=R}%
_{H}+\varepsilon$ with arbitrary $\varepsilon>0$, the probability of every
single leaf in the tree will be upper bounded by $p_{1}=2^{-2n\mathbf{R}%
}=\sqrt{12}^{-n}2^{-2\varepsilon n}$. In such a tree no more than $\sqrt
{12}^{n}$ leaves will be of a form that allows cheating without risking to be
caught, so the overall probability of cheating success is bounded from above by
\begin{equation}
p_{cheat}\leq2^{-2\varepsilon n},\label{p_cheat}%
\end{equation}
thus decreasing to zero exponentially with $n$. With this probability a bias of the output bit $\frac{1}{2}$
is achieved, whereas in all other bases the bias is $0$, so the resulting bias
of the output bit will be
\begin{equation}
bias(B)\leq2^{-(2\varepsilon n+1)}.\label{bias}%
\end{equation}

\begin{figure}[tb]
\begin{center}
\includegraphics[width=9cm,clip]{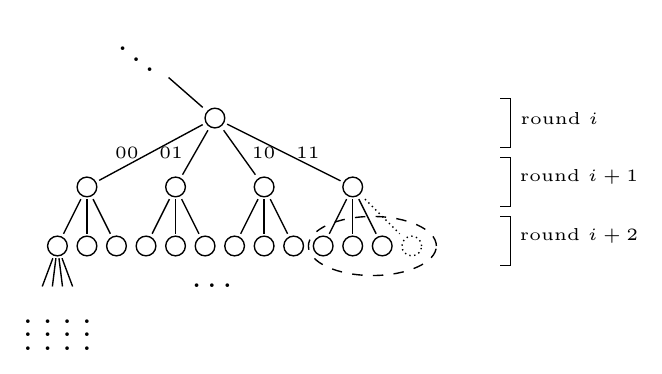}
\vspace*{-0.5cm}
\caption{\small\it By adding a fourth
child to a dishonest vertex, it is converted to an honest one in the zero-error scenario of the Pivoluska
and Plesch protocol (see Fig. \ref{fig:singleDeviceProtocol}). Therefore, all its children
(in the dashed oval)
have positive probability to appear -- therefore they all produce random outcomes in zero-error scenario.
In the risking scenario, the vertex stays dishonest,
the added (dotted) child leads to abortion of the protocol, however the other children remain deterministic.
}%
\label{randomtree_3}%
\end{center}
\end{figure}

It is worth to mention that
with growing min-entropy rate ${R}$ the number of cheatable leaves is in fact decreasing
and the actual cheating probability and consequently also the resulting bias will thus be strictly lower. This is due
to the fact that with every extra leave added to the probability tree, some
other leaves will convert from a fully biased to a perfectly random outcome.
This is due to the fact that the extra leaves can be added only by adding a
fourth child to a dishonest vertex, which is in this way converted to an
honest one, resulting into honestness of its leafs (for depiction see Fig.
\ref{randomtree_3}).

\begin{figure}[tb]
\begin{center}
\includegraphics[width=9cm,clip]{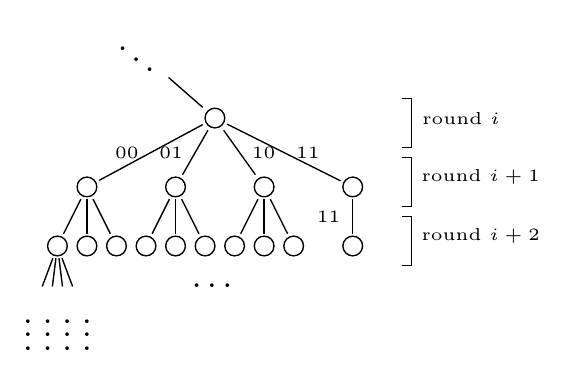}
\vspace*{-0.5cm}
\caption{\small\it The optimal achievable cheating strategy
in the Pivoluska and Plesch protocol (see Fig. \ref{fig:singleDeviceProtocol}) is
to repeat outcomes of the previous honest rounds. Random inputs
$\{001,010,100\}$ are interchangeable in the GHZ test, while input $\{111\}$ needs to be
exactly repeated in the dishonest round.}%
\label{randomtree2}%
\end{center}
\end{figure}

The rate ${R}_{H}=\log_{2}(12)/4$ is only an upper bound for the amount
of min-entropy for which the full cheating is possible. In fact, there is no
constructive attack that would be possible with such a min-entropy rate. As
we have also shown, the optimal implementable strategy is the one mentioned
earlier -- in every other round the boxes simple resend the outcomes of the
previous honest round. Such strategy can tolerate less min-entropy than
${R}_{H}$, as the dishonest vertex connected to it's honest parent by a
$11$ edge must have only one child, also labeled $11$ (see Fig.
\ref{randomtree2}).

Uniform distribution over the leaves of such tree has a
min--entropy rate
\begin{equation}
{R}_{\max}=1/4\log_{2}10,
\end{equation}
which is the highest rate for which full cheating is possible -- half of the
rounds are of type 1, quantum and honest, and half of the rounds are
of type 2, negating the bias of the output obtained of the previous runs. As soon as ${R}%
={R}_{\max}+\varepsilon$, the resulting bias
exponentially converges to zero with the same arguments as used for Hadamard extractor.

In a realistic scenario, if it would not be possible for Eve to limit the
inputs as needed for full cheating (\textit{i.e.} ${R}>\log_{2}(12)/4$), Eve
could simply try to use a classical strategy and guess the correct outcomes in
some of the honest rounds. Let us now analyze, what would be the probability
of successful cheating with such a strategy.

One can model such cheating strategy by adding extra leaves to the fully
cheatable tree (see Fig. \ref{randomtree_3}). This can be achieved by adding a fourth edge to some
of the dishonest vertices. In such round, Eve would simply use a classical
strategy, which is successful only in three out of four realizations.
Therefore, if this new added edge is actually realized by the random source, the
protocol fails by not satisfying the win condition of the GHZ game (Eq. (\ref{eq:GHZWinCon})).
The number of leaves for
which this strategy is successful stays exactly $\sqrt{12}^{n}$ and all the
other leaves lead to failure of the protocol. With a min-entropy rate
${R=R}_{H}+\varepsilon$ the minimal number of leaves in the tree is
$\sqrt{12}^{n}2^{2\varepsilon n}$, thus the probability of not failing the
protocol is $p_{guess}=2^{-2\varepsilon n}$.
Comparing to $p_{cheat}$ (\ref{p_cheat}) we see that the probability of successfully
cheating the protocol by risking is the same as the upper bound
of the probability of successful cheating of the protocol without risking.

%One might think about a more general attack where some reasonable bias is
%achieved with a very small probability of the protocol to fail by using a
%quantum strategy utilizing other than GHZ states. We limited ourselves to the
%analysis of quantum strategies that do not use entangled states across the
%individual rounds of the protocol. Using the SDP introduced in
%\cite{MironowiczPawlowski-Amplificationofarbitrarily-2013} we numerically showed that the bias
%of the output bit $b$ achievable by any quantum strategy, if the
%failure probability in every round is upper bounded by $p_{f}$, is upper bounded by $bias(b)^{2}\leq
%p_{f}$. Such strategy can be in fact realized
%by using states close to GHZ and by suitable changing the measurements used by
%the devices to POVM measurements with some pre-shared classical information.
%By utilizing this approach the adversary can slightly adjust results of the
%non-corrected quantum rounds with only a small probability of aborting the
%protocol. But due to the polynomial dependence between the bias and the
%failure probability, either the bias of the output bit or the probability of
%successful finishing of the protocol stays exponentially small.

%The real scenario, however, is somewhere in between the two presented
%scenarios. The adversary should be allowed a trade-off between the probability of aborting the protocol
%and the bias. Only preliminary analysis of such general attack has been carried out yet and it is
%still an open question whether this protocol is secure even in this case.

\setcounter{equation}{0} \setcounter{figure}{0} \setcounter{table}{0}\newpage
%%%%%%%%%%%%%%%%
%%%%%%%%%%%%%%%%%%%%%%%%%
\section{Conclusion}\label{sec:Conclusion}
%%%%%%%%%%%%%%%%%%%%%%%%%
%%%%%%%%%%%%%%%%
The aim of this paper was to present a thorough review on protocols for device independent randomness production.
These protocols provide a principal qualitative advantage in comparison with randomness generators
based on classical physical phenomena, as well as in comparison with protocols based on simple measurements of quantum states.

Randomness produced in a device independent way is certified by violation
of some Bell inequality. Such a violation guarantees that there is no deterministic model
for the observed correlations, which in turn guarantees that the outcomes
of measurement were not completely predetermined and therefore cannot be
fully correlated to any outside information. On the other hand, to certify violation of a Bell inequality
randomness is needed in the sense of ``free will'', the possibility to choose measurement settings independently
to the outside world. Thus even device-independent protocols are not able to produce randomness ``out of nowhere''.

Different protocols described in our paper fundamentally depend on the level and type of the accessible randomness. In general,
with better starting randomness simpler and more efficient protocols can be used, whereas in case where almost no randomness
is available, complicated protocols with many devices are necessary.

In all cases the general idea utilized in the protocols is the same. Available randomness is used to select measurement settings in a Bell type experiment and the resulting data are post-processed (possibly using part of the original data again)
into almost perfect randomness. This is defined as uniformly distributed bits that are uncorrelated to any other information
in the Universe.

Most of the protocols are technically rather simple and rely on measuring of few-partite entangled states in one of a few pre-determined basis states. This is in a sharp contradiction to most of other quantum protocols which rely on scaling of entanglement range with scaling of the problem. So the crutial obstacle preventing from application of device
independent randomness generators is the lack of a cheap, reliable on-demand source of multipartite entanglement and high efficiency
detectors.

%%%%%%%%%%%%%%%%%%%%%%%%%%%%%%%%%%%%%%%%%%%%%%%
\section*{Acknowledgments}
%%%%%%%%%%%%%%%%%%%%%%%%%%%%%%%%%%%%%%%%%%%%%%%
\addcontentsline{toc}{section}{Acknowledgement}
We would like to thank Jan Bouda,  Marcus Huber and Marcin Paw\l{}owski for valuable discussions.
This research was supported by the Czech Science
Foundation GA\v{C}R project P202/12/1142, EU project RAQUEL, as well as project VEGA 2/0043/15.

\newpage
\fancyhead[LO]{References}
\addcontentsline{toc}{section}{References}
\bibliography{DIrandomness}

\begin{thebibliography}{10}

\bibitem{-IDQuantique:-}
Id quantique:.
\newblock
  \href{http://www.idquantique.com/random-number-generators/products/products-overview.html}
  {Quantis}.

\bibitem{Ac'inMassarPironio-RandomnessversusNonlocality-2012}
A.~Ac\'in, S.~Massar, and S.~Pironio.
\newblock Randomness versus nonlocality and entanglement.
\newblock {\em Phys. Rev. Lett.}, 108:100402, Mar 2012.

\bibitem{AspectGrangierRoger-ExperimentalRealizationof-1982}
A.~Aspect, P.~Grangier, and G.~Roger.
\newblock Experimental realization of einstein-podolsky-rosen-bohm
  \textit{Gedankenexperiment} : A new violation of bell's inequalities.
\newblock {\em Phys. Rev. Lett.}, 49:91--94, 1982.

\bibitem{BancalScarani-MoreRandomnessFrom-2014}
J.-D. Bancal and V.~Scarani.
\newblock {More Randomness From Noisy Sources}.
\newblock In {\em 9th Conference on the Theory of Quantum Computation,
  Communication and Cryptography (TQC 2014)}, volume~27, pages 1--6, 2014.

\bibitem{BancalSheridanScarani-Morerandomnessfrom-2014}
J.-D. Bancal, L.~Sheridan, and V.~Scarani.
\newblock More randomness from the same data.
\newblock {\em New Journal of Physics}, 16(3):033011, 2014.

\bibitem{Barak2}
B.~Barak, G.~Kindler, R.~Shaltiel, B.~Sudakov, and A.~Wigderson.
\newblock Simulating independence: New constructions of condensers, ramsey
  graphs, dispersers, and extractors.
\newblock {\em J. ACM}, 57:20:1--20:52, 2010.

\bibitem{bb84}
C.~H. Bennett and G.~Brassard.
\newblock {Quantum Cryptography: Public Key Distribution and Coin Tossing}.
\newblock In {\em Proceedings of the IEEE International Conference on
  Computers, Systems and Signal Processing}, pages 175--179, 1984.

\bibitem{BosleyDodis-DoesPrivacyRequire-2006}
C.~Bosley and Y.~Dodis.
\newblock Does privacy require true randomness?
\newblock Cryptology ePrint Archive, Report 2006/283, 2006.
\newblock \url{http://eprint.iacr.org/}.

\bibitem{BoudaPawlowskiPivoluskaEtAl-Device-independentrandomnessextraction-2014}
J.~Bouda, M.~Paw\l{}owski, M.~Pivoluska, and M.~Plesch.
\newblock Device-independent randomness extraction from an arbitrarily weak
  min-entropy source.
\newblock {\em Phys. Rev. A}, 90:032313, 2014.

\bibitem{BoudaPivoluskaPlesch-ImprovingHadamardextractor-2012}
J.~Bouda, M.~Pivoluska, and M.~Plesch.
\newblock Improving the hadamard extractor.
\newblock {\em Theoretical Computer Science}, 459(0):69 -- 76, 2012.

\bibitem{BoudaPivoluskaPleschEtAl-Weakrandomnessseriously-2012}
J.~Bouda, M.~Pivoluska, M.~Plesch, and C.~Wilmott.
\newblock Weak randomness seriously limits the security of quantum key
  distribution.
\newblock {\em Phys. Rev. A}, 86:062308, 2012.

\bibitem{Bourgain}
J.~Bourgain.
\newblock More on the sum-product phenomenon in prime fields and its
  applications.
\newblock {\em International Journal of Number Theory}, 1:1--32, 2005.

\bibitem{BrandaoHarrow-QuantumDeFinetti-2013}
F.~G.~S.~L. Brand{\~a}o and A.~W. Harrow.
\newblock Quantum de finetti theorems under local measurements with
  applications.
\newblock In {\em Proceedings of the Forty-fifth Annual ACM Symposium on Theory
  of Computing}, STOC '13, pages 861--870, 2013.

\bibitem{BrandaoRamanathanGrudkaEtAl-RobustDevice-IndependentRandomness-2013}
F.~G.~S.~L. Brand{\~a}o, R.~{Ramanathan}, A.~{Grudka}, K.~{Horodecki},
  M.~{Horodecki}, and P.~{Horodecki}.
\newblock {Robust Device-Independent Randomness Amplification with Few
  Devices}.
\newblock 2013, quant-ph/1310.4544.

\bibitem{BraunsteinCaves-ChainedBellInequalities-1989}
S.~L. Braunstein and C.~M. Caves.
\newblock Chained bell inequalities.
\newblock In {\em Bell's Theorem, Quantum Theory and Conceptions of the
  Universe}, volume~37 of {\em Fundamental Theories of Physics}, pages 27--36.
  1989.

\bibitem{BrunnerCavalcantiPironioEtAl-Bellnonlocality-2014}
N.~Brunner, D.~Cavalcanti, S.~Pironio, V.~Scarani, and S.~Wehner.
\newblock Bell nonlocality.
\newblock {\em Rev. Mod. Phys.}, 86:419--478, Apr 2014.

\bibitem{CarterWegman-Universalclassesof-1979}
J.~L. Carter and M.~N. Wegman.
\newblock Universal classes of hash functions.
\newblock {\em Journal of Computer and System Sciences}, 18(2):143 -- 154,
  1979.

\bibitem{Chor}
B.~Chor and O.~Goldreich.
\newblock Unbiased bits from sources of weak randomness and probabilistic
  communication complexity.
\newblock {\em SIAM J. Comput.}, 17:230--261, 1988.

\bibitem{ChungShiWu-PhysicalRandomnessExtractors-2014}
K.-M. {Chung}, Y.~{Shi}, and X.~{Wu}.
\newblock {Physical Randomness Extractors}.
\newblock 2014, quant-ph/1402.4797.

\bibitem{ClauserHorneShimonyEtAl-ProposedExperimentto-1969}
J.~F. Clauser, M.~A. Horne, A.~Shimony, and R.~A. Holt.
\newblock Proposed experiment to test local hidden-variable theories.
\newblock {\em Phys. Rev. Lett.}, 23:880--884, 1969.

\bibitem{ColbeckKent-Privaterandomnessexpansion-2011}
R.~{Colbeck} and A.~{Kent}.
\newblock {Private randomness expansion with untrusted devices}.
\newblock {\em Journal of Physics A Mathematical General}, 44(9):095305, 2011.

\bibitem{ColbeckRenner-Freerandomnesscan-2012}
R.~{Colbeck} and R.~{Renner}.
\newblock {Free randomness can be amplified}.
\newblock {\em Nature Physics}, 8:450--454, 2012.

\bibitem{CoudronVidickYuen-RobustRandomnessAmplifiers:-2013}
M.~Coudron, T.~Vidick, and H.~Yuen.
\newblock Robust randomness amplifiers: Upper and lower bounds.
\newblock In {\em Approximation, Randomization, and Combinatorial Optimization.
  Algorithms and Techniques}, volume 8096, pages 468--483. 2013.

\bibitem{CoudronYuen-InfiniteRandomnessExpansion-2014}
M.~Coudron and H.~Yuen.
\newblock Infinite randomness expansion with a constant number of devices.
\newblock In {\em Proceedings of the 46th Annual ACM Symposium on Theory of
  Computing}, STOC '14, pages 427--436, 2014.

\bibitem{DePortmannVidickEtAl-Trevisan'sExtractorin-2012}
A.~De, C.~Portmann, T.~Vidick, and R.~Renner.
\newblock Trevisan's extractor in the presence of quantum side information.
\newblock {\em SIAM Journal on Computing}, 41(4):915--940, 2012.

\bibitem{DeVidick-Near-optimalExtractorsAgainst-2010}
A.~De and T.~Vidick.
\newblock Near-optimal extractors against quantum storage.
\newblock In {\em Proceedings of the Forty-second ACM Symposium on Theory of
  Computing}, STOC '10, pages 161--170, 2010.

\bibitem{Dodis}
Y.~Dodis, A.~Elbaz, R.~Oliveira, and R.~Raz.
\newblock {Improved randomness extraction from two independent sources}.
\newblock In {\em Approximation, Randomization, and Combinatorial Optimization.
  Algorithms and Techniques}, volume {3122}, pages {334--344}, {2004}.

\bibitem{Dodis2}
Y.~Dodis and R.~Oliveira.
\newblock On extracting private randomness over a public channel.
\newblock In {\em Approximation, Randomization, and Combinatorial Optimization.
  Algorithms and Techniques}, volume 2764, pages 827--836. 2003.

\bibitem{DodisOngPrabhakaranEtAl-(im)possibilityofcryptography-2004}
Y.~Dodis, Shien~Jin Ong, M.~Prabhakaran, and A.~Sahai.
\newblock On the (im)possibility of cryptography with imperfect randomness.
\newblock In {\em Foundations of Computer Science, 2004. Proceedings. 45th
  Annual IEEE Symposium on}, pages 196--205, 2004.

\bibitem{DodisSpencer-(non)universalityofone-time-2002}
Y.~Dodis and J.~Spencer.
\newblock On the (non)universality of the one-time pad.
\newblock In {\em Foundations of Computer Science, 2002. Proceedings. The 43rd
  Annual IEEE Symposium on}, pages 376--385, 2002.

\bibitem{EinsteinPodolskyRosen-CanQuantum-MechanicalDescription-1935}
A.~Einstein, B.~Podolsky, and N.~Rosen.
\newblock Can quantum-mechanical description of physical reality be considered
  complete?
\newblock {\em Phys. Rev.}, 47:777--780, 1935.

\bibitem{Ekert1991}
A.~K. Ekert.
\newblock Quantum cryptography based on bell's theorem.
\newblock {\em Phys. Rev. Lett.}, 67:661--663, 1991.

\bibitem{FehrGellesSchaffner-Securityandcomposability-2013}
S.~Fehr, R.~Gelles, and C.~Schaffner.
\newblock Security and composability of randomness expansion from bell
  inequalities.
\newblock {\em Phys. Rev. A}, 87:012335, 2013.

\bibitem{Fine-HiddenVariablesJoint-1982}
A.~Fine.
\newblock Hidden variables, joint probability, and the bell inequalities.
\newblock {\em Phys. Rev. Lett.}, 48:291--295, 1982.

\bibitem{FrauchigerRennerTroyer-Truerandomnessfrom-2013}
D.~{Frauchiger}, R.~{Renner}, and M.~{Troyer}.
\newblock {True randomness from realistic quantum devices}.
\newblock 2013, quant-ph/1311.4547.

\bibitem{GallegoMasanesEtAl-Fullrandomnessfrom-2013}
R.~{Gallego}, L.~{Masanes}, G.~{de la Torre}, C.~{Dhara}, L.~{Aolita}, and
  A.~{Ac{\'{\i}}n}.
\newblock {Full randomness from arbitrarily deterministic events}.
\newblock {\em Nature Communications}, 4, 2013.

\bibitem{GavinskyKempeKerenidisEtAl-ExponentialSeparationsOne-way-2007}
D.~Gavinsky, J.~Kempe, I.~Kerenidis, R.~Raz, and R.~de~Wolf.
\newblock Exponential separations for one-way quantum communication complexity,
  with applications to cryptography.
\newblock In {\em Proceedings of the Thirty-ninth Annual ACM Symposium on
  Theory of Computing}, STOC '07, pages 516--525, 2007.

\bibitem{GreenbergerHorneShimonyEtAl-Bellstheoremwithout-1990}
D.~M. Greenberger, M.~A. Horne, A.~Shimony, and A.~Zeilinger.
\newblock Bell's theorem without inequalities.
\newblock {\em American Journal of Physics}, 58(12):1131--1143, 1990.

\bibitem{GrudkaHorodeckiHorodeckiEtAl-Freerandomnessamplification-2014}
A.~Grudka, K.~Horodecki, M.~Horodecki, Pa. Horodecki, Ma. Paw\l{}owski, and
  R.~Ramanathan.
\newblock Free randomness amplification using bipartite chain correlations.
\newblock {\em Phys. Rev. A}, 90:032322, 2014.

\bibitem{Gruska-QuantumComputing-1999}
J.~Gruska.
\newblock {\em Quantum Computing}.
\newblock Osborne/McGraw-Hill, 1999.

\bibitem{GuruswamiUmansVadhan-Unbalancedexpandersand-2009}
V.~Guruswami, C.~Umans, and S.~Vadhan.
\newblock Unbalanced expanders and randomness extractors from parvaresh--vardy
  codes.
\newblock {\em J. ACM}, 56:20:1--20:34, 2009.

\bibitem{HuberPawlowski-Weakrandomnessin-2013}
M.~Huber and M.~Paw\l{}owski.
\newblock Weak randomness in device-independent quantum key distribution and
  the advantage of using high-dimensional entanglement.
\newblock {\em Phys. Rev. A}, 88:032309, 2013.

\bibitem{Jakobsson-TheoryMethodsand-2014}
Krister~Sune Jakobsson.
\newblock Theory, methods and tools for statistical testing of pseudo and
  quantum random number generators.
\newblock Master's thesis, 2014.
\newblock \url{http://www.icg.isy.liu.se/publications/}.

\bibitem{Kamp2}
J.~Kamp, A.~Rao, S.~Vadhan, and D.~Zuckerman.
\newblock Deterministic extractors for small-space sources.
\newblock {\em J. Comput. Syst. Sci.}, 77:191--220, 2011.

\bibitem{KasherKempe-Two-SourceExtractorsSecure-2010}
R.~Kasher and J.~Kempe.
\newblock Two-source extractors secure against quantum adversaries.
\newblock In {\em Approximation, Randomization, and Combinatorial Optimization.
  Algorithms and Techniques}, volume 6302 of {\em Lecture Notes in Computer
  Science}, pages 656--669. 2010.

\bibitem{KoflerPaterekBrukner-Experimenter'sfreedomin-2006}
J.~Kofler, T.~Paterek, and C.~Brukner.
\newblock Experimenter's freedom in bell's theorem and quantum cryptography.
\newblock {\em Phys. Rev. A}, 73:022104, 2006.

\bibitem{KonigMaurerRenner-powerofquantum-2005}
R.~Konig, U.~Maurer, and R.~Renner.
\newblock On the power of quantum memory.
\newblock {\em Information Theory, IEEE Transactions on}, 51(7):2391--2401,
  2005.

\bibitem{KonigRennerSchaffner-OperationalMeaningof-2009}
R.~K\"{o}nig, R.~Renner, and C.~Schaffner.
\newblock The operational meaning of min- and max-entropy.
\newblock {\em IEEE Trans. Inf. Theor.}, 55(9):4337--4347, 2009.

\bibitem{LawsonLindenPopescu-Biasednonlocalquantum-2010}
T.~{Lawson}, N.~{Linden}, and S.~{Popescu}.
\newblock {Biased nonlocal quantum games}.
\newblock {\em ArXiv e-prints}, 2010, quant-ph/1011.6245.

\bibitem{LenstraHughesAugierEtAl-Ronwaswrong-2012}
A.~K. Lenstra, J.~P. Hughes, M.~Augier, J.~W. Bos, T.~Kleinjung, and
  C.~Wachter.
\newblock Ron was wrong, whit is right.
\newblock Cryptology ePrint Archive, Report 2012/064, 2012.
\newblock \url{http://eprint.iacr.org/}.

\bibitem{LiVitnyi-IntroductiontoKolmogorov-2008}
M.~Li and P.~M.~B. Vitnyi.
\newblock {\em An Introduction to Kolmogorov Complexity and Its Applications}.
\newblock Springer Publishing Company, Incorporated, 3 edition, 2008.

\bibitem{LuReingoldVadhanEtAl-Extractors:optimalup-2003}
C.-J. Lu, O.~Reingold, S.~Vadhan, and A.~Wigderson.
\newblock Extractors: optimal up to constant factors.
\newblock In {\em Proceedings of the thirty-fifth annual ACM symposium on
  Theory of computing}, STOC '03, pages 602--611, 2003.

\bibitem{Masanes-Universallycomposableprivacy-2009}
L.~Masanes.
\newblock Universally composable privacy amplification from causality
  constraints.
\newblock {\em Physical review letters}, 102(14):140501, 2009.

\bibitem{MayersYao-Selftestingquantum-2004}
D.~Mayers and A.~Yao.
\newblock Self testing quantum apparatus.
\newblock {\em Quantum Information and Computation}, 4:273--268, 2004.

\bibitem{McInnes}
J.~L. McInnes and B.~Pinkas.
\newblock On the impossibility of private key cryptography with weakly random
  keys.
\newblock In {\em Proceedings of the 10th Annual International Cryptology
  Conference on Advances in Cryptology}, CRYPTO '90, pages 421--435, 1991.

\bibitem{McKague-Self-testinggraphstates-2010}
M.~{McKague}.
\newblock {Self-testing graph states}.
\newblock 2010, quant-ph/1010.1989.

\bibitem{MenezesHandbook}
A.~J. Menezes, S~A. Vanstone, and P.~C.~V. Oorschot.
\newblock {\em Handbook of Applied Cryptography}.
\newblock CRC Press, Inc., Boca Raton, FL, USA, 1996.

\bibitem{Mermin1990}
N.~D. Mermin.
\newblock Extreme quantum entanglement in a superposition of macroscopically
  distinct states.
\newblock {\em Phys. Rev. Lett.}, 65:1838--1840, 1990.

\bibitem{MillerShi-Optimalrobustquantum-2012}
C.~A. {Miller} and Y.~{Shi}.
\newblock {Optimal robust quantum self-testing by binary nonlocal XOR games}.
\newblock {\em ArXiv e-prints}, 2012, quant-ph/1207.1819.

\bibitem{MillerShi-RobustProtocolsSecurely-2014}
C.~A. Miller and Y.~Shi.
\newblock Robust protocols for securely expanding randomness and distributing
  keys using untrusted quantum devices.
\newblock In {\em Proceedings of the 46th Annual ACM Symposium on Theory of
  Computing}, STOC '14, pages 417--426, 2014.

\bibitem{MironowiczPawlowski-Amplificationofarbitrarily-2013}
P.~{Mironowicz} and M.~{Paw\l{}owski}.
\newblock {Amplification of arbitrarily weak randomness}.
\newblock 2013, quant-ph/1301.7722.

\bibitem{MotwaniRaghavan-Randomizedalgorithms-1995}
R.~Motwani and P.~Raghavan.
\newblock {\em Randomized algorithms}.
\newblock Cambridge University Press, New York, NY, USA, 1995.

\bibitem{NaorNaor-Small-BiasProbabilitySpaces:-1993}
J.~Naor and M.~Naor.
\newblock Small-bias probability spaces: Efficient constructions and
  applications.
\newblock {\em SIAM J. Comput.}, 22(4):838--856, 1993.

\bibitem{Navascues2008}
M.~{Navascu{\'e}s}, S.~{Pironio}, and A.~{Ac{\'{\i}}n}.
\newblock {A convergent hierarchy of semidefinite programs characterizing the
  set of quantum correlations}.
\newblock {\em New Journal of Physics}, 10(7):073013, 2008.

\bibitem{NielsenChuang-QuantumComputationand-2004}
M.~A. Nielsen and I.~L. Chuang.
\newblock {\em {Quantum Computation and Quantum Information (Cambridge Series
  on Information and the Natural Sciences)}}.
\newblock Cambridge University Press, 1 edition, 2000.

\bibitem{Nieto-SillerasPironioSilman-Usingcompletemeasurement-2014}
O.~Nieto-Silleras, S.~Pironio, and J.~Silman.
\newblock Using complete measurement statistics for optimal device-independent
  randomness evaluation.
\newblock {\em New Journal of Physics}, 16(1):013035, 2014.

\bibitem{Nisan}
N.~Nisan and A.~Ta-Shma.
\newblock Extracting randomness: a survey and new constructions.
\newblock {\em J. Comput. Syst. Sci.}, 58:148--173, 1999.

\bibitem{PawlowskiHorodeckiHorodeckiEtAl-QuantumCryptographyand-2010}
M.~Paw\l{}owski, K.~Horodecki, P.~Horodecki, and R.~Horodecki.
\newblock {\em Quantum Cryptography and Computing}, chapter Local bounds for
  general Bell inequalities with the reduced entropy of the settings, pages
  224--230.
\newblock IOS press, 2010.

\bibitem{Peres-IteratingVonNeumann's-1992}
Y.~Peres.
\newblock Iterating von neumann's procedure for extracting random bits.
\newblock {\em The Annals of Statistics}, 20(1):pp. 590--597, 1992.

\bibitem{PironioAc'inMassarEtAl-Randomnumberscertified-2010}
S.~{Pironio}, A.~{Ac{\'{\i}}n}, S.~{Massar}, A.~B. {de La Giroday}, D.~N.
  {Matsukevich}, P.~{Maunz}, S.~{Olmschenk}, D.~{Hayes}, L.~{Luo}, T.~A.
  {Manning}, and C.~{Monroe}.
\newblock {Random numbers certified by Bell's theorem}.
\newblock {\em Nature}, 464:1021--1024, 2010.

\bibitem{PironioMassar-Securityofpractical-2013}
S.~Pironio and S.~Massar.
\newblock Security of practical private randomness generation.
\newblock {\em Phys. Rev. A}, 87:012336, 2013.

\bibitem{PleschPivoluska-Device-independentrandomnessamplification-2014}
M.~Plesch and M.~Pivoluska.
\newblock Device-independent randomness amplification with a single device.
\newblock {\em Physics Letters A}, 378(40):2938 -- 2944, 2014.

\bibitem{RadhakrishnanTa-Shma-BoundsDispersersExtractors-2000}
J.~Radhakrishnan and A.~Ta-Shma.
\newblock Bounds for dispersers, extractors, and depth-two superconcentrators.
\newblock {\em SIAM JOURNAL ON DISCRETE MATHEMATICS}, 13:2000, 2000.

\bibitem{Raz}
R.~Raz.
\newblock Extractors with weak random seeds.
\newblock In {\em Proceedings of the thirty-seventh annual ACM symposium on
  Theory of computing}, STOC '05, pages 11--20, 2005.

\bibitem{ReichardtUngerVazirani-ClassicalLeashQuantum-2013}
B.~W. Reichardt, F.~Unger, and U.~Vazirani.
\newblock A classical leash for a quantum system: Command of quantum systems
  via rigidity of chsh games.
\newblock In {\em Proceedings of the 4th Conference on Innovations in
  Theoretical Computer Science}, ITCS '13, pages 321--322, 2013.

\bibitem{Renner2005a}
R.~Renner.
\newblock {\em Security of Quantum Key Distribution}.
\newblock PhD thesis, quant-ph/0512258v2, 2005.

\bibitem{RennerWolf-Unconditionalauthenticityand-2003}
R.~Renner and S.~Wolf.
\newblock Unconditional authenticity and privacy from an arbitrarily weak
  secret.
\newblock In {\em In Proc. CRYPTO'03}, pages 78--95. Springer-Verlag, 2003.

\bibitem{SanthaVazirani-Generatingquasi-randomsequences-1986}
M.~Santha and U.~Vazirani.
\newblock Generating quasi-random sequences from semi-random sources.
\newblock {\em Journal of Computer and System Sciences}, 33(1):75 -- 87, 1986.

\bibitem{Scarani-device-independentoutlookquantum-2012}
V.~Scarani.
\newblock The device-independent outlook on quantum physics (lecture notes on
  the power of bell's theorem).
\newblock {\em Acta Physica Sovaca}, 62, 2012.

\bibitem{Shaltiel1}
R.~Shaltiel.
\newblock Recent developments in explicit constructions of extractors.
\newblock {\em Bulletin of the EATCS}, (77):67--95, 2002.

\bibitem{Shannon49}
C.~E. Shannon.
\newblock {Communication Theory of Secrecy Systems}.
\newblock {\em Bell Systems Technical Journal}, 28:656--715, 1949.

\bibitem{SilmanPironioMassar-Device-IndependentRandomnessGeneration-2013}
J.~Silman, S.~Pironio, and S.~Massar.
\newblock Device-independent randomness generation in the presence of weak
  cross-talk.
\newblock {\em Phys. Rev. Lett.}, 110:100504, 2013.

\bibitem{Solcgravea-TestingofQuantum-2010}
R.~Solc\`{a}.
\newblock {\em Testing of a Quantum Random Number Generator}.
\newblock PhD thesis, Institute for Theoretical Physics, ETH Z\"{u}rich, 2010.

\bibitem{SrinivasanZuckerman-Computingwithvery-1994}
A.~Srinivasan and D.~Zuckerman.
\newblock Computing with very weak random sources.
\newblock In {\em Foundations of Computer Science, 1994 Proceedings., 35th
  Annual Symposium on}, pages 264--275, 1994.

\bibitem{ThinhSheridanScarani-Belltestswith-2013}
L.~P. Thinh, L.~Sheridan, and V.~Scarani.
\newblock Bell tests with min-entropy sources.
\newblock {\em Phys. Rev. A}, 87:062121, 2013.

\bibitem{TomamichelRennerSchaffnerEtAl-LeftoverHashingagainst-2010}
M.~Tomamichel, R.~Renner, C.~Schaffner, and A.~Smith.
\newblock Leftover hashing against quantum side information.
\newblock In {\em Information Theory Proceedings (ISIT), 2010 IEEE
  International Symposium on}, pages 2703--2707, 2010.

\bibitem{Trevisan}
L.~Trevisan and Vadhan. S.
\newblock Extracting randomness from samplable distributions.
\newblock {\em Foundations of Computer Science, Annual IEEE Symposium on},
  0:32, 2000.

\bibitem{Cirel'son-Quantumgeneralizationsof-1980}
B.~S. Tsirelson.
\newblock Quantum generalizations of bell's inequality.
\newblock {\em Letters in Mathematical Physics}, 4(2):93--100, 1980.

\bibitem{Vazirani}
U.~Vazirani.
\newblock Strong communication complexity or generating quasi-random sequences
  from two communicating semi-random sources.
\newblock {\em Combinatorica}, 7:375--392, 1987.

\bibitem{VaziraniVazirani-RandomPolynomialTime-1985}
U.~Vazirani and V.~Vazirani.
\newblock Random polynomial time is equal to slightly-random polynomial time.
\newblock In {\em FOCS'85}, pages 417--428, 1985.

\bibitem{vazirani2012certifiable}
U.~Vazirani and T.~Vidick.
\newblock Certifiable quantum dice: or, true random number generation secure
  against quantum adversaries.
\newblock In {\em Proceedings of the 44th symposium on Theory of Computing},
  pages 61--76. ACM, 2012.

\bibitem{vonNeumann}
J.~von Neumann.
\newblock Various techniques used in connection with random digits.
\newblock {\em Applied Math Series}, 12:36--38, 1951.

\bibitem{Zuckerman-SimulatingBPPusing-1996}
D.~Zuckerman.
\newblock Simulating bpp using a general weak random source.
\newblock {\em Algorithmica}, 16(4-5):367--391, 1996.

\end{thebibliography}
\bibliographystyle{hplain}

\end{document}